\documentclass{aa} 
\usepackage{graphicx}
\usepackage{threeparttable}
\usepackage[varg]{txfonts}
\usepackage{lipsum}
\usepackage{subcaption}         
\usepackage{lscape}
\usepackage{placeins}
\usepackage[colorlinks=true, citecolor=blue, linkcolor=black]{hyperref}
\usepackage{amsmath}
\usepackage{tablefootnote}
\usepackage{caption}
\usepackage{siunitx}
\usepackage{enumitem}
\usepackage{array}
\usepackage[dvipsnames]{xcolor}
\usepackage[normalem]{ulem}
\usepackage{placeins}

\begin{document} 
\title{\LARGE\textbf{Exploring the ultrahot Neptune LTT~9779~b with CRIRES$^+$}\\[1ex]
\Large Non-detection of helium transmission spectra and\\ detection of $\mathrm{H_2O}$ on the dayside}

\author{
    S.~Liu \inst{\ref{instUSTC}}
    \and F.~Yan\thanks{Corresponding author: \email{yanfei@ustc.edu.cn}} \inst{\ref{instUSTC}}
    \and L.~Nortmann \inst{\ref{instIAG}}
    \and B.~Guo \inst{\ref{instUSTC}}
    \and D.~Cont \inst{\ref{instLMU}, \ref{instExzO}}
    \and F.~Lesjak \inst{\ref{instAIP}}
    \and D.~Shulyak \inst{\ref{instIAA}}
    \and A.~Reiners \inst{\ref{instIAG}}
    \and N.~Piskunov \inst{\ref{instUppsala}}
    \and A.~P.~Hatzes \inst{\ref{instTLS}}
    \and U.~Seemann \inst{\ref{instESO}}
    \and A.~Lavail \inst{\ref{instIRAP}}
    \and A.~D.~Rains \inst{\ref{instIAC}, \ref{instUppsala}}
    \and H.\,L.~Ruh \inst{\ref{instIAG}}
    \and L.~Boldt-Christmas \inst{\ref{instUppsala}}
    \and M.~Rengel \inst{\ref{instMPS}}
    \and O.~Kochukhov \inst{\ref{instUppsala}}
    \and T.~Marquart \inst{\ref{instUppsala}}
}

\authorrunning{S.~Liu et al.}

\institute{
    Department of Astronomy, University of Science and Technology of China,
    Hefei 230026, People's Republic of China\label{instUSTC}
    \and
    Institut f\"ur Astrophysik und Geophysik,
    Georg-August-Universit\"at G\"ottingen,
    Friedrich-Hund-Platz 1, 37077 G\"ottingen, Germany\label{instIAG}
    \and
    Universit\"ats-Sternwarte,
    Ludwig-Maximilians-Universit\"at M\"unchen,
    Scheinerstrasse 1, 81679 M\"unchen, Germany\label{instLMU}
    \and
    Exzellenzcluster Origins,
    Boltzmannstrasse 2, 85748 Garching bei M\"unchen, Germany\label{instExzO}
    \and
    Leibniz-Institut f\"ur Astrophysik Potsdam (AIP),
    An der Sternwarte 16, 14482 Potsdam, Germany\label{instAIP}
    \and
    Instituto de Astrof{\'i}sica de Andaluc{\'i}a (IAA-CSIC),
    Glorieta de la Astronom{\'i}a s/n,
    18008 Granada, Spain\label{instIAA}
    \and
    Institutionen f\"or fysik och astronomi,
    Uppsala universitet, Box 516, 75120 Uppsala, Sweden\label{instUppsala}
    \and
    Th\"uringer Landessternwarte Tautenburg,
    Sternwarte 5, 07778 Tautenburg, Germany\label{instTLS}
    \and
    European Southern Observatory,
    Karl-Schwarzschild-Str. 2,
    85748 Garching bei M\"unchen, Germany\label{instESO}
    \and
    Institut de Recherche en Astrophysique et Plan\'etologie,
    Universit\'e de Toulouse, CNRS, IRAP/UMR 5277,
    14 avenue Edouard Belin, 31400 Toulouse, France\label{instIRAP}
    \and
    Instituto de Astrof{\'i}sica,
    Pontificia Universidad Cat{\'o}lica de Chile,
    Av. Vicu{\~n}a Mackenna 4860,
    782-0436 Macul, Santiago, Chile\label{instIAC}
    \and
    Max-Planck-Institut f\"ur Sonnensystemforschung,
    Justus-von-Liebig-Weg 3,
    37077 G\"ottingen, Germany\label{instMPS}
}

\date{Received 13 August 2026 / Accepted 28 September 2026}

\abstract
{Ultrahot Neptunes reside in the most extreme irradiation regime of the hot Neptune desert and remain challenging targets for atmospheric characterization because of their intrinsically low planet-to-star flux ratios. We present high-resolution transit and dayside thermal emission spectroscopy of the ultrahot Neptune LTT~9779~b obtained with CRIRES$^+$, aimed at probing both atmospheric escape and dayside molecular emission. No significant absorption is detected in the metastable \ion{He}{i} $\lambda 1083$\,nm triplet during transit, yielding a $3\sigma$ upper limit of 0.45\% on the excess helium absorption depth and supporting previous non-detections that suggest weak ongoing atmospheric escape. For the dayside emission spectra in the $K$ band, our cross-correlation analysis reveals an $\mathrm{H_2O}$ absorption signal and a tentative $\mathrm{CO}$ feature, with the combined signal reaching a detection significance of $4.8\sigma$. Both the CRIRES$^+$ and previous JWST data favor a non-inverted temperature structure, providing no evidence of a thermal inversion on the planetary dayside. We further perform a joint atmospheric retrieval combining the CRIRES$^+$ high-resolution spectra with JWST/NIRISS SOSS secondary-eclipse observations. The joint retrieval yields a low $\mathrm{H_2O}$ abundance, while the $\mathrm{CO}$ abundance remains weakly constrained. The inferred planetary equatorial rotational velocity, $v_{\rm eq}=4.9^{+4.4}_{-3.0}~\mathrm{km\,s^{-1}}$, is consistent within $1\sigma$ with the synchronous rotation velocity expected for a tidally locked planet ($\sim2.8~\mathrm{km\,s^{-1}}$). However, the large uncertainties prevent us from placing strong constraints on the planet's rotation state. Future phase-resolved CRIRES$^+$ observations with broader orbital-phase coverage and higher signal-to-noise ratios would improve the abundance constraints and probe the atmospheric dynamics of this rare ultrahot Neptune.}


\titlerunning{Exploring the ultrahot Neptune LTT~9779~b}
\maketitle
\nolinenumbers

\section{Introduction}
\begin{figure}[t]
    \centering
    \includegraphics[width=\columnwidth]{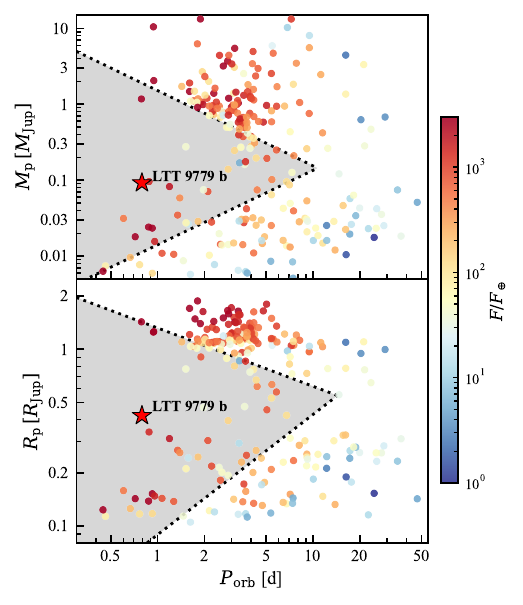}
    \caption{
    Distribution of confirmed exoplanets with masses and radii measured to better than 15\% in the period--mass plane (top) and period--radius plane (bottom). The shaded gray regions denote the hot Neptune desert, with the dotted lines indicating the empirical boundaries defined by \citet{ND_boundary}. LTT~9779~b is marked by the red star. Points are color-coded by the incident stellar flux relative to Earth, $F/F_{\oplus}$.
    }
    \label{fig:ND_population}
\end{figure}

With the growing number of discovered exoplanets, several distinct demographic structures have emerged. Among them, one of the most intriguing is the ``hot Neptune desert,'' a region where Neptune-sized planets are markedly absent at orbital periods of $\lesssim 4$ days \citep{ND1, ND_boundary}. Its location in the period--radius and period--mass planes is illustrated in Fig.~\ref{fig:ND_population}. The small number of planets in this region cannot be explained solely by observational biases \citep{ND3, ND4}. Instead, it is widely interpreted as evidence that strong evolutionary processes actively reshape the population boundaries of close-in planets. Several physical mechanisms have been proposed to carve out this desert:
(1) photoevaporation, where intense stellar X-ray and extreme-ultraviolet (XUV) irradiation heats and expands the upper atmosphere, driving substantial mass loss over timescales of millions to billions of years \citep{ND_Photoevaporation1, ND_Photoevaporation2, ND_Photoevaporation3};
(2) high-eccentricity migration and tidal disruption, during which planets arriving on highly eccentric orbits experience strong tidal forces that strip their envelopes or lead to partial disruption as the orbit circularizes \citep{ND_Eccentricity, ND_Photoevaporation3}; and
(3) Roche-lobe overflow, where atmospheric expansion pushes the upper layers beyond the Roche potential surface, resulting in sustained or runaway mass transfer from the planet \citep{ND_Roche1, ND_Roche2, ND_Roche3}. These mechanisms remain incompletely understood in their relative roles in sculpting the desert.

However, the desert is not strictly empty, with several Neptune-mass planets identified within this region \citep{ND_Planet1, ND_Planet2, ND_Planet3, ND_Planet4, LTT_Discovery, ND_Planet5, ND_Planet6, ND_Planet7}. Among these planets, LTT~9779~b, originally discovered by \citet{LTT_Discovery}, stands out as one of the most extreme survivors in this regime. With $M_\mathrm{p} = 29.32^{+0.78}_{-0.81}\,M_{\oplus}$, $R_\mathrm{p} = 4.72 \pm 0.23\,R_{\oplus}$, and $P_{\mathrm{orb}} \approx 0.79\,\mathrm{d}$, it lies deep within the hot Neptune desert in both the period--mass and period--radius planes (Fig.~\ref{fig:ND_population}). At such extreme irradiation levels and such a small orbital separation, the atmosphere of LTT~9779~b could be subject to vigorous photoevaporation and Roche-lobe overflow, as suggested by its restricted Jeans parameter, $\Lambda = GM_\mathrm{p} m_{\mathrm{H}}/(k_{\mathrm{B}} T_{\mathrm{eq}} R_\mathrm{p}) \approx 27$, its gravitational potential, $\Phi = GM_\mathrm{p}/R_\mathrm{p} \approx 4 \times 10^{12}\,\mathrm{erg\,g^{-1}}$, and its Roche-lobe filling factor, $R_\mathrm{p}/R_{\mathrm{Roche}} \approx 0.5$ \citep{LTT_WINERED}. Therefore, under such extreme irradiation and tidal forcing, LTT~9779~b is a particularly compelling target for transmission spectroscopy of the \ion{He}{i} $\lambda 1083$\,nm triplet, which has proven to be one of the most sensitive tracers of atmospheric escape across both space- and ground-based telescopes \citep{Seager_Sasselov_2000, HeI_intro1, HeI_HST, HeI_CARMENES, Helium_HATP11b}.

In addition, its extreme irradiation may significantly affect the planet's atmospheric thermal structure. First, the equilibrium temperature of LTT~9779~b is approximately 2000\,K \citep{LTT_Discovery}. Moreover, the follow-up \textit{Spitzer} and Transiting Exoplanet Survey Satellite (TESS) observations have measured dayside brightness temperatures of \mbox{$2300 \pm 140\,\mathrm{K}$} at 3.6\,$\mu$m and $1800 \pm 120\,\mathrm{K}$ at 4.5\,$\mu$m \citep{LTT_Spitzer_Tess, LTT_Spitzer}, which indicate a complex vertical thermal structure in its atmosphere. Such temperature levels are comparable to those of typical ultrahot Jupiters (UHJs, $T_\mathrm{eq} > 2000\,\mathrm{K}$), which, owing to their extreme stellar irradiation, are predicted to exhibit thermal inversions created by atomic or molecular absorbers such as iron, $\mathrm{TiO}$, and $\mathrm{VO}$ \citep{TP_UHJ_1, TP_UHJ_2, TP_UHJ_3}. These thermal inversions in UHJs have been commonly detected through chemical emission features across both the optical and the near-infrared bands using ground-based telescopes. For example, in the optical, \ion{Fe}{i} emission lines have been seen in KELT-9b \citep{TP1_KELT9, TP2_KELT9}, WASP-33b \citep{TP1_WASP33_Fe, TP2_WASP33_Fe, TP3_WASP33_Fe}, WASP-189b \citep{Yan_2020}, \mbox{MASCARA-1b} \citep{MASCARA1_PEPSI, Guo}, and HAT-P-70b \citep{HATP70_PEPSI}. In the near-infrared band, emission from molecules such as $\mathrm{CO}$ or $\mathrm{H_2O}$ has also been observed in several UHJs such as WASP-18b \citep{WASP18_IGRINS, CRIRES_Yan}, WASP-33b \citep{TP1_WASP33_CO, TP2_WASP33_CO, WASP33_Keck, TP3_WASP33_CO}, and \mbox{MASCARA-1b} \citep{TP1_MASCARA1, TP2_MASCARA1}. Notably, space-based observations have also revealed emission features in exoplanet spectra, for example in WASP-121b with the \textit{Hubble} Space Telescope (HST; \citealt{TP_WASP121}) and in WASP-18b with the \textit{James Webb} Space Telescope (JWST; \citealt{TP_WASP18}). Nevertheless, robust detections of stratospheric thermal inversions have so far been largely limited to UHJs orbiting A- and F-type stars. To date, no ultrahot gas giant orbiting a G-type star has shown unambiguous evidence of a thermal inversion, despite theoretical expectations that such inversions should be possible under sufficiently intense irradiation \citep{TP_Gstar}. In this context, LTT~9779~b, an ultrahot Neptune around a G-type star, offers a rare opportunity to test whether Neptune-sized planets can sustain Jupiter-like thermal inversions under extreme irradiation around cooler solar-type hosts.

More generally, atmospheric studies of exoplanets have traditionally focused on close-in hot Jupiters and UHJs, whose large atmospheric signals make them favorable targets for characterization. Neptune-sized planets, however, remain considerably more challenging owing to their smaller radii, lower atmospheric scale heights, and the possible presence of clouds or hazes. With the advent of JWST, atmospheric characterization of Neptunes and sub-Neptunes has begun to transition from a handful of individual detections to more systematic investigations. Space-based low-resolution spectroscopy has revealed atmospheric features in several such planets, providing evidence of molecular species such as $\mathrm{H_2O}$, $\mathrm{CH_4}$, $\mathrm{CO_2}$, and even $\mathrm{SO_2}$ \citep{JWST_K218, JWST_TOI270, JWST_WASP107, JWST_GJ3470, JWST_HATP26}. In contrast, detections of molecular signals in Neptune-sized exoplanets using ground-based high-resolution spectroscopy remain relatively scarce. One example is the reported signal of $\mathrm{H_2O}$ and $\mathrm{NH_3}$ in the atmosphere of HAT-P-11b obtained with GIANO at the 3.6\,m Telescopio Nazionale Galileo \citep{GIANO_HATP11}. Several additional Neptune-sized planets, including GJ~436b \citep{GJ436_CRIRES,GJ436_CRIRES_CARMENES,GJ436_Keck}, GJ~3470b \citep{GJ3470_CARMENES}, and WASP-166b \citep{WASP166_ESPRESSO}, as well as sub-Neptunes such as GJ~1214b \citep{GJ1214_CRIRES,GJ1214_CRIRES_Lisa} and GJ~3090b \citep{GJ3090_CRIRES}, have also been investigated with ground-based high-resolution spectroscopy, although most studies have yielded only tentative detections or upper limits. Expanding this sample is therefore essential for complementing JWST observations and for probing atmospheric composition, dynamics, and diversity across the Neptune and sub-Neptune population.

So far, LTT~9779~b has been the focus of extensive atmospheric investigations through transmission and emission spectroscopy, as well as reflected-light observations. Common tracers of atmospheric escape, such as the \ion{He}{i} $\lambda 1083$\,nm triplet and $\mathrm{H}\alpha$, showed no detectable signals in ground-based transit observations obtained with the Warm INfrared Echelle spectrograph to REalize Decent high-resolution spectroscopy (WINERED) on the 6.5\,m Clay/Magellan II Telescope and the Echelle SPectrograph for Rocky Exoplanets and Stable Spectroscopic Observations (ESPRESSO) at the Very Large Telescope (VLT), respectively \citep{LTT_WINERED, LTT_ESPRESSO_Transit}. These non-detections could indicate a weak mass-loss rate and point toward a high-metallicity atmosphere. Furthermore, high-resolution reflected-light observations with ESPRESSO in the four-unit telescope mode did not detect $\mathrm{TiO}$ features and further suggested the presence of high-altitude silicate clouds in the atmosphere \citep{LTT_ESPRESSO_Reflection}. Moving to space-based observations, previous transit observations with the HST Wide Field Camera 3 (WFC3) detected only weak spectral features \citep{LTT_HST_Transit}. Consistent with these findings, recent transmission spectra obtained with the JWST Near Infrared Imager and Slitless Spectrograph (NIRISS) also revealed muted molecular features and favored an atmosphere with extremely high metallicity (20--850$\times$ solar) and millibar-level clouds at the terminator \citep{LTT_JWST_Transit}. The dayside atmosphere has likewise been explored through phase-curve observations. Phase-curve observations with JWST/NIRISS in the single object slitless spectroscopy (SOSS) mode revealed a weak $\mathrm{H_2O}$ absorption feature \citep{ltt_jwst_emission}. At longer near-infrared wavelengths, recent phase-curve observations obtained with the JWST Near Infrared Spectrograph (NIRSpec) using the G395H grating showed that the obscuring effect of high-altitude clouds becomes less significant, revealing a carbon-rich atmosphere dominated by $\mathrm{CO}$ and $\mathrm{CO_2}$, with retrieved mixing ratios of $\sim10^{-1}$ and $\sim10^{-4}$, respectively, together with weaker $\mathrm{H_2O}$ absorption features \citep{LTT_JWST_NIRSpec}. Their atmospheric retrievals further favored a near-unity carbon-to-oxygen ratio and a metallicity exceeding $500\times$ solar.

The unusually high geometric albedo of LTT~9779~b has been investigated through several independent observations. Based on CHaracterising ExOPlanet Satellite (CHEOPS) secondary-eclipse measurements, \citet{LTT_CHEOPS} reported an extremely high geometric albedo of $A_\mathrm{g}=0.80^{+0.10}_{-0.17}$, ranking LTT~9779~b among the most reflective exoplanets known. Subsequent JWST NIRISS/SOSS phase-curve observations confirmed the planet's high geometric albedo ($A_\mathrm{g}=0.50\pm0.07$; \citealt{ltt_jwst_emission}) and revealed a pronounced longitudinal asymmetry, with the western hemisphere exhibiting a higher reflectivity ($A_{\mathrm{g, west}}=0.79\pm0.15$) than the eastern hemisphere ($A_{\mathrm{g, east}}=0.41\pm0.10$). A subsequent reanalysis of the ESPRESSO reflected-light dataset detected a broadband planetary signal at a signal-to-noise ratio (S/N) of $\sim5.4$ and inferred a geometric albedo of $A_\mathrm{g}=0.88\pm0.25$, with enhanced blue reflectivity consistent with efficient scattering or high-altitude clouds \citep{LTT_ESPRESSO_Reflection_Detection}. Although the CHEOPS and ESPRESSO measurements yield higher central values than the JWST result, the differences are modest ($\sim1.6\sigma$ and $\sim1.5\sigma$, respectively, based on the reported uncertainties) and do not indicate a significant discrepancy. These measurements were obtained over different wavelength ranges and rely on different approaches to estimating the reflected-light contribution. In particular, the CHEOPS albedo was inferred assuming a negligible thermal contribution, whereas the JWST phase-curve analysis separated reflected light from thermal emission. The wavelength-dependent reflectivity indicated by ESPRESSO further highlights the importance of accounting for differences in bandpass when comparing the reported albedos.

Here, we present high-resolution transit and emission spectroscopy of LTT~9779~b obtained with the upgraded Cryogenic Infrared Echelle Spectrograph (CRIRES$^+$), with the latter yielding the first ground-based detection of dayside thermal emission from a Neptune-sized exoplanet. No significant helium absorption features are detected during a single transit event. In contrast, the dayside emission spectra reveal the presence of $\mathrm{H_2O}$ and provide tentative evidence of $\mathrm{CO}$. By further performing a joint atmospheric retrieval with JWST NIRISS/SOSS data, we confirm the presence of a weak $\mathrm{H_2O}$ absorption in the atmosphere of LTT~9779~b. Our results demonstrate the feasibility of using ground-based high-resolution spectroscopy to probe ultrahot Neptunes and highlight the complementarity between ground-based and space-based facilities for infrared atmospheric characterization.

The paper is organized as follows. In Sect.~\ref{obs}, we describe the observations and data reduction. In Sect.~\ref{sec:analysis_methods}, we introduce the methods for the dayside molecular search and atmospheric retrievals. In Sect.~\ref{sec:results_discussion}, we present and discuss the results of the transmission spectroscopy, dayside molecular search, and atmospheric retrieval. Finally, we summarize our conclusions in Sect.~\ref{conclusion}.

\section{Observations and data}\label{obs}
\subsection{Observations}
The CRIRES$^+$ instrument is a cryogenic, cross-dispersed, high-resolution echelle spectrograph covering the near-infrared band, installed at the 8 m-class Unit Telescope 3 of the VLT \citep{CRIRES_Dorn}. As part of our guaranteed time observations with CRIRES$^+$, we observed the thermal emission spectrum of LTT~9779~b on 9 August 2022 between 05:56 UT and 08:40 UT, as well as the transmission spectrum on 12 August 2023 between 06:07 UT and 08:28 UT.

An ABBA nodding pattern was applied to both our observations. In this pattern, the target was observed at two different positions along the slit (A and B), which allows for an efficient removal of the sky background and detector artifacts. As illustrated in \citet{CRIRES_Yan}, in the data reduction procedure, we treated the A and B spectra as two independent datasets and combined them in the final cross-correlation results. The thermal emission observations cover the orbital phases between 0.29 and 0.43, where most of the dayside of the planet is visible to the observer before secondary eclipse. The transmission observations cover a full transit event with orbital phases between -0.09 and 0.03.

We used the Y1029 wavelength setting, covering \SI{950}{\nano\meter} to \SI{1120}{\nano\meter}, for the transmission observations to search for the \ion{He}{i} $\lambda 1083$\,nm triplet. The K2166 wavelength setting, with a wavelength coverage between \SI{1921}{\nano\meter} and \SI{2472}{\nano\meter}, was used for the thermal emission observations, as this range is sensitive to some major molecular spectral lines (e.g., $\mathrm{CO}$ and $\mathrm{H_2O}$). The Y1029 wavelength setting consists of nine echelle orders with two narrow gaps within each order, yielding 27 wavelength segments in total. The K2166 wavelength setting includes seven echelle orders with two narrow gaps within each order, yielding 21 wavelength segments in total. We used the \SI{0.2}{\arcsecond} slit of CRIRES$^+$, corresponding to a nominal resolving power of $R \sim 100\,000$. For the thermal emission observations, the adaptive-optics (AO)-corrected point spread function (PSF) was narrower than the slit, resulting in an effective resolving power higher than the nominal slit-limited value. This super-resolution effect has also been reported in recent CRIRES$^+$ studies \citep{CRIRES_WASP127, CRIRES_WASP107, CRIRES_51Peg}. We estimated the resolving power by fitting each wavelength segment separately with \texttt{molecfit}, obtaining values ranging from approximately $R=95\,000$ to $160\,000$ across different segments. We therefore adopted a representative resolving power of $R=120\,000$ throughout this work. The airmass and the mean S/N per pixel during the observations are illustrated in Fig. \ref{fig:airmass_snr}. 

\subsection{Raw data extraction}
The raw spectra were reduced by the European Southern Observatory (ESO) CRIRES$^+$ data reduction software pipeline \texttt{cr2res}, including dark subtraction, flat-field correction, echelle order localization, and wavelength calibration.  Bad pixels were masked after the raw data reduction, and wavelength calibration was performed using a uranium-neon lamp together with a Fabry--P\'erot etalon. The time series was organized into pairs of consecutive A and B nodding positions, yielding 33 pairs for transmission spectra and 38 pairs for thermal emission spectra. Then the pipeline was run on each pair, in which A and B spectra were subtracted from each other. Subsequently, the pipeline produced one-dimensional spectra, along with associated uncertainty estimates for each wavelength segment and nodding position. We discarded the first and last 20 pixels for every wavelength segment to mitigate edge effects from the detector.

During our observations, the PSF was smaller than the slit and not perfectly centered, resulting in wavelength shifts of a few $\mathrm{km\,s^{-1}}$ between the spectral data points and the pipeline-delivered wavelength solutions. These offsets typically differ between the A and B nodding positions due to imperfect alignment with the slit \citep{WASP178_David}. To account for the wavelength shifts, we used the \texttt{molecfit} software \citep{molecfit} to fit telluric absorption features and derive improved wavelength solutions for the A and B spectra, respectively. We adopted the first nodding pair of the transmission spectra in the \texttt{molecfit} calculation. In contrast, for the thermal emission spectra, we used the third nodding pair due to its higher S/N. The resulting wavelength solution was then applied to all exposures.

To assess the additional drifts between different exposures and examine the temporal stability of wavelength solutions derived from \texttt{molecfit}, we further calculated the radial velocity (RV) drifts between the median spectrum and individual exposures of the time series using the cross-correlation technique. Fig.~\ref{fig:rv_drifts} shows the distribution of RV drifts across all exposures and wavelength segments. For both the transmission and thermal emission observations, all the RV drifts are $ < 0.6  ~\mathrm{km\,s^{-1}}$, with most clustered around $0.2~\mathrm{km\,s^{-1}}$. Since these shifts are smaller than the velocity width of a single CRIRES$^+$ resolution element ($\sim2.5~\mathrm{km\,s^{-1}}$), the wavelength solutions from \texttt{molecfit} are sufficiently stable for our analysis, and we therefore did not apply further corrections to different exposures.


\begin{table}
\renewcommand{\arraystretch}{1.35}
\caption{Parameters of the LTT~9779 system.}
\label{tab:ltt_params}
\centering
\normalsize
\begin{tabular*}{\columnwidth}{@{\extracolsep{\fill}}lcl@{}}
\hline\hline
Parameter & Symbol (unit) & Value \\
\hline
\multicolumn{3}{l}{\textit{Star}} \\
Radius & $R_\star\, (R_\odot)$ & $0.949 \pm 0.006$ \\
Mass & $M_\star\, (M_\odot)$ & $1.02^{+0.02}_{-0.03}$ \\
Effective temperature & $T_{\mathrm{eff}}\, (\mathrm{K})$ & $5480 \pm 42$ \\
Metallicity & $[\mathrm{Fe}/\mathrm{H}]\, (\mathrm{dex})$ & $0.25 \pm 0.04$ \\
Systemic velocity & $\gamma\, (\mathrm{km\,s^{-1}})$ & $-10.6 \pm 1.0 \,^{(b)}$ \\
\hline
\multicolumn{3}{l}{\textit{Planet}} \\
Radius & $R_\mathrm{p}\, (R_\oplus)$ & $4.72 \pm 0.23$ \\
& $R_\mathrm{p}/R_\star$ & $0.0455^{+0.0022}_{-0.0017}$ \\
Mass & $M_\mathrm{p}\, (M_\oplus)$ & $29.32^{+0.78}_{-0.81}$ \\
Surface gravity & $\log g_{\mathrm{p}}\, (\mathrm{cgs})$ & $3.11$ \\
Equilibrium temperature &  $T_\mathrm{eq} \,(\mathrm{K})$ & $2333 \,^{(c)}$ \, ($A_\mathrm{B} = 0$) \\
~ & ~ & 1900 \, ($A_\mathrm{B} = 0.56$) \\
Orbital inclination & $i_\mathrm{p}\, (\mathrm{deg})$ & $76.39 \pm 0.43$ \\
Semimajor axis & $a\, (\mathrm{au})$ & $0.01679^{+0.00014}_{-0.00012}$ \\
& $a/R_\star$ & $3.877^{+0.090}_{-0.091}$ \\
Orbital period & $P_{\mathrm{orb}}\, (\mathrm{d})$ & $0.792\,064\,10$ \\
~ & ~ & $\pm 0.000\,000\,14 \,^{(a)}$ \\
Transit epoch & $T_0\,(\mathrm{BJD}_{\mathrm{TDB}})$ & $2\,459\,043.310\,602$ \\
~ & ~ & $\pm 0.000\,090\,^{(a)}$ \\
Transit duration & $T_{14}\, (\mathrm{h})$ & $0.82 \pm 0.11\,^{(d)}$ \\
Ingress duration & $T_{\mathrm{ingress}}\, (\mathrm{h})$ & $0.186 \pm 0.040$ \\
RV semi-amplitude & $K_\mathrm{p}\, (\mathrm{km\,s^{-1}})$ & $225.3^{+1.5}_{-2.2}\,^{(e)}$ \\
\hline
\end{tabular*}
\tablebib{
All parameters are from \citet{LTT_Discovery}, except for:
\textsuperscript{(a)}~\citet{LTT_HST_Transit};
\textsuperscript{(b)}~\textit{Gaia} DR2 \citep{Gaia_DR2}.
\textsuperscript{(c)} Eq.~(2) of \citet{EA_Sara} assumes dayside-only reradiation.
\textsuperscript{(d)} Eq.~(14) from \citet{Transit_equation}, with uncertainties estimated through first-order error propagation. The quoted uncertainty in $T_{14}$ may be optimistic if the uncertainty in $R_\star$ reported by \citet{LTT_Discovery} is underestimated.
\textsuperscript{(e)} Eq.~(\ref{Kp}) of this work, with uncertainties estimated through first-order error propagation. The quoted uncertainty in $K_\mathrm{p}$ may be optimistic if the uncertainty in $M_\star$ reported by \citet{LTT_Discovery} is underestimated.
}
\end{table}


\subsection{Data reduction} \label{data_reduction}
To extract the faint signal of the exoplanet atmosphere, we must remove the telluric absorption, the stellar spectrum, and any contribution from instrumental effects. These data reduction procedures are described in the following subsections and were applied consistently to both the thermal emission and transmission observations. 

\subsubsection{Pre-processing the spectra} \label{sec:preprocessing}
\begin{enumerate}[label=(\arabic*)]

\item Normalization. During the observations, the flux levels of individual exposures varied, primarily because of variable slit losses caused by changes in seeing and airmass-dependent atmospheric extinction, with additional contributions from instrumental effects such as variations in the blaze function. To account for these variations, all spectra were normalized to a common continuum level. We first divided the wavelength range into 100 equal-width bins and selected the 90th percentile flux within each bin as a representative point on the continuum. A second-order polynomial was then fitted to these points to model the continuum. For every wavelength segment and every frame, the spectrum was divided by its fitted continuum. 

\item Clean up the spectral matrix. In this step, we first calculated a master spectrum by taking the mean of all spectra along each pixel column, then computed the standard deviation for each pixel column, following the same procedure as in \citet{CRIRES_Yan}. We subsequently obtained the empirical S/N of each pixel column by dividing the master spectrum by the standard deviation. Based on this empirical S/N spectrum, the lower S/N pixel columns were set to \text{NaN} values. For both of our observations, about 10$\%$ of the pixel columns were masked. This procedure ensures that strong telluric lines are effectively masked without requiring prior knowledge of their positions. Notably, normalization has to be performed before this step since continuum variations can significantly affect the computed standard deviation across pixel columns, ultimately resulting in an inaccurate estimate of the empirical S/N.

\item Sigma clipping. To correct bad pixels, we further performed $3\sigma$ clipping on the time evolution of each pixel column and then replaced the bad pixels with the mean values of each pixel column.

\end{enumerate}

\subsubsection{Removal of telluric and stellar lines} \label{telluric_removal}

To date, two main methods have been developed to correct the stationary telluric and stellar lines in the spectral matrix. One approach involves accurately modeling the telluric spectrum using tools such as \texttt{molecfit}, which uses a line-by-line radiative transfer model together with atmospheric profiles from the Global Data Assimilation System to generate synthetic transmission spectra of the Earth's atmosphere across the observed wavelength range \citep[e.g.,][]{KELT9_HARPS}. Another approach employs detrending algorithms such as \texttt{SYSREM} \citep{sysrem}, which iteratively identifies and removes systematic trends shared across the spectral matrix while accounting for the uncertainty of each data point. In this work, we adopted \texttt{SYSREM}, as it has been demonstrated to be a robust tool in high-resolution spectroscopy \citep{CCF_2013, sysrem2}. Moreover, \citet{sysrem_and_molecfit} showed that \texttt{SYSREM} outperforms \texttt{molecfit} for close-in planets ($P = 2$ days in their simulations), making it particularly suitable for our short-period target LTT~9779~b.

The \texttt{SYSREM} algorithm was applied iteratively to the observed spectral matrix. 
At each iteration $k$, one dominant wavelength-dependent ($\mathbf{a}^{(k)}$) and time-dependent ($\mathbf{c}^{(k)}$) systematic component was fitted and removed from the residual spectra, as summarized by
\begin{equation}
\mathbf{R}^{(k+1)} = \mathbf{R}^{(k)} - \mathbf{c}^{(k)}\mathbf{a}^{(k)\mathrm{T}},
\end{equation}
where $\mathbf{R}^{(0)}$ denotes the original spectral matrix and $\mathbf{R}^{(k+1)}$ represents the residual matrix after the $k$-th iteration. The detailed description of the algorithm is provided in Appendix~\ref{appendix:sysrem}.

We further computed the cumulative model matrix, $\mathbf{M}^{(N)}$, after applying \texttt{SYSREM} for N iterations, which can be written as
\begin{equation}
\mathbf{M}^{(N)} = {\sum_{k=0}^{N-1} \mathbf{M}^{(k)}} = {\sum_{k=0}^{N-1} \mathbf{c}^{(k)}\mathbf{a}^{(k)\mathrm{T}}}.
\end{equation} 
Instead of subtracting the cumulative model from the original observed spectra, we divided the original spectra by the cumulative model. This approach preserves the relative depths of planetary spectral features and ensures accurate uncertainty propagation. This method is particularly important for identifying planetary spectral lines that overlap with telluric absorption lines, such as those from $\mathrm{H_2O}$.

Before performing \texttt{SYSREM}, we re-estimated the uncertainties using an approach similar to that described in \citet{Yan_2020}, so that they more closely reflect the empirical scatter in the observed spectra. This procedure involves first applying one iteration of \texttt{SYSREM} using the original pipeline uncertainties to obtain a noise-dominated residual spectrum. Subsequently, we calculated the standard deviation for each pixel bin (vertical dimension) and for each frame bin (horizontal dimension) of the residual spectral matrix. The final uncertainty assigned to each data point was determined by averaging the respective row-wise and column-wise standard deviations.

\begin{figure}[htbp]
    \centering
    \includegraphics[width=\columnwidth]{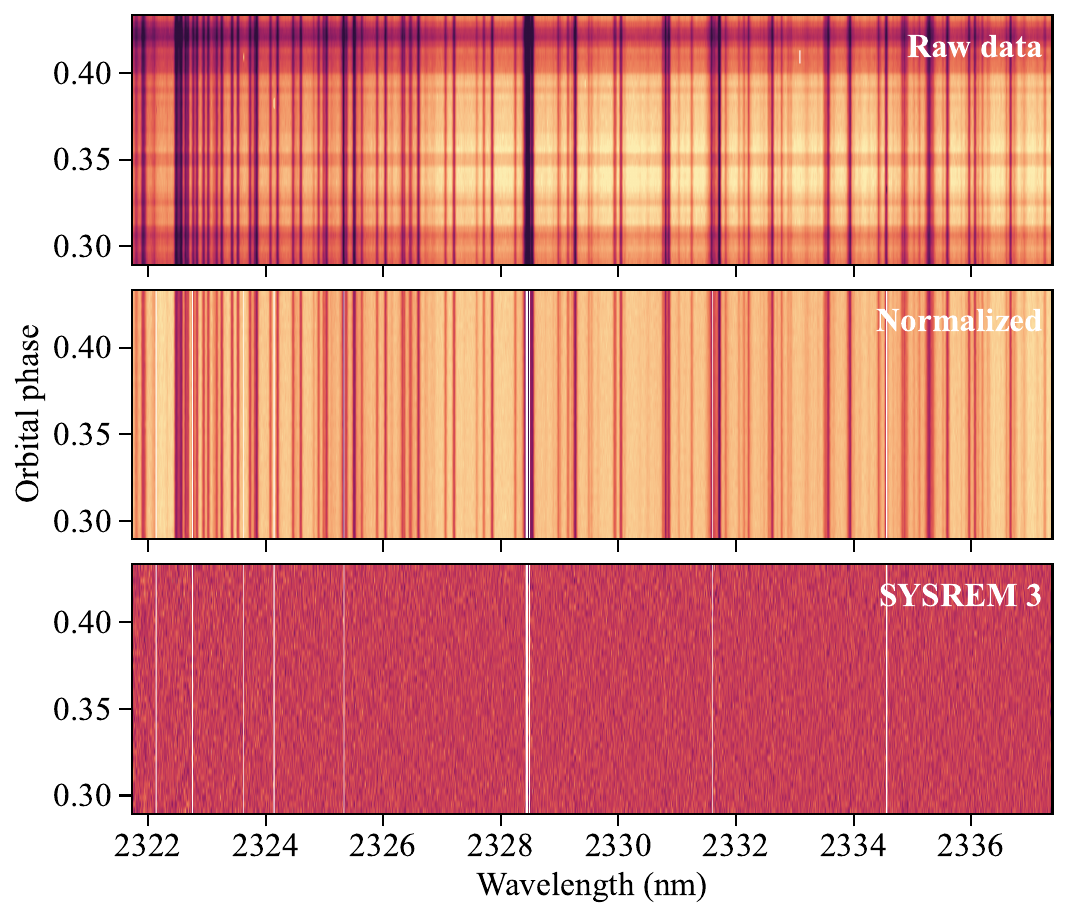}
    \caption{Illustration of the data reduction process for a representative wavelength interval. Top panel: Raw spectra extracted by the CRIRES$^+$ pipeline. Middle panel: Spectra after continuum normalization and masking of deep telluric absorption lines. Bottom panel: Spectra after three iterations of \texttt{SYSREM} for the removal of stellar and telluric systematics.
    }
    \label{data-reduction}
\end{figure}

After re-estimating the uncertainties, we ran \texttt{SYSREM} for fifteen iterations to generate a set of residual spectra. To remove any broadband features, we further applied a Gaussian high-pass filter with a standard deviation of 31 points. The final residual spectra represent the normalized planetary signals: 
$1 + F_\mathrm{p} / F_\mathrm{s}$ in the case of thermal emission, and 
$F_{\mathrm{in}} / F_{\mathrm{out}}$ for transmission, where $F_{\mathrm{in}}$ and $F_{\mathrm{out}}$ denote the in-transit and out-of-transit stellar flux, respectively. An overview of the entire data reduction procedure is shown in Fig. \ref{data-reduction}.

\section{Methods} \label{sec:analysis_methods}
\subsection{Search for atmospheric species on the dayside} \label{emission}
To investigate the presence of chemical species on the dayside, we performed the cross-correlation analysis to extract the weak planetary spectral lines \citep{CCF_2010, CCF_2012, CCF_2013}. This analysis computes the cross-correlation function (CCF) between the template and the residual spectra, coherently combining the many weak molecular lines that are shifted due to the Doppler effect into a detectable signal peak. We searched for dominant molecular species in the \mbox{$K$ band}, such as $\mathrm{CO}$ and $\mathrm{H_2O}$, as well as other molecules including $\mathrm{CO_2}$, $\mathrm{OH}$, $\mathrm{CH_4}$, and $\mathrm{HCN}$, although these latter species are not expected to be abundant at such high temperatures. We also searched for $\mathrm{Fe}$, which has been detected in some UHJs with CRIRES$^+$ (e.g., \citealt{CCF_Fe1, W189_Fabio}).

\subsubsection{Model spectra} \label{model_spec}
The model template was constructed following a procedure similar to that described in \citet{Yan_2020} and \citet{WASP178_David}. Following the retrieval framework detailed in Sect.~\ref{joint_framework}, we defined a pressure grid of 51 layers sampled logarithmically from $10^{-8}$ to $10^{2}~\mathrm{bar}$ and used a ten-point parametric temperature–pressure ($T$-$p$) profile. We adopted the best-fit $T$-$p$ profile derived from the joint retrieval (see Sect.~\ref{fixed_framework}) to generate the final model spectra.

The volume mixing ratios (VMRs) of $\mathrm{H_2O}$ and $\mathrm{CO}$ used for the template calculation were adopted from the best-fit values corresponding to the maximum posterior probability in the CRIRES$^+$-only retrieval (Sect.~\ref{fixed_discussion}). The corresponding mean molecular weight (MMW) was calculated by assuming that the remaining atmosphere consists of $\mathrm{H_2}$ and $\mathrm{He}$ in solar proportions \citep{solar_abundance}. However, when generating the individual templates of $\mathrm{CO_2}$, $\mathrm{OH}$, $\mathrm{CH_4}$, $\mathrm{HCN}$, and $\mathrm{Fe}$, whose abundances are not tightly constrained by the CRIRES$^+$-only retrieval, their VMRs and the corresponding MMWs were computed using \texttt{FastChem} \citep{fastchem} under the assumption of chemical equilibrium, adopting a carbon-to-oxygen ratio of $\mathrm{C/O}=1$ and a metallicity of $\sim 500\times$ solar as reported by \citet{LTT_JWST_NIRSpec}.

These abundances, together with the temperature-, \mbox{pressure-,} and wavelength-dependent opacities, were then used in the radiative transfer calculations to generate the synthetic spectra. The high-resolution opacity data adopted in this study are obtained from the following sources: $\mathrm{CO}$ opacity data from \citet{CO_opacity}, $\mathrm{H_2O}$ line list from \citet{H2O_opacity}, $\mathrm{CO_2}$ data from \citet{CO2_opacity}, $\mathrm{OH}$ line information from \citet{OH_opacity}, $\mathrm{CH_4}$ opacity from \citet{CH4_opacity}, $\mathrm{HCN}$ from \citet{HCN_opacity}, and $\mathrm{Fe}$ from \citet{Fe_opacity}. Finally, we ran the radiative transfer code \texttt{petitRADTRANS} \citep{pRT} to generate the high-resolution  ($R = 10^6$) model spectra for the corresponding chemical species. 

The template fluxes generated by \texttt{petitRADTRANS} correspond to the planetary emission, $F_\mathrm{p}$. To be consistent with the data reduction procedure, we first assumed a stellar blackbody spectrum ($F_\mathrm{s}$) and regarded the observed spectrum as $F_\mathrm{p}$ + $F_\mathrm{s}$. Then we normalized the model spectra by dividing them by $F_\mathrm{s}$ to obtain the stellar-normalized model spectrum: $1 + F_\mathrm{p}/F_\mathrm{s}$. We applied the same Gaussian filtering used in Sect.~\ref{telluric_removal} to remove the planetary continuum, and the normalized spectra were subsequently convolved with the Gaussian instrumental profile at the observational resolving power and with the rotational broadening profile, both described in Sect.~\ref{joint_framework}. The final model spectra used in the cross-correlation analysis are shown in Fig.~\ref{CCF-results}.

\subsubsection{Cross-correlation} \label{ccf_function}
We used the cross-correlation method to detect chemical species. First, we created a Doppler-shifted model grid of RV from $-500\,\mathrm{km\,s^{-1}}$ to $500\,\mathrm{km\,s^{-1}}$ with a cadence of $1~\mathrm{km\,s^{-1}}$. We subtracted unity from both the residual spectra and the model template, and computed the weighted CCF at frame $i$ as
\begin{equation}
\label{eq:ccf}
\mathrm{CCF}_i(v) \;=\; \sum_{j} \frac{R_{ij}\, m_j(v)}{\sigma_{ij}^{2}},
\end{equation}
where $R_{ij}$ is the matrix of residual spectra, $\sigma_{ij}$ the corresponding uncertainties, and $m_j(v)$ the template shifted by the RV, $v$. For each wavelength segment, the time series of $\mathrm{CCF}_i(v)$ was arranged into a two-dimensional array. Arrays from different spectral segments were co-added to form CCF maps for the A and B nod positions separately; the two nodding maps were then combined into a single CCF map for each species.

The original CCF was calculated in the observer’s rest frame. We aligned the CCF map to the planetary rest frame, assuming the RV as
\begin{equation}
v_\mathrm{p} = \gamma - v_\mathrm{bary}
+ K_\mathrm{p} \sin(2\pi \phi) + \Delta v ,
\label{rv}
\end{equation}
where $\gamma$ is the systemic velocity, $v_{\mathrm{bary}}$ the barycentric velocity, $\phi$ the orbital phase, and $K_\mathrm{p}$ the orbital RV semi-amplitude of the planet. This equation assumes a circular planetary orbit, and $\Delta v$ serves as a parameter to account for possible planetary rotation and atmospheric circulation. A phase-dependent weighting factor, described in Sect.~\ref{joint_framework}, was also applied to account for the varying visibility of the planetary dayside during the observations. For a certain $K_\mathrm{p}$ value, the weighted two-dimensional CCF map was collapsed along the orbital phase, resulting in a one-dimensional CCF. We then repeated this step for a range of $K_\mathrm{p}$ values from $100\,\mathrm{km\,s^{-1}}$ to $300\,\mathrm{km\,s^{-1}}$ with a cadence of $1~\mathrm{km\,s^{-1}}$. Eventually, we obtained a two-dimensional CCF map as a function of $K_\mathrm{p}$ and $\Delta v$, known as the $K_\mathrm{p}$-$\Delta v$ map.

To estimate the noise level in the $K_\mathrm{p}$-$\Delta v$ map, we excluded the region around the expected planetary signal and treated all remaining CCF values as noise-dominated. The noise dispersion was quantified using the median absolute deviation, which is robust against outliers and potential weak residual signals. As a final step, the $K_\mathrm{p}$-$\Delta v$ map was normalized by this noise estimate to obtain the S/N map. Any planetary atmospheric signal is expected to appear as a localized peak near the expected $K_\mathrm{p}$ and $\Delta v = 0$. For this system, the expected $K_\mathrm{p}$ is $225~\mathrm{km\,s^{-1}}$, calculated with
\begin{equation}
K_\mathrm{p}
= \left( \frac{2 \pi \mathrm{G} M_\star}{P} \right)^{1/3}
\sin i_\mathrm{p},
\label{Kp}
\end{equation}
where $M_\star$ is the stellar mass, $P$ is the orbital period, and $i_\mathrm{p}$ is the orbital inclination of the planet.
\subsection{Framework of atmospheric retrieval} \label{retrieval}
Over the past few years, atmospheric retrieval techniques tailored to high-resolution spectroscopy have undergone rapid development (e.g., \citealt{RF_1, RF_3, Yan_2020}).
Using retrieval frameworks, statistical constraints on atmospheric properties are obtained by comparing parameterized model spectra with observations.
In this work, the atmospheric conditions of LTT~9779~b are quantitatively investigated following the approach of \citet{WASP178_David} and \citet{Yan_GJ1132}.

In all three retrievals described below, we fixed $R_\mathrm{p}$ and $R_\star$ to the nominal values in Table~\ref{tab:ltt_params} rather than sampling them. The reported posterior intervals are therefore conditional on the adopted $R_\mathrm{p}/R_\star$ and do not include uncertainty in this ratio.

\subsubsection{Joint retrieval} \label{joint_framework}
Given that the molecular signals detected in the cross-correlation analysis of the CRIRES$^+$ data are not highly significant, we adopted a joint atmospheric retrieval framework that combines ground-based high-resolution spectroscopy with space-based low-resolution observations. The inclusion of the JWST NIRISS/SOSS data, whose spectral coverage overlaps with the CRIRES$^+$ $K$-band region, provides complementary information on the continuum level and absolute flux scale, thereby enabling tighter constraints on the atmospheric properties. For the NIRISS/SOSS data, we adopted the secondary-eclipse spectrum reported by \citet{ltt_jwst_emission}. In addition, we used the CRIRES$^+$ residual spectra after the seventh iteration of \texttt{SYSREM}, as this iteration yields the strongest CCF signal when combining $\mathrm{CO}$ and $\mathrm{H_2O}$ (Fig.~\ref{CCF-sysrem}).

Within the joint retrieval framework, a single atmospheric model with a common thermal structure, chemical composition, and gray cloud parameterization was mapped to each dataset through resolution- and instrument-specific forward models. Here, the thermal structure is described by a free $T$-$p$ profile,
parameterized by ten temperature nodes uniformly distributed in log-pressure space between $10^{-8}$~bar and $10^{2}$~bar. We imposed a uniform prior between 100 K and 5000 K on each temperature point, while a Gaussian prior was applied as a penalty to mitigate nonphysical oscillations in the $T$-$p$ structure \citep{Free_TP}:
\begin{equation}
\ln \pi(\mathbf{T}) =
\frac{-1}{2\sigma_\mathrm{s}^2 \log\left
(p_{\mathrm{bot}}/p_{\mathrm{top}}\right)}
\sum_{i=2}^{N-1}
\frac{\left(T_{i+1} - 2T_i + T_{i-1}\right)^2}
{\left(\Delta \log p\right)_i^{3}} ,
\label{eq:Tp_smoothness_prior}
\end{equation}
where $p_{\mathrm{top}}$ and $p_{\mathrm{bot}}$ denote the pressures at the top and bottom of the atmosphere, $\sigma_{\mathrm{s}}$ controls the strength of the smoothness penalty, and the second-order finite difference approximates the local curvature of the temperature structure in log-pressure space. We adopted a value of $\sigma_{\mathrm{s}} = 300~\mathrm{K~dex^{-2}}$, consistent with \citet{ltt_jwst_emission}. The parameterized $T$-$p$ profile was subsequently interpolated onto 51 logarithmically spaced pressure layers for the forward radiative transfer calculations.

The VMRs of all chemical species were assumed to be constant with pressure. We included only $\mathrm{CO}$ and $\mathrm{H_2O}$ as spectroscopically active trace species, as the cross-correlation analysis detected $\mathrm{H_2O}$ and provided tentative evidence of $\mathrm{CO}$. The rest of the atmosphere was assumed to consist of $\mathrm{H_2}$ and $\mathrm{He}$ in solar proportions \citep{solar_abundance}. The MMW was computed self-consistently from the abundances of all atmospheric species, including $\mathrm{H_2}$ and $\mathrm{He}$, while the continuum opacity was dominated by $\mathrm{H_2}$--$\mathrm{H_2}$ and $\mathrm{H_2}$--$\mathrm{He}$ collision-induced absorption\footnote{\url{https://petitradtrans.readthedocs.io/en/latest/content/available_opacities.html}}. In the light of previous observational indications of a cloudy atmosphere \citep{LTT_CHEOPS, LTT_JWST_Transit, LTT_ESPRESSO_Transit, ltt_jwst_emission}, a gray cloud deck was included, below which (i.e., at higher pressures) the atmosphere was assumed to be isothermal. With this atmospheric model in place, forward atmospheric spectra were computed using \texttt{petitRADTRANS} at both high and low spectral resolutions. The resulting spectra are expressed as planet-to-star flux ratios and compared with the CRIRES$^+$ and JWST NIRISS/SOSS observations in the likelihood calculation.

For the high-resolution CRIRES$^+$ observations, an accurate normalization and alignment of the model spectral lines are required to match the relative line strengths. Since the observations span a wide range of orbital phases, a phase-dependent correction factor, $k(\phi)$, was included in the forward model to account for variations in the planet's visible dayside \citep{Phase_factor}:
\begin{equation}
k(\phi) = \frac{1 - \cos \left( 2\pi \phi \right)}{2}.
\end{equation}

To accurately model the observed line profiles, we included several sources of spectral line broadening. Thermal and pressure broadening were already included in the opacity calculations. Instrumental broadening was applied by convolving the model spectrum with a Gaussian kernel corresponding to the spectrograph's spectral resolution, using the \texttt{broadGaussFast} function from \texttt{PyAstronomy} \citep{pyastronomy}. We additionally included rotational broadening by convolving the model spectrum with the analytic rotational kernel described by \citet{Rotation}:
\begin{equation}
G(x) =
\frac{2(1-\epsilon)\sqrt{1-x^2} + \frac{\pi\epsilon}{2}(1-x^2)}
{\pi\left(1-\frac{\epsilon}{3}\right)},
\end{equation}
where
\begin{equation}
x = \ln\left(\frac{\lambda}{\lambda_0}\right)
\frac{c}{v_{\rm eq}\sin i_\mathrm{p}}.
\end{equation}
Here $\lambda_0$ is the central wavelength of the spectral line, $\epsilon$ is the linear limb-darkening coefficient, $v_{\rm eq}$ is the planetary equatorial rotational velocity, and $i_\mathrm{p}$ is the orbital inclination of the planet. We adopted a linear limb-darkening law and set $\epsilon = 1$. 

In addition, to account for the distortion of the planetary signal introduced by \texttt{SYSREM} during the data reduction, we adopted the fast \texttt{SYSREM} filtering approach proposed by \citet{SYSREM_Filter} and applied the resulting filter operator to the initial phase-corrected two-dimensional forward model matrix, $\mathbf{M}_{\mathrm{initial}}$:
\begin{equation}
\mathbf{F}
=
\mathbf{U}
\left( \boldsymbol{\Lambda}\mathbf{U} \right)^{\dagger}
\left( \boldsymbol{\Lambda}\mathbf{M}_{\mathrm{initial}} \right),
\end{equation}
where $\mathbf{U}$ corresponds to the column vector, $\mathbf{c}$, defined in Sect.~\ref{telluric_removal}, $\boldsymbol{\Lambda}$ is a diagonal matrix containing the inverse of the mean wavelength-dependent uncertainties, and $(\cdot)^{\dagger}$ denotes the Moore–Penrose pseudoinverse, defined as $\mathbf{U}^{\dagger} = (\mathbf{U}^{T}\mathbf{U})^{-1}\mathbf{U}^{T}$. 
The resulting filtered model matrix is given by $\mathbf{M_{\mathrm{filtered}}} = \mathbf{M}_{\mathrm{initial}} \,/\, \mathbf{F}$. The high-resolution model spectra were subsequently Doppler-shifted according to the orbital phase using Eq.~(\ref{rv}). Finally, the same Gaussian filter with a width of 31 points was applied to the filtered model matrix before comparison with the data.

For the low-resolution JWST NIRISS/SOSS data, the observed eclipse depths represent the weighted planet-to-star flux ratios within each spectral bin. We followed an approach similar to that proposed by \citet{KELT20_Yan}. For a given passband $\lambda$, the eclipse depth is modeled as
\begin{equation}
\delta_{\mathrm{ecl}} =
A_\mathrm{g} \left( \frac{R_\mathrm{p}}{a} \right)^2
+ \left( \frac{R_\mathrm{p}}{R_\mathrm{s}} \right)^2
\frac{\int f(\lambda)\, \eta(\lambda)\, \mathrm{d}\lambda}
     {\int \eta(\lambda)\, \mathrm{d}\lambda},
\label{eq:eclipse_depth}
\end{equation}
where $A_\mathrm{g}$ denotes the geometric albedo, $a$ is the orbital separation, and $R_\mathrm{p}$ and $R_\mathrm{s}$ are the planetary and stellar radii, respectively.
The function $f(\lambda) \equiv F_\mathrm{p}(\lambda) / F_\mathrm{s}(\lambda)$ represents the ratio of the fluxes originating from the planetary and stellar atmospheres per unit surface area and per unit wavelength, computed as described above. The quantity $\eta(\lambda)$ denotes the instrumental response function within the considered passband, which was approximated as a rectangular function in this work. The first term of the equation represents the contribution of reflected stellar light to the planet-to-star flux ratio, while the second term represents the contribution of the planet’s thermal emission to the planet-to-star flux ratio. In our calculations, the geometric albedo, $A_\mathrm{g}$, was assumed to be wavelength-independent, such that the reflected-light contribution remains constant across all spectral bins. The resulting eclipse depth, $\delta_{\mathrm{ecl}}$, was then compared with the observed secondary-eclipse spectrum.

We performed the atmospheric retrieval by directly sampling the posterior probability distribution using the Markov chain Monte Carlo (MCMC) approach with the ensemble sampler \texttt{emcee}. The logarithmic likelihood function is defined as the sum of the log-likelihood functions corresponding to the high- and low-resolution data, and is given by
\begin{equation}
\ln \mathcal{L}
=
\ln \mathcal{L}_{\mathrm{high}}
+
\ln \mathcal{L}_{\mathrm{low}}.
\label{eq:total_likelihood}
\end{equation}
Assuming Gaussian-distributed residuals, the high-resolution log-likelihood is given by
\begin{equation}
\ln \mathcal{L}_{\mathrm{high}}
=
-\frac{1}{2}
\sum_{i,j}
\left[
\frac{(R_{ij} - M_{ij})^2}{(\beta\,\sigma_{ij})^2}
+
\ln \! 2\pi (\beta\,\sigma_{ij})^2
\right],
\label{eq:likelihood_high}
\end{equation}
where $R_{ij}$ denotes the residual spectral matrix, $\sigma_{ij}$ represents the corresponding uncertainties obtained through error propagation, $M_{ij}$ is the filtered model matrix, $\mathbf{M}_{\mathrm{filtered}}$, and $\beta$ is the scale factor of the noise level.
The low-resolution log-likelihood is defined as
\begin{equation}
\ln \mathcal{L}_{\mathrm{low}}
=
-\frac{1}{2}
\sum_{i}
\left[
\frac{(R^{\mathrm{ecl}}_{\mathrm{i}} - \delta^{\mathrm{ecl}}_{\mathrm{i}})^2}{(\sigma{^{\mathrm{ecl}}_{\mathrm{i}}})^{2}}
+
\ln \! 2\pi (\sigma{^{\mathrm{ecl}}_{\mathrm{i}}})^{2}
\right],
\label{eq:likelihood_low}
\end{equation}
where $R^{\mathrm{ecl}}_{\mathrm{i}}$ denotes the binned observed eclipse depth, $\sigma{^{\mathrm{ecl}}_{\mathrm{i}}}$ is the corresponding uncertainty, and $\delta^{\mathrm{ecl}}_{\mathrm{i}}$ represents the modeled eclipse depth. We performed the MCMC sampling using 100 walkers with 20\,000 steps. For the final retrieval results, the first 10\,000 steps were discarded as burn-in.

\subsubsection{JWST-only retrieval} 
\label{jwst_framework}
To assess the contribution of the CRIRES$^+$ observations to the joint retrieval, we performed an independent retrieval using only the JWST NIRISS/SOSS secondary-eclipse spectrum. The atmospheric model, parameter priors, low-resolution forward model, and MCMC configuration were identical to those adopted for the JWST component of the joint retrieval described in Sect.~\ref{joint_framework}, while the CRIRES$^+$ likelihood and parameters specific to the high-resolution observations were excluded. The posterior distributions obtained from this retrieval are shown in Fig.~\ref{fig:jwst}, and the retrieved parameters are listed in Table~\ref{tab:retrieved_parameter}.

\subsubsection{High-resolution retrieval with fixed parameters} \label{fixed_framework}
To further explore the capability of CRIRES$^+$ high-resolution spectroscopy to constrain the atmosphere of a Neptune-sized planet, we performed an atmospheric retrieval using the high-resolution data alone. In addition to $\mathrm{H_2O}$ and $\mathrm{CO}$ included in the joint retrieval, we also incorporated $\mathrm{CO_2}$, $\mathrm{OH}$, $\mathrm{CH_4}$, $\mathrm{HCN}$, and $\mathrm{Fe}$ as additional free chemical species to investigate whether the CRIRES$^+$ data alone could place any meaningful constraints on their abundances. However, the S/N of the single high-resolution observation is insufficient to independently constrain all free parameters. We therefore fixed the $T$-$p$ profile and the atmospheric dynamical parameters ($K_\mathrm{p}$ and $\Delta v$) in this retrieval. These parameters were adopted from the best-fit model corresponding to the maximum a posteriori (MAP) solution of the joint retrieval. All remaining retrieval settings were kept identical to those adopted in Sect.~\ref{joint_framework}, and the adopted fixed parameter values are listed in Table~\ref{tab:retrieved_parameter}. 

\section{Results and discussion} \label{sec:results_discussion}
\subsection{Transmission spectroscopy of He\,\textsc{i}} \label{trans_reuslts}
The goal of the transmission observations of LTT~9779~b using the Y1029 wavelength setting is to search for possible absorption in the metastable \ion{He}{i} $\lambda 1083$\,nm triplet at air wavelengths of 1082.909\, nm, 1083.025\, nm, and 1083.034\, nm. For illustration purposes, instead of treating the A and B spectra as two independent datasets, we combined them by sorting in the order of phase and interpolating the \texttt{molecfit} wavelength solution of B onto A. We discarded one spectrum of A due to data loss around 1083\, nm, most likely resulting from a pipeline reduction artifact, while all other exposures remained unaffected. In total, we retained 21 in-transit spectra across all observations, based on the first and fourth contact times (T1 and T4) calculated from the parameters listed in Table~\ref{tab:ltt_params}.

We performed the full data reduction procedure (Sect.~\ref{data_reduction}) on the observed transmission spectra in the observer’s frame. We found that after three iterations, the \texttt{SYSREM} algorithm could effectively remove both stellar and telluric lines around the \ion{He}{i} $\lambda 1083$\,nm triplet. We also inspected the strong telluric OH emission line at 1083.1322\, nm (air wavelength; \citealt{OH_line}), but found no significant emission in our data. Therefore, no masking was applied to this region. The residual spectral matrix was subsequently shifted into the stellar rest frame, accounting for the barycentric Earth RV (\(v_\mathrm{bary}\)) and the stellar systemic velocity (\(\gamma = -10.6\pm1.0~\mathrm{km\,s^{-1}}\), from \textit{Gaia} DR2; \citealt{Gaia_DR2}).

\begin{figure}
    \centering
    \includegraphics[width=\columnwidth]{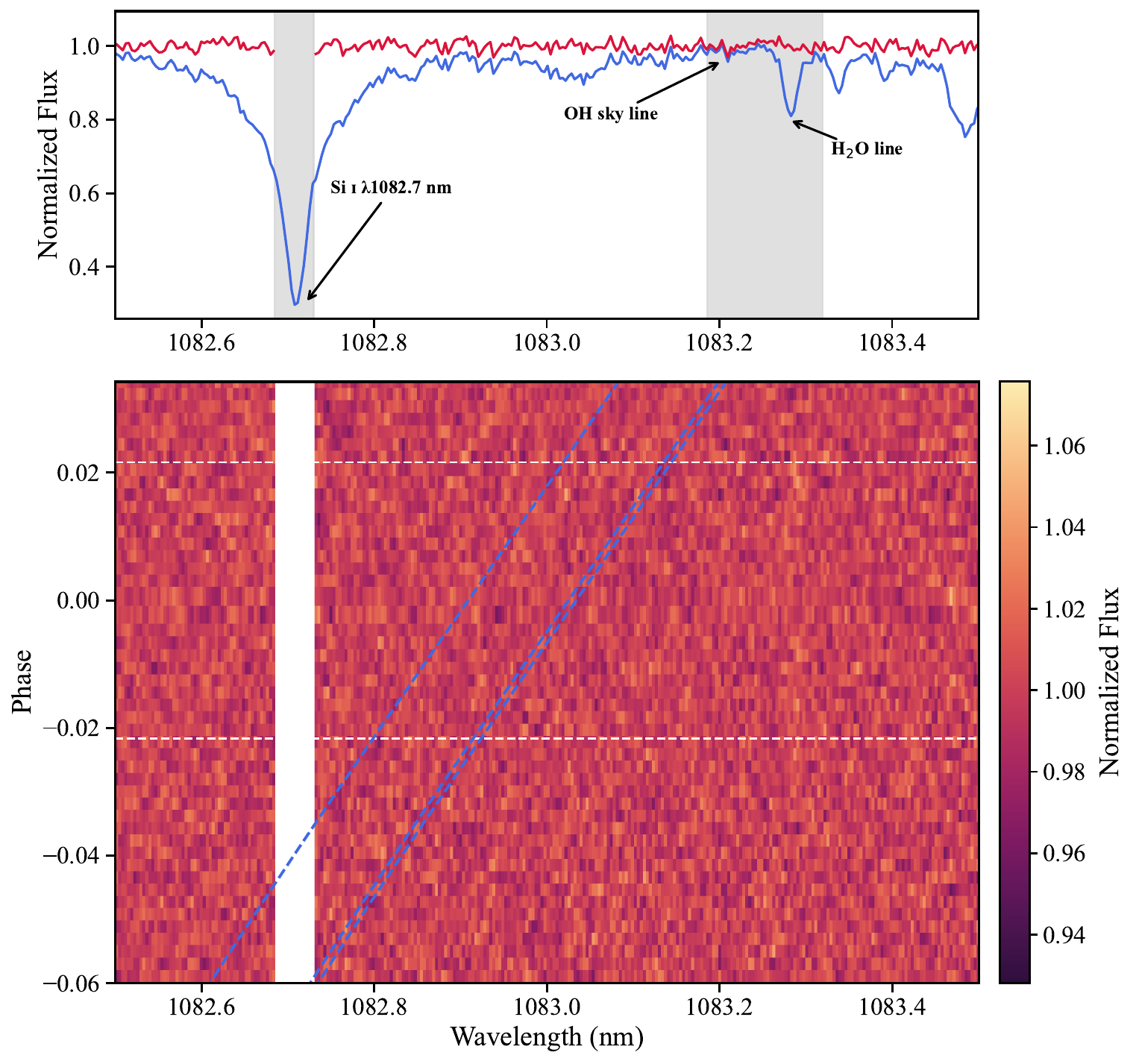}
    \caption{Transmission spectral matrix around the \ion{He}{i} $\lambda 1083$\,nm triplet in the stellar rest frame.
    Upper panel: Example of the original spectrum in the stellar rest frame (blue), together with a normalized spectrum after data reduction (red). The shaded gray regions indicate wavelength ranges strongly contaminated by stellar and telluric residuals, including the Si~{\sc i} $\lambda$1082.7\,nm, OH sky emission, and a strong $\mathrm{H_2O}$ absorption feature. The Si~{\sc i} line was automatically masked because the corresponding pixel columns had low empirical S/N, while the telluric $\mathrm{H_2O}$ absorption was removed using \texttt{SYSREM}. 
    Lower panel: Two-dimensional transmission spectral matrix as a function of orbital phase and wavelength. The horizontal dashed white lines mark the ingress and egress of the transit. The three dashed blue lines indicate the expected Doppler trail of the planetary \ion{He}{i} $\lambda 1083$\,nm triplet absorption during transit. The wavelengths in both panels are given in air.
    }
    \label{He-matrix}
\end{figure}

\begin{figure}
    \centering
    \includegraphics[width=\columnwidth]{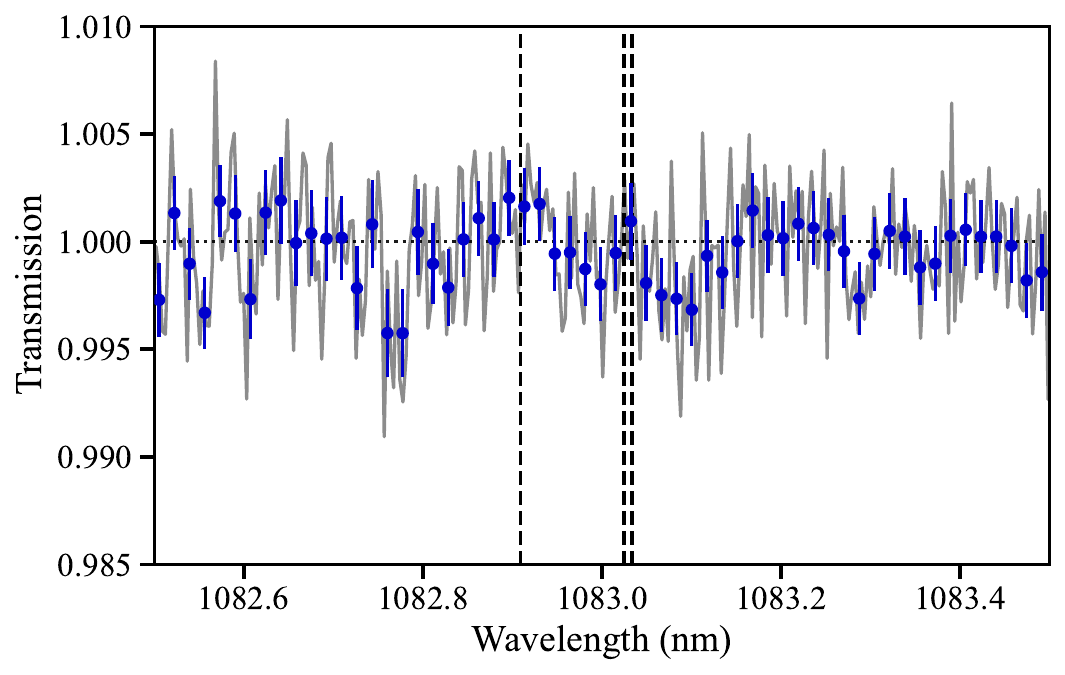}
    \caption{Combined transmission spectrum around the \ion{He}{i} $\lambda 1083$\,nm triplet. The positions of the \ion{He}{i} $\lambda 1083$\,nm triplet are indicated by the vertical dashed black lines. The horizontal dotted black line marks the unity level of the transmission spectrum. The gray line shows the original transmission spectrum, while the blue dots represent the spectrum binned over 5 pixels (\mbox{\(\sim0.017\,\mathrm{nm}\)}). The wavelengths are given in air.
    }
    \label{He-tran}
\end{figure}

The final transmission spectral matrix is shown in Fig. \ref{He-matrix}. There is no significant signal along the expected \ion{He}{i} absorption track. No influence from the OH emission line is apparent in our data, and the deep $\mathrm{Si}$ and $\mathrm{H_2O}$ absorption features are effectively masked and removed, respectively, by our data reduction procedure and do not affect the in-transit spectra. Next, we shifted the residual matrix into the planetary rest frame using the semi-amplitude, $K_\mathrm{p} = 225~\mathrm{km\,s^{-1}}$ (Table \ref{tab:ltt_params}). We then computed the error-weighted mean in-transit spectrum, where each spectrum was weighted by its inverse variance, to produce the one-dimensional transmission spectrum (Fig. \ref{He-tran}). Again, no significant absorption is observed around the \ion{He}{i} $\lambda 1083$\,nm triplet.

We further rebinned the one-dimensional planetary transmission spectrum into 5-pixel bins (0.017\,nm), and the resulting binned spectrum is presented in Fig. \ref{He-tran}. The transmission spectrum has a standard deviation of $0.15\%$, leading to a $3\sigma$ upper limit of $0.45\%$ for helium absorption. This result is consistent with that of \citet{LTT_WINERED}, who reported a $3\sigma$ upper limit of $0.20\%$ from observations with the WINERED spectrograph ($R \sim 68\,000$) on the 6.5\,m Clay/Magellan~II Telescope. Similarly, when we applied the same 0.75\,\AA\ binning, our CRIRES$^+$ data yielded a comparable value of $0.25\%$.

\citet{LTT_WINERED} further showed that, within an isothermal H/He Parker-wind framework, the mass-loss-rate upper limits inferred from the \ion{He}{i} non-detection are lower than the mass-loss rates predicted by pure H/He photoevaporation models, even after accounting for the host star's weak XUV irradiation \citep{LTT_Xray}. The non-detection of H$\alpha$ in ESPRESSO observations likewise provides no evidence of an extended atmosphere traced by excited neutral hydrogen \citep{LTT_ESPRESSO_Transit}. Taken together, the non-detections of \ion{He}{i} and H$\alpha$ disfavor a vigorous, extended H/He-dominated outflow. A high atmospheric metallicity provides a plausible explanation, as the resulting higher MMW and enhanced metal-line cooling could reduce the atmospheric scale height and suppress hydrodynamic escape, although weaker or compositionally different outflows cannot be excluded.

\begin{figure*}[htbp]
    \includegraphics[width=\textwidth]{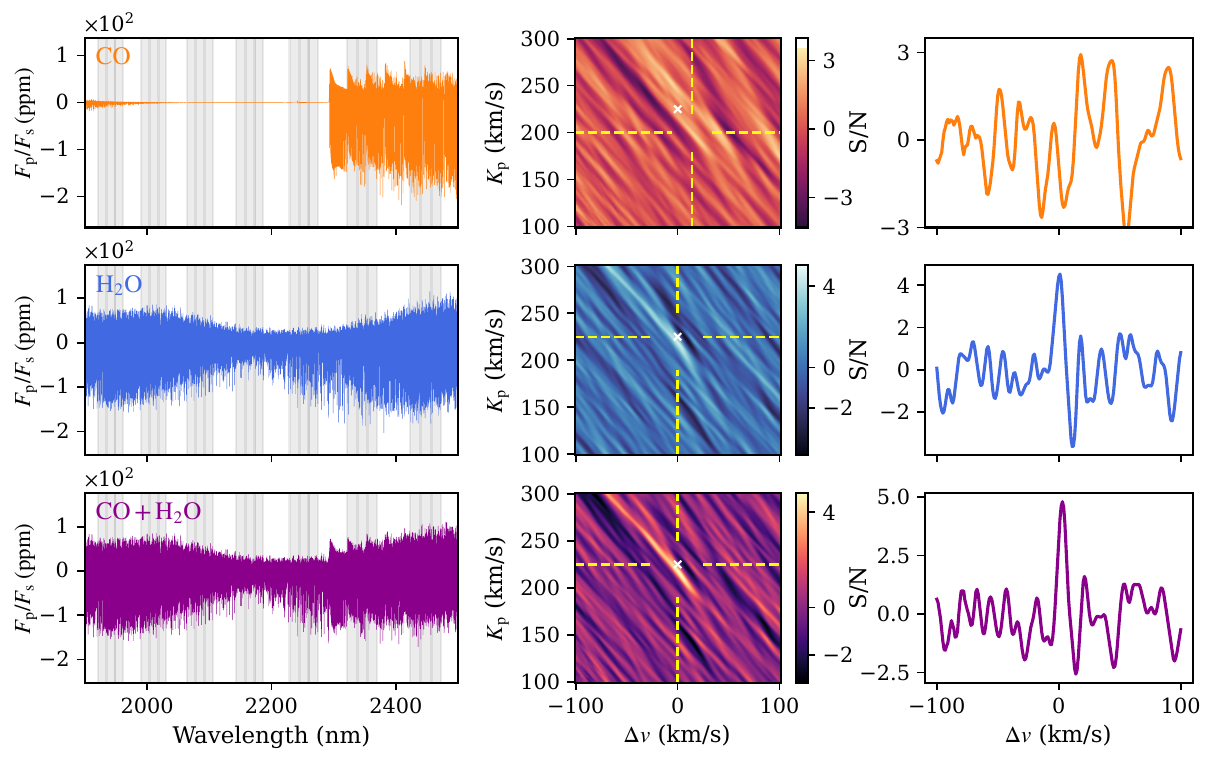}
    \caption{
    Model spectra and S/N maps for $\mathrm{CO}$, $\mathrm{H_2O}$, and both species.
    Left panels: Continuum-normalized model spectra. The shaded gray regions indicate the wavelength coverage of the CRIRES$^+$ K2166 setting used in this work.
    Middle panels: Corresponding S/N maps as a function of $K_\mathrm{p}$ and  $\Delta v$. The dashed yellow lines indicate the location of the maximum S/N, while the white crosses mark the theoretically expected planetary signal position.
    Right panels: One-dimensional CCFs extracted at the $K_\mathrm{p}$ values corresponding to the maximum S/N.
    }
    \label{CCF-results}
\end{figure*}

\begin{figure}[htbp]
    \centering
    \includegraphics[width=\columnwidth]{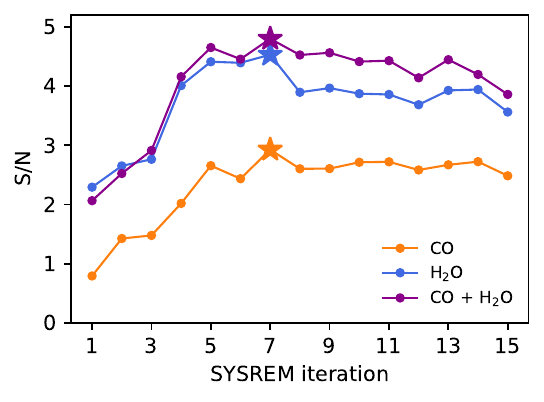}
    \caption{
    Detection significance as a function of \texttt{SYSREM} iterations for $\mathrm{CO}$, $\mathrm{H_2O}$, and both species. The S/N values correspond to the maximum S/N measured in the $K_\mathrm{p}$-$\Delta v$ maps for each \texttt{SYSREM} iteration. The star symbols indicate the \texttt{SYSREM} iteration at which each species reaches its highest S/N.
    }
    \label{CCF-sysrem}
\end{figure}
\subsection{Search for atmospheric species on the dayside} \label{emission_results}
We detect $\mathrm{H_2O}$ and find tentative evidence of $\mathrm{CO}$. The final results are shown in Fig.~\ref{CCF-results}. The $\mathrm{CO}$ signal achieves a maximum $S/N \sim 2.8$ at $K_\mathrm{p} = 200 \pm 20 ~\mathrm{km\,s^{-1}}$ and $\Delta v = 10 \pm 20 ~\mathrm{km\,s^{-1}}$, while the $\mathrm{H_2O}$ signal reaches a maximum $S/N \sim 4.5$ at $K_\mathrm{p} = 218^{+16}_{-15}~\mathrm{km\,s^{-1}}$ and $\Delta v = 1^{+8}_{-14}~\mathrm{km\,s^{-1}}$. When combining both species, we obtain a joint signal with $S/N \sim 4.8$ at $K_\mathrm{p} = 214^{+20}_{-13}~\mathrm{km\,s^{-1}}$ and $\Delta v = 3^{+9}_{-20}~\mathrm{km\,s^{-1}}$. The presence of absorption features in our detection suggests the absence of a thermal inversion in the $T$-$p$ profile.

Fig.~\ref{CCF-sysrem} shows the evolution of the maximum $S/N$ as a function of the \texttt{SYSREM} iterations for our detected signals. The $\mathrm{H_2O}$, $\mathrm{CO}$, and combined signals show similar behavior, all reaching their maximum $S/N$ after seven iterations. Comparison between the combined and individual signals indicates that $\mathrm{H_2O}$ dominated the combined signal, which is also evident in the model template (Fig.~\ref{CCF-results}). Theoretically, UHJs exhibiting thermal inversions are expected to show stronger $\mathrm{CO}$ signals than $\mathrm{H_2O}$, as water molecules are thermally dissociated in the inverted atmospheres. In contrast, our detection reveals a stronger $\mathrm{H_2O}$ signal. We speculate that this behavior arises from the absence of a thermal inversion, which prevents water from being largely dissociated.

The absorption signal of $\mathrm{H_2O}$ is consistent with previous observations obtained with JWST NIRISS/SOSS \citep{ltt_jwst_emission}. In addition, the JWST result proposed an atmospheric circulation regime with an equatorial jet that transports heat eastward from the dayside toward the cold nightside. If such an equatorial jet exists, it should manifest as a blueshifted signal in our observations since our data were taken before the secondary eclipse. However, due to the relatively large $1\sigma$ range in the $K_\mathrm{p}$-$\Delta v$ map, our single pre-eclipse observation is insufficient to place strong constraints on the atmospheric dynamics of LTT~9779~b. Future observations are expected to yield a stronger $\mathrm{H_2O}$ signal, thereby improving our ability to probe the atmospheric dynamics of LTT~9779~b.

No significant signals of $\mathrm{CO_2}$, $\mathrm{OH}$, $\mathrm{CH_4}$, $\mathrm{HCN}$, or $\mathrm{Fe}$ were detected, which may be attributed to a combination of observational and physical factors. First, the observed planet-to-star flux ratio ($F_\mathrm{p}/F_\mathrm{s}$) scales with the square of the radius ratio ($R_\mathrm{p}/R_\mathrm{s}$); therefore, for a Neptune-sized planet, the signal is intrinsically much weaker than that of a Jupiter-sized planet. Second, these molecules may lack sufficiently dense or strong spectral features within the CRIRES$^+$ $K$ band. Third, the limited S/N of our single observation may prevent the detection of weak spectral features. Finally, for species such as $\mathrm{CH_4}$ and $\mathrm{HCN}$, their atmospheric abundances are expected to be intrinsically low at the high temperatures probed by our observations.

\subsection{Joint retrieval} \label{joint_discussion}
\begin{figure*}[htbp]
    \centering
    \includegraphics[width=0.95\textwidth]{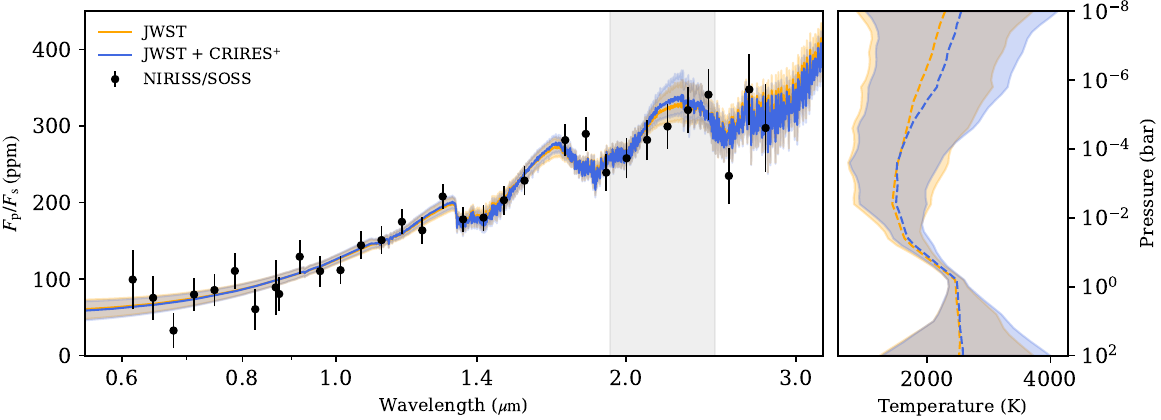}
    \caption{
    Retrieved spectra and $T$-$p$ profiles from the JWST-only retrieval (orange curves) and the joint retrieval (blue curves).
    Left panel: Retrieved model spectra derived from the median of 10\,000 random posterior draws. The black points show the JWST/NIRISS/SOSS observations, and the shaded gray region marks the wavelength coverage of the CRIRES$^+$ K2166 setting.
    Right panel: Corresponding $T$-$p$ profiles. Dashed lines indicate the median profiles, and shaded regions represent the $1\sigma$ uncertainties. The two retrievals yield broadly consistent spectra and $T$-$p$ profiles within the $1\sigma$ uncertainties.
    }
    \label{JOINT-spec}
\end{figure*}

We present the posterior probability distributions obtained from the joint retrieval in Fig.~\ref{fig:joint}, with a summary of the retrieved parameters listed in Table~\ref{tab:retrieved_parameter}. The retrieved spectra and corresponding $T$-$p$ profiles from the joint and JWST-only retrievals are compared in Fig.~\ref{JOINT-spec}. The $T$-$p$ structure and $\mathrm{H_2O}$ abundance obtained from the joint retrieval are broadly consistent within $1\sigma$ with those from the JWST-only retrieval, indicating that the low-resolution JWST data primarily drive the constraints on these quantities in the joint analysis.

The $T$-$p$ profile is relatively well constrained over the approximate pressure range 0.05--0.6~bar, while the constraints at other atmospheric levels remain weak. The $T$-$p$ profile shows that the detected signal originates from a non-inverted atmospheric region, in agreement with the detection from the CRIRES$^+$ CCF results. Thermal inversions are generally favored when incident stellar radiation is absorbed at low pressures, particularly when the short-wavelength absorption opacity is large relative to the thermal infrared opacity \citep{Guillot2010}. Self-consistent models of ultrahot giant planets further show that the absorption of short-wavelength stellar radiation by atomic metals can produce thermal inversions even for planets orbiting G-type stars \citep{TP_Gstar}. The absence of an inversion over the pressure range probed here is therefore noteworthy. The relatively high reflectivity of LTT~9779~b may partly explain this result. Assuming Lambertian scattering, for which the Bond and geometric albedos are related through $A_\mathrm{B}=3/2\,A_\mathrm{g}$, the retrieved geometric albedo of $A_\mathrm{g}=0.37 \pm 0.10$ implies an equilibrium temperature of $T_\mathrm{eq}\sim1900$~K (Table~\ref{tab:ltt_params}). This places LTT~9779~b closer to the temperature regime of typical hot Jupiters, for which no robust evidence of thermal inversions has been reported to date \citep{EA_Review_Madhu, HR_Review_Snellen}. Several cloud-related processes could contribute to this cooler, non-inverted atmosphere. First, \citet{LTT_JWST_Transit} proposed a positive feedback mechanism in which clouds increase the planetary albedo by reflecting incident stellar radiation. The resulting reduction in absorbed stellar flux cools the atmosphere, which in turn facilitates further cloud formation. Second, preliminary HELIOS calculations for LTT~9779~b suggest that, for the adopted cloud properties, clouds warm the atmospheric layers below the cloud deck while cooling those above it. This behavior is consistent with the effects of enhanced long-wavelength cloud opacity predicted by semi-analytical cloudy-atmosphere models \citep{Heng2012_cloud}. By modifying the vertical temperature gradient, this radiative effect may weaken or suppress a thermal inversion driven by high-altitude short-wavelength absorbers. Finally, condensation or cold trapping may deplete iron and other atomic metals from the upper atmosphere, removing some of the principal short-wavelength absorbers thought to drive thermal inversions in ultrahot giant planets \citep{TP_UHJ_3, Parmentier2018_TP_inversion}. The combined effects of enhanced reflection, cloud-induced radiative redistribution, and the depletion of gaseous optical absorbers may therefore make the atmosphere of LTT~9779~b cooler and less favorable for the development of a thermal inversion than would otherwise be expected.

\begin{figure}[htbp]
    \centering
    \includegraphics[width=\columnwidth]{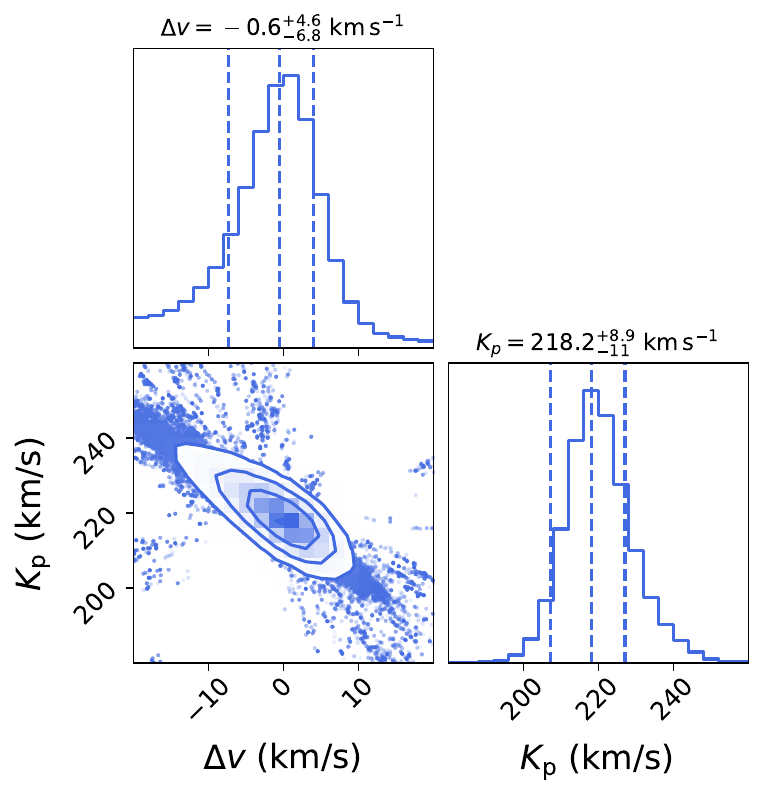}
    \caption{
   Corner plot of the posterior distributions of the planetary orbital velocity semi-amplitude, $K_\mathrm{p}$, and the velocity offset, $\Delta v$, obtained from the joint retrieval.
    }
    \label{JOINT-Kp}
\end{figure}

Although the joint retrieval is primarily driven by the JWST NIRISS/SOSS data, the consistency of the retrieved parameters across different datasets indicates good agreement between the ground-based and space-based telescopes. Notably, although the planetary orbital velocity semi-amplitude, $K_\mathrm{p}$, in the joint retrieval is constrained solely by the high-resolution likelihood, the resulting posterior distribution is well constrained, yielding $K_\mathrm{p}=218.2^{+8.9}_{-11}~\mathrm{km\,s^{-1}}$ and $\Delta v=-0.6^{+4.6}_{-6.8}~\mathrm{km\,s^{-1}}$. The retrieved $K_\mathrm{p}$ is consistent within $1\sigma$ with the expected orbital value ($225~\mathrm{km\,s^{-1}}$; Fig.~\ref{JOINT-Kp}), while the velocity offset is consistent with zero. This consistency further supports the planetary origin of the high-resolution signal.

Due to the large uncertainties in the two-dimensional posterior distribution of $K_\mathrm{p}$ and $\Delta v$ (Fig.~\ref{JOINT-Kp}), the S/N of a single pre-eclipse observation is insufficient to place meaningful constraints on the atmospheric dynamics traced by $\mathrm{H_2O}$. If the posterior distribution in the $K_\mathrm{p}$-$\Delta v$ plane were to shift by more than $2\sigma$ from the theoretically expected location, such a deviation may indicate a departure from the local slope of the Keplerian RV curve, which could plausibly be attributed to atmospheric circulation or spatial inhomogeneities in the planetary atmosphere \citep{WASP18_IGRINS}. Consequently, for the eastward equatorial jet scenario proposed by \citet{ltt_jwst_emission}, as discussed in Sect.~\ref{emission_results}, the current observations do not allow for a definitive conclusion. 

\begin{figure*}[htbp]
    \sidecaption
    \includegraphics[width=12cm]{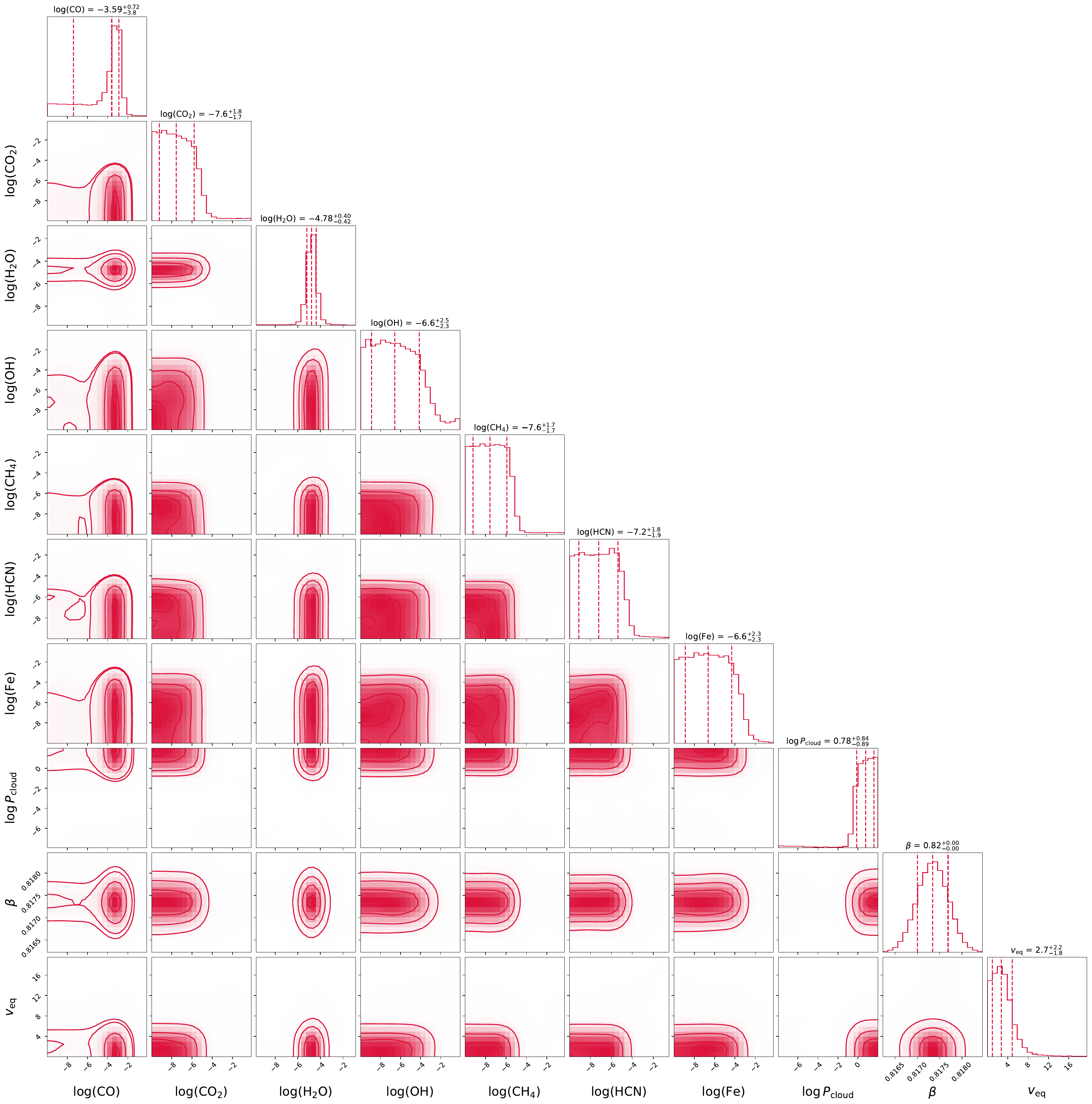}
    \caption{
    Corner plot of the posterior distributions obtained from the CRIRES$^+$-only retrieval, with the $T$-$p$ profile, $K_\mathrm{p}$, and $\Delta v$ fixed to the MAP values derived from the joint retrieval.
    }
    \label{CRIRES-fixed}
\end{figure*}
\subsection{High-resolution retrieval with fixed parameters} \label{fixed_discussion}
The results of the CRIRES$^+$-only retrieval are shown in Fig.~\ref{CRIRES-fixed}. In this analysis, the abundance of $\mathrm{H_2O}$ is relatively well constrained, whereas only an upper limit is obtained for $\mathrm{CO}$. Combined with the consistency between the expected orbital parameters and the retrieved $K_\mathrm{p}$ from the joint retrieval, we conclude that the high-resolution data support the presence of $\mathrm{H_2O}$ in the planetary atmosphere, while only tentative evidence of $\mathrm{CO}$ is found.

Our retrieval yields a subsolar $\mathrm{H_2O}$ abundance of $\mathrm{log\,(H_2O)} = -4.78^{+0.40}_{-0.42}$, compared with a solar-composition equilibrium value of $\mathrm{log\,(H_2O)} = -3.46$. This reference value was calculated with \texttt{FastChem} from 10\,000 joint-posterior $T$-$p$ profiles, averaged over the approximate pressure range 0.05--0.6~bar. The low $\mathrm{H_2O}$ abundance, together with a relatively deep cloud-top pressure ($\log P_{\mathrm{cloud}} > -0.11$ at $1\sigma$), naturally explains the comparatively weak water line depths observed in our data. However, as the retrieval probes the disk-integrated dayside emission, the retrieved cloud-top pressure should be interpreted as a flux-weighted average over the visible hemisphere and does not exclude the presence of localized high-altitude cloud patches. This result is also consistent within the $1\sigma$ uncertainties with the retrieval performed using the JWST NIRISS/SOSS data alone. In contrast, the $\mathrm{CO}$ abundance is only weakly constrained by the CRIRES$^+$ data alone, with $\mathrm{log\,(CO)} = -3.59^{+0.72}_{-3.8}$. Despite this weak constraint, the retrieved abundance remains consistent at the $1\sigma$ level with the CO-rich dayside solution inferred from the JWST/NIRSpec retrieval analysis of \citet{LTT_JWST_NIRSpec} ($\mathrm{log\,(CO)} = -1.44^{+0.99}_{-1.59}$). Taken together, the weak water features found in our analysis and the CO-rich dayside atmosphere inferred from JWST/NIRSpec support a picture of a metal-enriched atmosphere with a carbon-to-oxygen ratio near unity for LTT~9779~b.

For the remaining species, we derive only upper limits on their abundances. The posterior distributions of $\mathrm{CO_2}$, $\mathrm{OH}$, $\mathrm{CH_4}$, $\mathrm{HCN}$, and $\mathrm{Fe}$ are broad, indicating that the CRIRES$^+$ data alone do not provide significant constraints on their abundances. From the marginalized posteriors, we obtain approximate $1\sigma$ upper limits of 
$\mathrm{log\,(CO_2)} \lesssim -5.8$, 
$\mathrm{log\,(OH)} \lesssim -4.1$, 
$\mathrm{log\,(CH_4)} \lesssim -5.9$, 
$\mathrm{log\,(HCN)} \lesssim -5.4$, and 
$\mathrm{log\,(Fe)} \lesssim -4.3$. 
These weak constraints are consistent with the absence of significant CCF peaks for these species, indicating that their abundances remain largely unconstrained by the current CRIRES$^+$ data.

The retrieved planetary equatorial rotational velocities, 
$v_{\rm eq}=4.9^{+4.4}_{-3.0}~\mathrm{km\,s^{-1}}$ from the joint retrieval and 
$v_{\rm eq}=2.7^{+2.2}_{-1.8}~\mathrm{km\,s^{-1}}$ from the CRIRES$^+$-only retrieval, 
are both consistent within $1\sigma$ with the synchronous rotation velocity expected for tidal locking 
($\sim2.8~\mathrm{km\,s^{-1}}$). However, given the large uncertainties, the current data do not allow us to robustly constrain any jet-induced broadening in the planetary spectrum. Recent high-resolution emission spectroscopy has nevertheless revealed atmospheric wind structures in UHJs \citep[e.g.,][]{W189_Fabio, KELT9_KPF}. Future high-resolution observations of LTT~9779~b may therefore help to further probe its atmospheric dynamics.

These observational ambiguities highlight the importance of future high-resolution ground-based observations spanning multiple orbital phases: on the one hand, phase-resolved measurements would help to simultaneously constrain the water abundance and cloud locations, thereby improving our understanding of the feedback of clouds on the atmospheric thermal structure; on the other hand, they would enable tests of the presence of equatorial jets and thus provide a direct test of current general circulation model predictions \citep{GCM_1, GCM_2}.


\section{Conclusions} \label{conclusion}
In this work, we presented high-resolution CRIRES$^+$ transit and dayside emission spectroscopy of the ultrahot Neptune LTT~9779~b. During the transit observation, we searched for absorption in the metastable \ion{He}{i} $\lambda 1083$\,nm triplet and found no significant signal. From the combined transmission spectrum, we derived a $3\sigma$ upper limit of 0.45\% for the helium absorption depth, consistent with previous ground-based non-detections and suggesting that the planet is unlikely to be undergoing strong, extended H/He-dominated atmospheric escape at present.

For the dayside observation, we performed a cross-correlation analysis in the CRIRES$^+$ $K$-band spectra to search for molecular emission or absorption signatures. We found an absorption signal of $\mathrm{H_2O}$ with a S/N of $\sim 4.5$ and a tentative signal of $\mathrm{CO}$.
The combined signal of $\mathrm{H_2O}$ and CO reaches a S/N of $\sim 4.8$. The location of the signal in the $K_\mathrm{p}$-$\Delta v$ map is broadly consistent with the expected planetary trail.

To further assess the robustness of the high-resolution signal, we performed atmospheric retrievals combining the CRIRES$^+$ data with JWST NIRISS/SOSS secondary-eclipse observations. The joint retrieval yields a non-inverted $T$-$p$ profile and a relatively low $\mathrm{H_2O}$ abundance, consistent with the weak water features seen in both the low- and high-resolution data. In addition, the retrieved $K_\mathrm{p}$ is consistent with the theoretical orbital value within the uncertainties, further supporting the interpretation that the high-resolution signal arises from the dayside atmosphere of LTT~9779~b. A CRIRES$^+$-only retrieval with the $T$-$p$ profile, $K_\mathrm{p}$, and $\Delta v$ fixed to the joint-retrieval values independently recovers the $\mathrm{H_2O}$ signal, while the evidence of $\mathrm{CO}$ remains marginal.

Overall, our results indicate the presence of weak $\mathrm{H_2O}$ absorption in the dayside atmosphere of LTT~9779~b and suggest the absence of a strong thermal inversion in the pressure range probed by the observations. This picture is consistent with a cloudy, high-albedo atmosphere, where enhanced reflection may reduce the absorbed stellar energy. Cloud-induced radiative effects may also modify the vertical temperature gradient in a way that disfavors the development of a strong thermal inversion.

Our study demonstrates that ground-based high-resolution spectroscopy can probe molecular signatures in Neptune-sized exoplanets, even when the planet-to-star flux ratio is extremely small. More importantly, these results highlight the complementary strengths of joint retrievals combining ground-based and space-based observations. Although the continuum information is removed from the high-resolution spectra during normalization, ground-based spectroscopy remains highly sensitive to the relative depths and contrasts of individual molecular lines and provides direct constraints on the planetary orbital velocity and possible atmospheric dynamics. In contrast, space-based low-resolution observations retain the continuum information and are therefore better suited to constraining the broadband spectral shape, reflected-light contribution, and thermal structure of deeper atmospheric layers. However, the current constraints from the CRIRES$^+$ data are still limited by the S/N of a single pre-eclipse observation. Future phase-resolved CRIRES$^+$ observations, with broader orbital-phase coverage and higher cumulative S/N, will be essential for confirming the molecular detections, improving abundance constraints, and probing the atmospheric dynamics of this rare ultrahot Neptune.


\begin{acknowledgements}
We acknowledge the support by the National Natural Science Foundation of China (grant no. 42375118). 
CRIRES$^+$ is an ESO upgrade project carried out by Th\"uringer Landessternwarte Tautenburg, Georg-August Universit\"at G\"ottingen, and Uppsala University. The project is funded by the Federal Ministry of Education and Research (Germany) through Grants 05A11MG3, 05A14MG4, 05A17MG2 and the Knut and Alice Wallenberg Foundation. Based on observations collected at the European Organisation for Astronomical Research in the Southern Hemisphere under ESO programmes 109.23HN.003 and 111.254J.003.
This work was co-funded by the European Union (ERC-CoG, EVAPORATOR, Grant agreement No. 101170037). Views and opinions expressed are however those of the author(s) only and do not necessarily reflect those of the European Union or the European Research Council.
D.S. acknowledges support from the project
PID2021-126365NB-C21 (MCI/AEI/FEDER, UE) and the Severo Ochoa grant
CEX2021-001131-S funded by MCIN/AEI/10.13039/501100011033.
O.K. acknowledges support by the Swedish Research Council (grant agreement no. 2023-03667) and by the Swedish National Space Agency.
\end{acknowledgements}
\bibliographystyle{aa}
\bibliography{ref}

@ARTICLE{ND1,
       author = {{Szab{\'o}}, Gy. M. and {Kiss}, L.~L.},
        title = "{A Short-period Censor of Sub-Jupiter Mass Exoplanets with Low Density}",
      journal = {\apjl},
         year = 2011,
        month = feb,
       volume = {727},
       number = {2},
          eid = {L44},
        pages = {L44},
          doi = {10.1088/2041-8205/727/2/L44},
archivePrefix = {arXiv},
       eprint = {1012.4791},
 primaryClass = {astro-ph.EP},
       adsurl = {https://ui.adsabs.harvard.edu/abs/2011ApJ...727L..44S}
}

@ARTICLE{ND_boundary,
       author = {{Mazeh}, T. and {Holczer}, T. and {Faigler}, S.},
        title = "{Dearth of short-period Neptunian exoplanets: A desert in period-mass and period-radius planes}",
      journal = {\aap},
         year = 2016,
        month = may,
       volume = {589},
          eid = {A75},
        pages = {A75},
          doi = {10.1051/0004-6361/201528065},
archivePrefix = {arXiv},
       eprint = {1602.07843},
 primaryClass = {astro-ph.EP},
       adsurl = {https://ui.adsabs.harvard.edu/abs/2016A&A...589A..75M}
}

@ARTICLE{ND3,
       author = {{Lecavelier Des Etangs}, A.},
        title = "{A diagram to determine the evaporation status of extrasolar planets}",
      journal = {\aap},
         year = 2007,
        month = jan,
       volume = {461},
       number = {3},
        pages = {1185-1193},
          doi = {10.1051/0004-6361:20065014},
archivePrefix = {arXiv},
       eprint = {astro-ph/0609744},
 primaryClass = {astro-ph},
       adsurl = {https://ui.adsabs.harvard.edu/abs/2007A&A...461.1185L}
}

@ARTICLE{ND4,
       author = {{Beaug{\'e}}, C. and {Nesvorn{\'y}}, D.},
        title = "{Emerging Trends in a Period-Radius Distribution of Close-in Planets}",
      journal = {\apj},
         year = 2013,
        month = jan,
       volume = {763},
       number = {1},
          eid = {12},
        pages = {12},
          doi = {10.1088/0004-637X/763/1/12},
archivePrefix = {arXiv},
       eprint = {1211.4533},
 primaryClass = {astro-ph.EP},
       adsurl = {https://ui.adsabs.harvard.edu/abs/2013ApJ...763...12B}
}

@ARTICLE{ND_Photoevaporation1,
       author = {{Owen}, James E. and {Wu}, Yanqin},
        title = "{Kepler Planets: A Tale of Evaporation}",
      journal = {\apj},
         year = 2013,
        month = oct,
       volume = {775},
       number = {2},
          eid = {105},
        pages = {105},
          doi = {10.1088/0004-637X/775/2/105},
archivePrefix = {arXiv},
       eprint = {1303.3899},
 primaryClass = {astro-ph.EP},
       adsurl = {https://ui.adsabs.harvard.edu/abs/2013ApJ...775..105O}
}

@ARTICLE{ND_Photoevaporation2,
       author = {{Lopez}, Eric D. and {Fortney}, Jonathan J.},
        title = "{The Role of Core Mass in Controlling Evaporation: The Kepler Radius Distribution and the Kepler-36 Density Dichotomy}",
      journal = {\apj},
         year = 2013,
        month = oct,
       volume = {776},
       number = {1},
          eid = {2},
        pages = {2},
          doi = {10.1088/0004-637X/776/1/2},
archivePrefix = {arXiv},
       eprint = {1305.0269},
 primaryClass = {astro-ph.EP},
       adsurl = {https://ui.adsabs.harvard.edu/abs/2013ApJ...776....2L}
}

@ARTICLE{ND_Photoevaporation3,
       author = {{Owen}, James E. and {Lai}, Dong},
        title = "{Photoevaporation and high-eccentricity migration created the sub-Jovian desert}",
      journal = {\mnras},
         year = 2018,
        month = oct,
       volume = {479},
       number = {4},
        pages = {5012-5021},
          doi = {10.1093/mnras/sty1760},
archivePrefix = {arXiv},
       eprint = {1807.00012},
 primaryClass = {astro-ph.EP},
       adsurl = {https://ui.adsabs.harvard.edu/abs/2018MNRAS.479.5012O}
}

@ARTICLE{ND_Roche1,
       author = {{Kurokawa}, H. and {Nakamoto}, T.},
        title = "{Mass-loss Evolution of Close-in Exoplanets: Evaporation of Hot Jupiters and the Effect on Population}",
      journal = {\apj},
         year = 2014,
        month = mar,
       volume = {783},
       number = {1},
          eid = {54},
        pages = {54},
          doi = {10.1088/0004-637X/783/1/54},
archivePrefix = {arXiv},
       eprint = {1401.2511},
 primaryClass = {astro-ph.EP},
       adsurl = {https://ui.adsabs.harvard.edu/abs/2014ApJ...783...54K}
}

@ARTICLE{ND_Roche2,
       author = {{Koskinen}, Tommi T. and {Lavvas}, Panayotis and {Huang}, Chenliang and {Bergsten}, Galen and {Fernandes}, Rachel B. and {Young}, Mitchell E.},
        title = "{Mass Loss by Atmospheric Escape from Extremely Close-in Planets}",
      journal = {\apj},
         year = 2022,
        month = apr,
       volume = {929},
       number = {1},
          eid = {52},
        pages = {52},
          doi = {10.3847/1538-4357/ac4f45},
archivePrefix = {arXiv},
       eprint = {2203.06302},
 primaryClass = {astro-ph.EP},
       adsurl = {https://ui.adsabs.harvard.edu/abs/2022ApJ...929...52K}
}

@ARTICLE{ND_Roche3,
       author = {{Thorngren}, Daniel P. and {Lee}, Eve J. and {Lopez}, Eric D.},
        title = "{Removal of Hot Saturns in Mass-Radius Plane by Runaway Mass Loss}",
      journal = {\apjl},
         year = 2023,
        month = mar,
       volume = {945},
       number = {2},
          eid = {L36},
        pages = {L36},
          doi = {10.3847/2041-8213/acbd35},
archivePrefix = {arXiv},
       eprint = {2211.11770},
 primaryClass = {astro-ph.EP},
       adsurl = {https://ui.adsabs.harvard.edu/abs/2023ApJ...945L..36T}
}

@ARTICLE{ND_Eccentricity,
       author = {{Matsakos}, Titos and {K{\"o}nigl}, Arieh},
        title = "{On the Origin of the Sub-Jovian Desert in the Orbital-period-Planetary-mass Plane}",
      journal = {\apjl},
         year = 2016,
        month = mar,
       volume = {820},
       number = {1},
          eid = {L8},
        pages = {L8},
          doi = {10.3847/2041-8205/820/1/L8},
archivePrefix = {arXiv},
       eprint = {1603.00414},
 primaryClass = {astro-ph.EP},
       adsurl = {https://ui.adsabs.harvard.edu/abs/2016ApJ...820L...8M}
}

@ARTICLE{ND_planet1,
       author = {{Berger}, Travis A. and {Huber}, Daniel and {Gaidos}, Eric and {van Saders}, Jennifer L.},
        title = "{Revised Radii of Kepler Stars and Planets Using Gaia Data Release 2}",
      journal = {\apj},
         year = 2018,
        month = oct,
       volume = {866},
       number = {2},
          eid = {99},
        pages = {99},
          doi = {10.3847/1538-4357/aada83},
archivePrefix = {arXiv},
       eprint = {1805.00231},
 primaryClass = {astro-ph.EP},
       adsurl = {https://ui.adsabs.harvard.edu/abs/2018ApJ...866...99B}
}

@ARTICLE{ND_planet2,
       author = {{Livingston}, John H. and {Crossfield}, Ian J.~M. and {Petigura}, Erik A. and {Gonzales}, Erica J. and {Ciardi}, David R. and {Beichman}, Charles A. and {Christiansen}, Jessie L. and {Dressing}, Courtney D. and {Henning}, Thomas and {Howard}, Andrew W. and {Isaacson}, Howard and {Fulton}, Benjamin J. and {Kosiarek}, Molly and {Schlieder}, Joshua E. and {Sinukoff}, Evan and {Tamura}, Motohide},
        title = "{Sixty Validated Planets from K2 Campaigns 5-8}",
      journal = {\aj},
         year = 2018,
        month = dec,
       volume = {156},
       number = {6},
          eid = {277},
        pages = {277},
          doi = {10.3847/1538-3881/aae778},
archivePrefix = {arXiv},
       eprint = {1810.04074},
 primaryClass = {astro-ph.EP},
       adsurl = {https://ui.adsabs.harvard.edu/abs/2018AJ....156..277L}
}

@ARTICLE{ND_planet3,
       author = {{Barrag{\'a}n}, O. and {Aigrain}, S. and {Kubyshkina}, D. and {Gandolfi}, D. and {Livingston}, J. and {Fridlund}, M.~C.~V. and {Fossati}, L. and {Korth}, J. and {Parviainen}, H. and {Malavolta}, L. and {Palle}, E. and {Deeg}, H.~J. and {Nowak}, G. and {Rajpaul}, V.~M. and {Zicher}, N. and {Antoniciello}, G. and {Narita}, N. and {Albrecht}, S. and {Bedin}, L.~R. and {Cabrera}, J. and {Cochran}, W.~D. and {de Leon}, J. and {Eigm{\"u}ller}, Ph and {Fukui}, A. and {Granata}, V. and {Grziwa}, S. and {Guenther}, E. and {Hatzes}, A.~P. and {Kusakabe}, N. and {Latham}, D.~W. and {Libralato}, M. and {Luque}, R. and {Monta{\~n}{\'e}s-Rodr{\'\i}guez}, P. and {Murgas}, F. and {Nardiello}, D. and {Pagano}, I. and {Piotto}, G. and {Persson}, C.~M. and {Redfield}, S. and {Tamura}, M.},
        title = "{Radial velocity confirmation of K2-100b: a young, highly irradiated, and low-density transiting hot Neptune}",
      journal = {\mnras},
         year = 2019,
        month = nov,
       volume = {490},
       number = {1},
        pages = {698-708},
          doi = {10.1093/mnras/stz2569},
archivePrefix = {arXiv},
       eprint = {1909.05252},
 primaryClass = {astro-ph.EP},
       adsurl = {https://ui.adsabs.harvard.edu/abs/2019MNRAS.490..698B}
}

@ARTICLE{ND_planet4,
       author = {{Armstrong}, David J. and {Lopez}, Th{\'e}o A. and {Adibekyan}, Vardan and {Booth}, Richard A. and {Bryant}, Edward M. and {Collins}, Karen A. and {Deleuil}, Magali and {Emsenhuber}, Alexandre and {Huang}, Chelsea X. and {King}, George W. and {Lillo-Box}, Jorge and {Lissauer}, Jack J. and {Matthews}, Elisabeth and {Mousis}, Olivier and {Nielsen}, Louise D. and {Osborn}, Hugh and {Otegi}, Jon and {Santos}, Nuno C. and {Sousa}, S{\'e}rgio G. and {Stassun}, Keivan G. and {Veras}, Dimitri and {Ziegler}, Carl and {Acton}, Jack S. and {Almenara}, Jose M. and {Anderson}, David R. and {Barrado}, David and {Barros}, Susana C.~C. and {Bayliss}, Daniel and {Belardi}, Claudia and {Bouchy}, Francois and {Brice{\~n}o}, C{\'e}sar and {Brogi}, Matteo and {Brown}, David J.~A. and {Burleigh}, Matthew R. and {Casewell}, Sarah L. and {Chaushev}, Alexander and {Ciardi}, David R. and {Collins}, Kevin I. and {Col{\'o}n}, Knicole D. and {Cooke}, Benjamin F. and {Crossfield}, Ian J.~M. and {D{\'\i}az}, Rodrigo F. and {Delgado Mena}, Elisa and {Demangeon}, Olivier D.~S. and {Dorn}, Caroline and {Dumusque}, Xavier and {Eigm{\"u}ller}, Philipp and {Fausnaugh}, Michael and {Figueira}, Pedro and {Gan}, Tianjun and {Gandhi}, Siddharth and {Gill}, Samuel and {Gonzales}, Erica J. and {Goad}, Michael R. and {G{\"u}nther}, Maximilian N. and {Helled}, Ravit and {Hojjatpanah}, Saeed and {Howell}, Steve B. and {Jackman}, James and {Jenkins}, James S. and {Jenkins}, Jon M. and {Jensen}, Eric L.~N. and {Kennedy}, Grant M. and {Latham}, David W. and {Law}, Nicholas and {Lendl}, Monika and {Lozovsky}, Michael and {Mann}, Andrew W. and {Moyano}, Maximiliano and {McCormac}, James and {Meru}, Farzana and {Mordasini}, Christoph and {Osborn}, Ares and {Pollacco}, Don and {Queloz}, Didier and {Raynard}, Liam and {Ricker}, George R. and {Rowden}, Pamela and {Santerne}, Alexandre and {Schlieder}, Joshua E. and {Seager}, Sara and {Sha}, Lizhou and {Tan}, Thiam-Guan and {Tilbrook}, Rosanna H. and {Ting}, Eric and {Udry}, St{\'e}phane and {Vanderspek}, Roland and {Watson}, Christopher A. and {West}, Richard G. and {Wilson}, Paul A. and {Winn}, Joshua N. and {Wheatley}, Peter and {Villasenor}, Jesus Noel and {Vines}, Jose I. and {Zhan}, Zhuchang},
        title = "{A remnant planetary core in the hot-Neptune desert}",
      journal = {\nat},
         year = 2020,
        month = jul,
       volume = {583},
       number = {7814},
        pages = {39-42},
          doi = {10.1038/s41586-020-2421-7},
archivePrefix = {arXiv},
       eprint = {2003.10314},
 primaryClass = {astro-ph.EP},
       adsurl = {https://ui.adsabs.harvard.edu/abs/2020Natur.583...39A}
}

@ARTICLE{ND_Planet5,
       author = {{Persson}, Carina M. and {Georgieva}, Iskra Y. and {Gandolfi}, Davide and {Acuna}, Lorena and {Aguichine}, Artem and {Muresan}, Alexandra and {Guenther}, Eike and {Livingston}, John and {Collins}, Karen A. and {Dai}, Fei and {Fridlund}, Malcolm and {Goffo}, Elisa and {Jenkins}, James S. and {Kab{\'a}th}, Petr and {Korth}, Judith and {Levine}, Alan M. and {Serrano}, Luisa M. and {Vines}, Jos{\'e} and {Barragan}, Oscar and {Carleo}, Ilaria and {Colon}, Knicole D. and {Cochran}, William D. and {Christiansen}, Jessie L. and {Deeg}, Hans J. and {Deleuil}, Magali and {Dragomir}, Diana and {Esposito}, Massimiliamo and {Gan}, Tianjun and {Grziwa}, Sascha and {Hatzes}, Artie P. and {Hesse}, Katharine and {Horne}, Keith and {Jenkins}, Jon M. and {Kielkopf}, John F. and {Klagyivik}, P. and {Lam}, Kristine W.~F. and {Latham}, David W. and {Luque}, Rafa and {Orell-Miquel}, Jaume and {Mortier}, Annelies and {Mousis}, Olivier and {Narita}, Noria and {Osborne}, Hannah L.~M. and {Palle}, Enric and {Papini}, Riccardo and {Ricker}, George R. and {Schmerling}, Hendrik and {Seager}, Sara and {Stassun}, Keivan G. and {Van Eylen}, Vincent and {Vanderspek}, Roland and {Wang}, Gavin and {Winn}, Joshua N. and {Wohler}, Bill and {Zambelli}, Roberto and {Ziegler}, Carl},
        title = "{TOI-2196 b: Rare planet in the hot Neptune desert transiting a G-type star}",
      journal = {\aap},
         year = 2022,
        month = oct,
       volume = {666},
          eid = {A184},
        pages = {A184},
          doi = {10.1051/0004-6361/202244118},
archivePrefix = {arXiv},
       eprint = {2208.05797},
 primaryClass = {astro-ph.EP},
       adsurl = {https://ui.adsabs.harvard.edu/abs/2022A&A...666A.184P}
}

@ARTICLE{ND_Planet6,
       author = {{Naponiello}, Luca and {Mancini}, Luigi and {Sozzetti}, Alessandro and {Bonomo}, Aldo S. and {Morbidelli}, Alessandro and {Dou}, Jingyao and {Zeng}, Li and {Leinhardt}, Zoe M. and {Biazzo}, Katia and {Cubillos}, Patricio E. and {Pinamonti}, Matteo and {Locci}, Daniele and {Maggio}, Antonio and {Damasso}, Mario and {Lanza}, Antonino F. and {Lissauer}, Jack J. and {Collins}, Karen A. and {Carter}, Philip J. and {Jensen}, Eric L.~N. and {Bignamini}, Andrea and {Boschin}, Walter and {Bouma}, Luke G. and {Ciardi}, David R. and {Cosentino}, Rosario and {Crossfield}, Ian and {Desidera}, Silvano and {Dumusque}, Xavier and {Fiorenzano}, Aldo F.~M. and {Fukui}, Akihiko and {Giacobbe}, Paolo and {Gnilka}, Crystal L. and {Ghedina}, Adriano and {Guilluy}, Gloria and {Harutyunyan}, Avet and {Howell}, Steve B. and {Jenkins}, Jon M. and {Lund}, Michael B. and {Kielkopf}, John F. and {Lester}, Katie V. and {Malavolta}, Luca and {Mann}, Andrew W. and {Matson}, Rachel A. and {Matthews}, Elisabeth C. and {Nardiello}, Domenico and {Narita}, Norio and {Pace}, Emanuele and {Pagano}, Isabella and {Palle}, Enric and {Pedani}, Marco and {Seager}, Sara and {Schlieder}, Joshua E. and {Schwarz}, Richard P. and {Shporer}, Avi and {Twicken}, Joseph D. and {Winn}, Joshua N. and {Ziegler}, Carl and {Zingales}, Tiziano},
        title = "{A super-massive Neptune-sized planet}",
      journal = {\nat},
         year = 2023,
        month = oct,
       volume = {622},
       number = {7982},
        pages = {255-260},
          doi = {10.1038/s41586-023-06499-2},
archivePrefix = {arXiv},
       eprint = {2309.01464},
 primaryClass = {astro-ph.EP},
       adsurl = {https://ui.adsabs.harvard.edu/abs/2023Natur.622..255N}
}

@ARTICLE{ND_Planet7,
       author = {{Osborn}, Ares and {Armstrong}, David J. and {Fern{\'a}ndez Fern{\'a}ndez}, Jorge and {Knierim}, Henrik and {Adibekyan}, Vardan and {Collins}, Karen A. and {Delgado-Mena}, Elisa and {Fridlund}, Malcolm and {Gomes da Silva}, Jo{\~a}o and {Hellier}, Coel and {Jackson}, David G. and {King}, George W. and {Lillo-Box}, Jorge and {Matson}, Rachel A. and {Matthews}, Elisabeth C. and {Santos}, Nuno C. and {Sousa}, S{\'e}rgio G. and {Stassun}, Keivan G. and {Tan}, Thiam-Guan and {Ricker}, George R. and {Vanderspek}, Roland and {Latham}, David W. and {Seager}, Sara and {Winn}, Joshua N. and {Jenkins}, Jon M. and {Bayliss}, Daniel and {Bouma}, Luke G. and {Ciardi}, David R. and {Collins}, Kevin I. and {Col{\'o}n}, Knicole D. and {Crossfield}, Ian J.~M. and {Demangeon}, Olivier D.~S. and {D{\'\i}az}, Rodrigo F. and {Dorn}, Caroline and {Dumusque}, Xavier and {Keniger}, Marcelo Aron Fetzner and {Figueira}, Pedro and {Gan}, Tianjun and {Goeke}, Robert F. and {Hadjigeorghiou}, Andreas and {Hawthorn}, Faith and {Helled}, Ravit and {Howell}, Steve B. and {Nielsen}, Louise D. and {Osborn}, Hugh P. and {Quinn}, Samuel N. and {Sefako}, Ramotholo and {Shporer}, Avi and {Str{\o}m}, Paul A. and {Twicken}, Joseph D. and {Vanderburg}, Andrew and {Wheatley}, Peter J.},
        title = "{TOI-332 b: a super dense Neptune found deep within the Neptunian desert}",
      journal = {\mnras},
         year = 2023,
        month = nov,
       volume = {526},
       number = {1},
        pages = {548-566},
          doi = {10.1093/mnras/stad2575},
archivePrefix = {arXiv},
       eprint = {2308.12137},
 primaryClass = {astro-ph.EP},
       adsurl = {https://ui.adsabs.harvard.edu/abs/2023MNRAS.526..548O}
}

@ARTICLE{LTT_WINERED,
       author = {{Vissapragada}, Shreyas and {McCreery}, Patrick and {Dos Santos}, Leonardo A. and {Espinoza}, N{\'e}stor and {McWilliam}, Andrew and {Matsunaga}, Noriyuki and {Redai}, J{\'e}a Adams and {Behr}, Patrick and {France}, Kevin and {Hamano}, Satoshi and {Hull}, Charlie and {Ikeda}, Yuji and {Katoh}, Haruki and {Kawakita}, Hideyo and {L{\'o}pez-Morales}, Mercedes and {Ortiz Ceballos}, Kevin N. and {Otsubo}, Shogo and {Sarugaku}, Yuki and {Takeuchi}, Tomomi},
        title = "{A High-resolution Non-detection of Escaping Helium in the Ultrahot Neptune LTT 9779b: Evidence for Weakened Evaporation}",
      journal = {\apjl},
         year = 2024,
        month = feb,
       volume = {962},
       number = {1},
          eid = {L19},
        pages = {L19},
          doi = {10.3847/2041-8213/ad23cf},
archivePrefix = {arXiv},
       eprint = {2401.16474},
 primaryClass = {astro-ph.EP},
       adsurl = {https://ui.adsabs.harvard.edu/abs/2024ApJ...962L..19V}
}

@ARTICLE{Seager_Sasselov_2000,
       author = {{Seager}, S. and {Sasselov}, D.~D.},
        title = "{Theoretical Transmission Spectra during Extrasolar Giant Planet Transits}",
      journal = {\apj},
         year = 2000,
        month = jul,
       volume = {537},
       number = {2},
        pages = {916-921},
          doi = {10.1086/309088},
archivePrefix = {arXiv},
       eprint = {astro-ph/9912241},
 primaryClass = {astro-ph},
       adsurl = {https://ui.adsabs.harvard.edu/abs/2000ApJ...537..916S}
}

@ARTICLE{HeI_intro1,
       author = {{Oklop{\v{c}}i{\'c}}, Antonija and {Hirata}, Christopher M.},
        title = "{A New Window into Escaping Exoplanet Atmospheres: 10830 {\r{A}} Line of Helium}",
      journal = {\apjl},
         year = 2018,
        month = mar,
       volume = {855},
       number = {1},
          eid = {L11},
        pages = {L11},
          doi = {10.3847/2041-8213/aaada9},
archivePrefix = {arXiv},
       eprint = {1711.05269},
 primaryClass = {astro-ph.EP},
       adsurl = {https://ui.adsabs.harvard.edu/abs/2018ApJ...855L..11O}
}

@ARTICLE{HeI_HST,
       author = {{Spake}, J.~J. and {Sing}, D.~K. and {Evans}, T.~M. and {Oklop{\v{c}}i{\'c}}, A. and {Bourrier}, V. and {Kreidberg}, L. and {Rackham}, B.~V. and {Irwin}, J. and {Ehrenreich}, D. and {Wyttenbach}, A. and {Wakeford}, H.~R. and {Zhou}, Y. and {Chubb}, K.~L. and {Nikolov}, N. and {Goyal}, J.~M. and {Henry}, G.~W. and {Williamson}, M.~H. and {Blumenthal}, S. and {Anderson}, D.~R. and {Hellier}, C. and {Charbonneau}, D. and {Udry}, S. and {Madhusudhan}, N.},
        title = "{Helium in the eroding atmosphere of an exoplanet}",
      journal = {\nat},
         year = 2018,
        month = may,
       volume = {557},
       number = {7703},
        pages = {68-70},
          doi = {10.1038/s41586-018-0067-5},
archivePrefix = {arXiv},
       eprint = {1805.01298},
 primaryClass = {astro-ph.EP},
       adsurl = {https://ui.adsabs.harvard.edu/abs/2018Natur.557...68S}
}

@ARTICLE{HeI_CARMENES,
       author = {{Nortmann}, Lisa and {Pall{\'e}}, Enric and {Salz}, Michael and {Sanz-Forcada}, Jorge and {Nagel}, Evangelos and {Alonso-Floriano}, F. Javier and {Czesla}, Stefan and {Yan}, Fei and {Chen}, Guo and {Snellen}, Ignas A.~G. and {Zechmeister}, Mathias and {Schmitt}, J{\"u}rgen H.~M.~M. and {L{\'o}pez-Puertas}, Manuel and {Casasayas-Barris}, N{\'u}ria and {Bauer}, Florian F. and {Amado}, Pedro J. and {Caballero}, Jos{\'e} A. and {Dreizler}, Stefan and {Henning}, Thomas and {Lamp{\'o}n}, Manuel and {Montes}, David and {Molaverdikhani}, Karan and {Quirrenbach}, Andreas and {Reiners}, Ansgar and {Ribas}, Ignasi and {S{\'a}nchez-L{\'o}pez}, Alejandro and {Schneider}, P. Christian and {Zapatero Osorio}, Mar{\'\i}a R.},
        title = "{Ground-based detection of an extended helium atmosphere in the Saturn-mass exoplanet WASP-69b}",
      journal = {Science},
         year = 2018,
        month = dec,
       volume = {362},
       number = {6421},
        pages = {1388-1391},
          doi = {10.1126/science.aat5348},
archivePrefix = {arXiv},
       eprint = {1812.03119},
 primaryClass = {astro-ph.EP},
       adsurl = {https://ui.adsabs.harvard.edu/abs/2018Sci...362.1388N}
}

@ARTICLE{Helium_HATP11b,
       author = {{Allart}, R. and {Bourrier}, V. and {Lovis}, C. and {Ehrenreich}, D. and {Spake}, J.~J. and {Wyttenbach}, A. and {Pino}, L. and {Pepe}, F. and {Sing}, D.~K. and {Lecavelier des Etangs}, A.},
        title = "{Spectrally resolved helium absorption from the extended atmosphere of a warm Neptune-mass exoplanet}",
      journal = {Science},
         year = 2018,
        month = dec,
       volume = {362},
       number = {6421},
        pages = {1384-1387},
          doi = {10.1126/science.aat5879},
archivePrefix = {arXiv},
       eprint = {1812.02189},
 primaryClass = {astro-ph.EP},
       adsurl = {https://ui.adsabs.harvard.edu/abs/2018Sci...362.1384A}
}

@ARTICLE{LTT_Spitzer_Tess,
       author = {{Crossfield}, Ian J.~M. and {Dragomir}, Diana and {Cowan}, Nicolas B. and {Daylan}, Tansu and {Wong}, Ian and {Kataria}, Tiffany and {Deming}, Drake and {Kreidberg}, Laura and {Mikal-Evans}, Thomas and {Gorjian}, Varoujan and {Jenkins}, James S. and {Benneke}, Bj{\"o}rn and {Collins}, Karen A. and {Burke}, Christopher J. and {Henze}, Christopher E. and {McDermott}, Scott and {Mireles}, Ismael and {Watanabe}, David and {Wohler}, Bill and {Ricker}, George and {Vanderspek}, Roland and {Seager}, Sara and {Jenkins}, Jon M.},
        title = "{Phase Curves of Hot Neptune LTT 9779b Suggest a High-metallicity Atmosphere}",
      journal = {\apjl},
         year = 2020,
        month = nov,
       volume = {903},
       number = {1},
          eid = {L7},
        pages = {L7},
          doi = {10.3847/2041-8213/abbc71},
archivePrefix = {arXiv},
       eprint = {2010.12745},
 primaryClass = {astro-ph.EP},
       adsurl = {https://ui.adsabs.harvard.edu/abs/2020ApJ...903L...7C}
}

@ARTICLE{LTT_Spitzer,
       author = {{Dragomir}, Diana and {Crossfield}, Ian J.~M. and {Benneke}, Bj{\"o}rn and {Wong}, Ian and {Daylan}, Tansu and {Diaz}, Matias and {Deming}, Drake and {Molliere}, Paul and {Kreidberg}, Laura and {Jenkins}, James S. and {Berardo}, David and {Christiansen}, Jessie L. and {Dressing}, Courtney D. and {Gorjian}, Varoujan and {Kane}, Stephen R. and {Mikal-Evans}, Thomas and {Morales}, Farisa Y. and {Werner}, Michael and {Ricker}, George R. and {Vanderspek}, Roland and {Seager}, S. and {Winn}, Joshua N. and {Jenkins}, Jon M. and {Col{\'o}n}, Knicole D. and {Fong}, Willie and {Guerrero}, Natalia and {Hesse}, Katharine and {Osborn}, Hugh P. and {E. Rose}, Mark and {Smith}, Jeffrey C. and {Ting}, Eric B.},
        title = "{Spitzer Reveals Evidence of Molecular Absorption in the Atmosphere of the Hot Neptune LTT 9779b}",
      journal = {\apjl},
         year = 2020,
        month = nov,
       volume = {903},
       number = {1},
          eid = {L6},
        pages = {L6},
          doi = {10.3847/2041-8213/abbc70},
archivePrefix = {arXiv},
       eprint = {2010.12744},
 primaryClass = {astro-ph.EP},
       adsurl = {https://ui.adsabs.harvard.edu/abs/2020ApJ...903L...6D}
}

@ARTICLE{TP_UHJ_1,
       author = {{Hubeny}, Ivan and {Burrows}, Adam and {Sudarsky}, David},
        title = "{A Possible Bifurcation in Atmospheres of Strongly Irradiated Stars and Planets}",
      journal = {\apj},
         year = 2003,
        month = sep,
       volume = {594},
       number = {2},
        pages = {1011-1018},
          doi = {10.1086/377080},
archivePrefix = {arXiv},
       eprint = {astro-ph/0305349},
 primaryClass = {astro-ph},
       adsurl = {https://ui.adsabs.harvard.edu/abs/2003ApJ...594.1011H}
}

@ARTICLE{TP_UHJ_2,
       author = {{Fortney}, J.~J. and {Lodders}, K. and {Marley}, M.~S. and {Freedman}, R.~S.},
        title = "{A Unified Theory for the Atmospheres of the Hot and Very Hot Jupiters: Two Classes of Irradiated Atmospheres}",
      journal = {\apj},
         year = 2008,
        month = may,
       volume = {678},
       number = {2},
        pages = {1419-1435},
          doi = {10.1086/528370},
archivePrefix = {arXiv},
       eprint = {0710.2558},
 primaryClass = {astro-ph},
       adsurl = {https://ui.adsabs.harvard.edu/abs/2008ApJ...678.1419F}
}

@ARTICLE{TP_UHJ_3,
       author = {{Lothringer}, Joshua D. and {Barman}, Travis and {Koskinen}, Tommi},
        title = "{Extremely Irradiated Hot Jupiters: Non-oxide Inversions, H$^{-}$ Opacity, and Thermal Dissociation of Molecules}",
      journal = {\apj},
         year = 2018,
        month = oct,
       volume = {866},
       number = {1},
          eid = {27},
        pages = {27},
          doi = {10.3847/1538-4357/aadd9e},
archivePrefix = {arXiv},
       eprint = {1805.00038},
 primaryClass = {astro-ph.EP},
       adsurl = {https://ui.adsabs.harvard.edu/abs/2018ApJ...866...27L}
}

@ARTICLE{TP1_KELT9,
       author = {{Pino}, Lorenzo and {D{\'e}sert}, Jean-Michel and {Brogi}, Matteo and {Malavolta}, Luca and {Wyttenbach}, Aur{\'e}lien and {Line}, Michael and {Hoeijmakers}, Jens and {Fossati}, Luca and {Bonomo}, Aldo Stefano and {Nascimbeni}, Valerio and {Panwar}, Vatsal and {Affer}, Laura and {Benatti}, Serena and {Biazzo}, Katia and {Bignamini}, Andrea and {Borsa}, Franscesco and {Carleo}, Ilaria and {Claudi}, Riccardo and {Cosentino}, Rosario and {Covino}, Elvira and {Damasso}, Mario and {Desidera}, Silvano and {Giacobbe}, Paolo and {Harutyunyan}, Avet and {Lanza}, Antonino Francesco and {Leto}, Giuseppe and {Maggio}, Antonio and {Maldonado}, Jesus and {Mancini}, Luigi and {Micela}, Giuseppina and {Molinari}, Emilio and {Pagano}, Isabella and {Piotto}, Giampaolo and {Poretti}, Ennio and {Rainer}, Monica and {Scandariato}, Gaetano and {Sozzetti}, Alessandro and {Allart}, Romain and {Borsato}, Luca and {Bruno}, Giovanni and {Di Fabrizio}, Luca and {Ehrenreich}, David and {Fiorenzano}, Aldo and {Frustagli}, Giuseppe and {Lavie}, Baptiste and {Lovis}, Christophe and {Magazz{\`u}}, Antonio and {Nardiello}, Domenico and {Pedani}, Marco and {Smareglia}, Riccardo},
        title = "{Neutral Iron Emission Lines from the Dayside of KELT-9b: The GAPS Program with HARPS-N at TNG XX}",
      journal = {\apjl},
         year = 2020,
        month = may,
       volume = {894},
       number = {2},
          eid = {L27},
        pages = {L27},
          doi = {10.3847/2041-8213/ab8c44},
archivePrefix = {arXiv},
       eprint = {2004.11335},
 primaryClass = {astro-ph.EP},
       adsurl = {https://ui.adsabs.harvard.edu/abs/2020ApJ...894L..27P}
}

@ARTICLE{TP2_KELT9,
       author = {{Kasper}, David and {Bean}, Jacob L. and {Line}, Michael R. and {Seifahrt}, Andreas and {St{\"u}rmer}, Julian and {Pino}, Lorenzo and {D{\'e}sert}, Jean-Michel and {Brogi}, Matteo},
        title = "{Confirmation of Iron Emission Lines and Nondetection of TiO on the Dayside of KELT-9b with MAROON-X}",
      journal = {\apjl},
         year = 2021,
        month = nov,
       volume = {921},
       number = {1},
          eid = {L18},
        pages = {L18},
          doi = {10.3847/2041-8213/ac30e1},
archivePrefix = {arXiv},
       eprint = {2108.08389},
 primaryClass = {astro-ph.EP},
       adsurl = {https://ui.adsabs.harvard.edu/abs/2021ApJ...921L..18K}
}

@ARTICLE{TP1_WASP33_Fe,
       author = {{Nugroho}, Stevanus K. and {Gibson}, Neale P. and {de Mooij}, Ernst J.~W. and {Herman}, Miranda K. and {Watson}, Chris A. and {Kawahara}, Hajime and {Merritt}, Stephanie R.},
        title = "{Detection of Fe I Emission in the Dayside Spectrum of WASP-33b}",
      journal = {\apjl},
         year = 2020,
        month = aug,
       volume = {898},
       number = {2},
          eid = {L31},
        pages = {L31},
          doi = {10.3847/2041-8213/aba4b6},
archivePrefix = {arXiv},
       eprint = {2007.05508},
 primaryClass = {astro-ph.EP},
       adsurl = {https://ui.adsabs.harvard.edu/abs/2020ApJ...898L..31N}
}

@ARTICLE{TP2_WASP33_Fe,
       author = {{Cont}, D. and {Yan}, F. and {Reiners}, A. and {Casasayas-Barris}, N. and {Molli{\`e}re}, P. and {Pall{\'e}}, E. and {Henning}, Th. and {Nortmann}, L. and {Stangret}, M. and {Czesla}, S. and {L{\'o}pez-Puertas}, M. and {S{\'a}nchez-L{\'o}pez}, A. and {Rodler}, F. and {Ribas}, I. and {Quirrenbach}, A. and {Caballero}, J.~A. and {Amado}, P.~J. and {Carone}, L. and {Khaimova}, J. and {Kreidberg}, L. and {Molaverdikhani}, K. and {Montes}, D. and {Morello}, G. and {Nagel}, E. and {Oshagh}, M. and {Zechmeister}, M.},
        title = "{Detection of Fe and evidence for TiO in the dayside emission spectrum of WASP-33b}",
      journal = {\aap},
         year = 2021,
        month = jul,
       volume = {651},
          eid = {A33},
        pages = {A33},
          doi = {10.1051/0004-6361/202140732},
archivePrefix = {arXiv},
       eprint = {2105.10230},
 primaryClass = {astro-ph.EP},
       adsurl = {https://ui.adsabs.harvard.edu/abs/2021A&A...651A..33C}
}

@ARTICLE{TP3_WASP33_Fe,
       author = {{Herman}, Miranda K. and {de Mooij}, Ernst J.~W. and {Nugroho}, Stevanus K. and {Gibson}, Neale P. and {Jayawardhana}, Ray},
        title = "{Dayside Fe I Emission, Day-Night Brightness Contrast and Phase Offset of the Exoplanet WASP-33b}",
      journal = {\aj},
         year = 2022,
        month = jun,
       volume = {163},
       number = {6},
          eid = {248},
        pages = {248},
          doi = {10.3847/1538-3881/ac5f4d},
archivePrefix = {arXiv},
       eprint = {2203.11188},
 primaryClass = {astro-ph.EP},
       adsurl = {https://ui.adsabs.harvard.edu/abs/2022AJ....163..248H}
}

@ARTICLE{HATP70_PEPSI,
       author = {{Guo}, B. and {Yan}, F. and {Henning}, Th. and {Nortmann}, L. and {Stangret}, M. and {Cont}, D. and {Pall{\'e}}, E. and {Shulyak}, D. and {Strassmeier}, K.~G. and {Ilyin}, I. and {Lesjak}, F. and {Reiners}, A. and {Liu}, S. and {Molaverdikhani}, K. and {Scandariato}, G. and {Keles}, E. and {Caballero}, J.~A. and {Amado}, P.~J. and {Quirrenbach}, A. and {Ribas}, I. and {G{\'o}ngora}, S. and {Hatzes}, A.~P. and {L{\'o}pez-Puertas}, M. and {Montes}, D. and {Poppenhaeger}, K. and {Schlawin}, E. and {Schweitzer}, A. and {Sicilia}, D.},
        title = "{Elemental abundance pattern and temperature inversion on the dayside of HAT-P-70b observed with CARMENES and PEPSI}",
      journal = {\aap},
         year = 2026,
        month = feb,
       volume = {706},
          eid = {A126},
        pages = {A126},
          doi = {10.1051/0004-6361/202557869},
archivePrefix = {arXiv},
       eprint = {2512.21470},
 primaryClass = {astro-ph.EP},
       adsurl = {https://ui.adsabs.harvard.edu/abs/2026A&A...706A.126G}
}

@ARTICLE{MASCARA1_PEPSI,
       author = {{Scandariato}, G. and {Borsa}, F. and {Bonomo}, A.~S. and {Gaudi}, B.~S. and {Henning}, Th. and {Ilyin}, I. and {Johnson}, M.~C. and {Malavolta}, L. and {Mallonn}, M. and {Molaverdikhani}, K. and {Nascimbeni}, V. and {Patience}, J. and {Pino}, L. and {Poppenhaeger}, K. and {Schlawin}, E. and {Shkolnik}, E.~L. and {Sicilia}, D. and {Sozzetti}, A. and {Strassmeier}, K.~G. and {Veillet}, C. and {Wang}, J. and {Yan}, F.},
        title = "{The PEPSI Exoplanet Transit Survey (PETS). III. The detection of Fe I, Cr I, and Ti I in the atmosphere of MASCARA-1 b through high-resolution emission spectroscopy}",
      journal = {\aap},
         year = 2023,
        month = jun,
       volume = {674},
          eid = {A58},
        pages = {A58},
          doi = {10.1051/0004-6361/202245539},
archivePrefix = {arXiv},
       eprint = {2304.03328},
 primaryClass = {astro-ph.EP},
       adsurl = {https://ui.adsabs.harvard.edu/abs/2023A&A...674A..58S}
}

@ARTICLE{WASP18_IGRINS,
       author = {{Brogi}, Matteo and {Emeka-Okafor}, Vanessa and {Line}, Michael R. and {Gandhi}, Siddharth and {Pino}, Lorenzo and {Kempton}, Eliza M.-R. and {Rauscher}, Emily and {Parmentier}, Vivien and {Bean}, Jacob L. and {Mace}, Gregory N. and {Cowan}, Nicolas B. and {Shkolnik}, Evgenya and {Wardenier}, Joost P. and {Mansfield}, Megan and {Welbanks}, Luis and {Smith}, Peter and {Fortney}, Jonathan J. and {Birkby}, Jayne L. and {Zalesky}, Joseph A. and {Dang}, Lisa and {Patience}, Jennifer and {D{\'e}sert}, Jean-Michel},
        title = "{The Roasting Marshmallows Program with IGRINS on Gemini South I: Composition and Climate of the Ultrahot Jupiter WASP-18 b}",
      journal = {\aj},
         year = 2023,
        month = mar,
       volume = {165},
       number = {3},
          eid = {91},
        pages = {91},
          doi = {10.3847/1538-3881/acaf5c},
archivePrefix = {arXiv},
       eprint = {2209.15548},
 primaryClass = {astro-ph.EP},
       adsurl = {https://ui.adsabs.harvard.edu/abs/2023AJ....165...91B}
}

@ARTICLE{WASP33_Keck,
       author = {{Finnerty}, Luke and {Schofield}, Tobias and {Sappey}, Ben and {Xuan}, Jerry W. and {Ruffio}, Jean-Baptiste and {Wang}, Jason J. and {Delorme}, Jacques-Robert and {Blake}, Geoffrey A. and {Buzard}, Cam and {Fitzgerald}, Michael P. and {Baker}, Ashley and {Bartos}, Randall and {Bond}, Charlotte Z. and {Calvin}, Benjamin and {Cetre}, Sylvain and {Doppmann}, Greg and {Echeverri}, Daniel and {Jovanovic}, Nemanja and {Liberman}, Joshua and {L{\'o}pez}, Ronald A. and {Martin}, Emily C. and {Mawet}, Dimitri and {Morris}, Evan and {Pezzato}, Jacklyn and {Phillips}, Caprice L. and {Ragland}, Sam and {Skemer}, Andrew and {Venenciano}, Taylor and {Wallace}, J. Kent and {Wallack}, Nicole L. and {Wang}, Ji and {Wizinowich}, Peter},
        title = "{Keck Planet Imager and Characterizer Emission Spectroscopy of WASP-33b}",
      journal = {\aj},
         year = 2023,
        month = jul,
       volume = {166},
       number = {1},
          eid = {31},
        pages = {31},
          doi = {10.3847/1538-3881/acda91},
archivePrefix = {arXiv},
       eprint = {2305.19389},
 primaryClass = {astro-ph.EP},
       adsurl = {https://ui.adsabs.harvard.edu/abs/2023AJ....166...31F}
}

@ARTICLE{TP1_WASP33_CO,
       author = {{Yan}, F. and {Pall{\'e}}, E. and {Reiners}, A. and {Casasayas-Barris}, N. and {Cont}, D. and {Stangret}, M. and {Nortmann}, L. and {Molli{\`e}re}, P. and {Henning}, Th. and {Chen}, G. and {Molaverdikhani}, K.},
        title = "{Detection of CO emission lines in the dayside atmospheres of WASP-33b and WASP-189b with GIANO}",
      journal = {\aap},
         year = 2022,
        month = may,
       volume = {661},
          eid = {L6},
        pages = {L6},
          doi = {10.1051/0004-6361/202243503},
archivePrefix = {arXiv},
       eprint = {2204.10158},
 primaryClass = {astro-ph.EP},
       adsurl = {https://ui.adsabs.harvard.edu/abs/2022A&A...661L...6Y}
}

@ARTICLE{TP2_WASP33_CO,
       author = {{van Sluijs}, Lennart and {Birkby}, Jayne L. and {Lothringer}, Joshua and {Lee}, Elspeth K.~H. and {Crossfield}, Ian J.~M. and {Parmentier}, Vivien and {Brogi}, Matteo and {Kulesa}, Craig and {McCarthy}, Don and {Charbonneau}, David},
        title = "{Carbon monoxide emission lines reveal an inverted atmosphere in the ultra hot Jupiter WASP-33 b consistent with an eastward hot spot}",
      journal = {\mnras},
         year = 2023,
        month = jun,
       volume = {522},
       number = {2},
        pages = {2145-2170},
          doi = {10.1093/mnras/stad1103},
archivePrefix = {arXiv},
       eprint = {2203.13234},
 primaryClass = {astro-ph.EP},
       adsurl = {https://ui.adsabs.harvard.edu/abs/2023MNRAS.522.2145V}
}

@ARTICLE{TP3_WASP33_CO,
       author = {{Choi}, Yeon-Ho and {Jeong}, Ueejeong and {Lee}, Jae-Joon and {Kim}, Hyun-Jeong and {Oh}, Heeyoung and {Park}, Chan and {Kye}, Changwoo and {Finnerty}, Luke and {Line}, Micheal R. and {Kanumalla}, Krishna and {Sanchez}, Jorge A. and {Smith}, Peter C.~B. and {Kim}, Sanghyuk and {Lee}, Hye-In and {Park}, Woojin and {Yu}, Youngsam and {Kim}, Yunjong and {Chun}, Moo-Young and {Oh}, Jae Sok and {Lee}, Sungho and {Jang}, Jeong-Gyun and {Jang}, Bi-Ho and {Seong}, Hyeon Cheol and {Brooks}, Cynthia B. and {Mace}, Gregory N. and {Lee}, Hanshin and {Good}, John M. and {Jaffe}, Daniel T. and {Kim}, Kang-Min and {Yuk}, In-Soo and {Hwang}, Narae and {Park}, Byeong-Gon and {Kim}, Hwihyun and {Chinn}, Brian and {Ramos}, Francisco and {Prado}, Pablo and {Diaz}, Ruben and {White}, John and {Tapia}, Eduardo and {Olivares}, Andres and {Oyarzun}, Valentina and {Kurz}, Emma and {Stecher}, Hawi and {Quiroz}, Carlos and {Arriagada}, Ignacio and {Hayward}, Thomas L. and {Suh}, Hyewon and {Miller}, Jen and {Xu}, Siyi and {Farina}, Emanuele Paolo and {Figura}, Charlie and {Mocnik}, Teo and {Hartman}, Zachary and {Rawlings}, Mark and {Stephens}, Andrew and {Miller}, Bryan and {Labrie}, Kathleen and {Hirst}, Paul and {Lee}, Byeong-Cheol},
        title = "{An Early Look at the Performance of IGRINS-2 at Gemini-North with Application to the Ultrahot Jupiter, WASP-33 b}",
      journal = {\aj},
         year = 2025,
        month = oct,
       volume = {170},
       number = {4},
          eid = {238},
        pages = {238},
          doi = {10.3847/1538-3881/aded8c},
archivePrefix = {arXiv},
       eprint = {2503.12736},
 primaryClass = {astro-ph.EP},
       adsurl = {https://ui.adsabs.harvard.edu/abs/2025AJ....170..238C}
}

@ARTICLE{TP1_MASCARA1,
       author = {{Holmberg}, M{\r{a}}ns and {Madhusudhan}, Nikku},
        title = "{A First Look at CRIRES+: Performance Assessment and Exoplanet Spectroscopy}",
      journal = {\aj},
         year = 2022,
        month = sep,
       volume = {164},
       number = {3},
          eid = {79},
        pages = {79},
          doi = {10.3847/1538-3881/ac77eb},
archivePrefix = {arXiv},
       eprint = {2206.10621},
 primaryClass = {astro-ph.EP},
       adsurl = {https://ui.adsabs.harvard.edu/abs/2022AJ....164...79H}
}

@ARTICLE{TP2_MASCARA1,
       author = {{Ramkumar}, Swaetha and {Gibson}, Neale P. and {Nugroho}, Stevanus K. and {Fortune}, Mark and {Maguire}, Cathal},
        title = "{New perspectives on MASCARA-1b: A combined analysis of pre- and post-eclipse emission data using CRIRES+}",
      journal = {\aap},
         year = 2025,
        month = mar,
       volume = {695},
          eid = {A110},
        pages = {A110},
          doi = {10.1051/0004-6361/202453520},
archivePrefix = {arXiv},
       eprint = {2502.11180},
 primaryClass = {astro-ph.EP},
       adsurl = {https://ui.adsabs.harvard.edu/abs/2025A&A...695A.110R}
}

@ARTICLE{TP_WASP121,
       author = {{Evans}, Thomas M. and {Sing}, David K. and {Kataria}, Tiffany and {Goyal}, Jayesh and {Nikolov}, Nikolay and {Wakeford}, Hannah R. and {Deming}, Drake and {Marley}, Mark S. and {Amundsen}, David S. and {Ballester}, Gilda E. and {Barstow}, Joanna K. and {Ben-Jaffel}, Lotfi and {Bourrier}, Vincent and {Buchhave}, Lars A. and {Cohen}, Ofer and {Ehrenreich}, David and {Garc{\'\i}a Mu{\~n}oz}, Antonio and {Henry}, Gregory W. and {Knutson}, Heather and {Lavvas}, Panayotis and {Lecavelier Des Etangs}, Alain and {Lewis}, Nikole K. and {L{\'o}pez-Morales}, Mercedes and {Mandell}, Avi M. and {Sanz-Forcada}, Jorge and {Tremblin}, Pascal and {Lupu}, Roxana},
        title = "{An ultrahot gas-giant exoplanet with a stratosphere}",
      journal = {\nat},
         year = 2017,
        month = aug,
       volume = {548},
       number = {7665},
        pages = {58-61},
          doi = {10.1038/nature23266},
archivePrefix = {arXiv},
       eprint = {1708.01076},
 primaryClass = {astro-ph.EP},
       adsurl = {https://ui.adsabs.harvard.edu/abs/2017Natur.548...58E}
}

@ARTICLE{TP_WASP18,
       author = {{Coulombe}, Louis-Philippe and {Benneke}, Bj{\"o}rn and {Challener}, Ryan and {Piette}, Anjali A.~A. and {Wiser}, Lindsey S. and {Mansfield}, Megan and {MacDonald}, Ryan J. and {Beltz}, Hayley and {Feinstein}, Adina D. and {Radica}, Michael and {Savel}, Arjun B. and {Dos Santos}, Leonardo A. and {Bean}, Jacob L. and {Parmentier}, Vivien and {Wong}, Ian and {Rauscher}, Emily and {Komacek}, Thaddeus D. and {Kempton}, Eliza M.-R. and {Tan}, Xianyu and {Hammond}, Mark and {Lewis}, Neil T. and {Line}, Michael R. and {Lee}, Elspeth K.~H. and {Shivkumar}, Hinna and {Crossfield}, Ian J.~M. and {Nixon}, Matthew C. and {Rackham}, Benjamin V. and {Wakeford}, Hannah R. and {Welbanks}, Luis and {Zhang}, Xi and {Batalha}, Natalie M. and {Berta-Thompson}, Zachory K. and {Changeat}, Quentin and {D{\'e}sert}, Jean-Michel and {Espinoza}, N{\'e}stor and {Goyal}, Jayesh M. and {Harrington}, Joseph and {Knutson}, Heather A. and {Kreidberg}, Laura and {L{\'o}pez-Morales}, Mercedes and {Shporer}, Avi and {Sing}, David K. and {Stevenson}, Kevin B. and {Aggarwal}, Keshav and {Ahrer}, Eva-Maria and {Alam}, Munazza K. and {Bell}, Taylor J. and {Blecic}, Jasmina and {Caceres}, Claudio and {Carter}, Aarynn L. and {Casewell}, Sarah L. and {Crouzet}, Nicolas and {Cubillos}, Patricio E. and {Decin}, Leen and {Fortney}, Jonathan J. and {Gibson}, Neale P. and {Heng}, Kevin and {Henning}, Thomas and {Iro}, Nicolas and {Kendrew}, Sarah and {Lagage}, Pierre-Olivier and {Leconte}, J{\'e}r{\'e}my and {Lendl}, Monika and {Lothringer}, Joshua D. and {Mancini}, Luigi and {Mikal-Evans}, Thomas and {Molaverdikhani}, Karan and {Nikolov}, Nikolay K. and {Ohno}, Kazumasa and {Palle}, Enric and {Piaulet}, Caroline and {Redfield}, Seth and {Roy}, Pierre-Alexis and {Tsai}, Shang-Min and {Venot}, Olivia and {Wheatley}, Peter J.},
        title = "{A broadband thermal emission spectrum of the ultra-hot Jupiter WASP-18b}",
      journal = {\nat},
         year = 2023,
        month = aug,
       volume = {620},
       number = {7973},
        pages = {292-298},
          doi = {10.1038/s41586-023-06230-1},
archivePrefix = {arXiv},
       eprint = {2301.08192},
 primaryClass = {astro-ph.EP},
       adsurl = {https://ui.adsabs.harvard.edu/abs/2023Natur.620..292C}
}

@ARTICLE{TP_Gstar,
       author = {{Lothringer}, Joshua D. and {Barman}, Travis},
        title = "{The Influence of Host Star Spectral Type on Ultra-hot Jupiter Atmospheres}",
      journal = {\apj},
         year = 2019,
        month = may,
       volume = {876},
       number = {1},
          eid = {69},
        pages = {69},
          doi = {10.3847/1538-4357/ab1485},
archivePrefix = {arXiv},
       eprint = {1903.12183},
 primaryClass = {astro-ph.EP},
       adsurl = {https://ui.adsabs.harvard.edu/abs/2019ApJ...876...69L}
}

@ARTICLE{JWST_WASP107,
       author = {{Welbanks}, Luis and {Bell}, Taylor J. and {Beatty}, Thomas G. and {Line}, Michael R. and {Ohno}, Kazumasa and {Fortney}, Jonathan J. and {Schlawin}, Everett and {Greene}, Thomas P. and {Rauscher}, Emily and {McGill}, Peter and {Murphy}, Matthew and {Parmentier}, Vivien and {Tang}, Yao and {Edelman}, Isaac and {Mukherjee}, Sagnick and {Wiser}, Lindsey S. and {Lagage}, Pierre-Olivier and {Dyrek}, Achr{\`e}ne and {Triantafillides}, Anastasia},
        title = "{A high internal heat flux and large core in a warm Neptune exoplanet}",
      journal = {\nat},
         year = 2024,
        month = jun,
       volume = {630},
       number = {8018},
        pages = {836-840},
          doi = {10.1038/s41586-024-07514-w},
archivePrefix = {arXiv},
       eprint = {2405.11018},
 primaryClass = {astro-ph.EP},
       adsurl = {https://ui.adsabs.harvard.edu/abs/2024Natur.630..836W}
}

@ARTICLE{JWST_HATP26,
       author = {{Gressier}, Am{\'e}lie and {Batalha}, Natasha E. and {Wogan}, Nicholas and {Alderson}, Lili and {Doud}, Dominic and {Espinoza}, N{\'e}stor and {MacDonald}, Ryan J. and {Wakeford}, Hannah R. and {Valenti}, Jeff A. and {Lewis}, Nikole K. and {Seager}, Sara and {Stevenson}, Kevin B. and {Allen}, Natalie H. and {Ca{\~n}as}, Caleb I. and {Challener}, Ryan C. and {Glidden}, Ana and {Huang}, Jingcheng and {Lin}, Zifan and {Louie}, Dana R. and {Maguire}, Cathal and {Mullens}, Elijah and {Sotzen}, Kristin and {Valentine}, Daniel and {Clampin}, Mark and {Pueyo}, Laurent and {van der Marel}, Roeland P. and {Mountain}, C. Matt},
        title = "{JWST-TST DREAMS: Sulfur Dioxide in the Atmosphere of the Neptune-mass Planet HAT-P-26 b from NIRSpec G395H Transmission Spectroscopy}",
      journal = {\aj},
         year = 2025,
        month = nov,
       volume = {170},
       number = {5},
          eid = {292},
        pages = {292},
          doi = {10.3847/1538-3881/ae0929},
archivePrefix = {arXiv},
       eprint = {2509.16082},
 primaryClass = {astro-ph.EP},
       adsurl = {https://ui.adsabs.harvard.edu/abs/2025AJ....170..292G}
}

@ARTICLE{JWST_GJ3470,
       author = {{Beatty}, Thomas G. and {Welbanks}, Luis and {Schlawin}, Everett and {Bell}, Taylor J. and {Line}, Michael R. and {Murphy}, Matthew and {Edelman}, Isaac and {Greene}, Thomas P. and {Fortney}, Jonathan J. and {Henry}, Gregory W. and {Mukherjee}, Sagnick and {Ohno}, Kazumasa and {Parmentier}, Vivien and {Rauscher}, Emily and {Wiser}, Lindsey S. and {Triantafillides}, Anastasia},
        title = "{Sulfur Dioxide and Other Molecular Species in the Atmosphere of the Sub-Neptune GJ 3470 b}",
      journal = {\apjl},
         year = 2024,
        month = jul,
       volume = {970},
       number = {1},
          eid = {L10},
        pages = {L10},
          doi = {10.3847/2041-8213/ad55e9},
archivePrefix = {arXiv},
       eprint = {2406.04450},
 primaryClass = {astro-ph.EP},
       adsurl = {https://ui.adsabs.harvard.edu/abs/2024ApJ...970L..10B}
}

@ARTICLE{JWST_K218,
       author = {{Madhusudhan}, Nikku and {Sarkar}, Subhajit and {Constantinou}, Savvas and {Holmberg}, M{\r{a}}ns and {Piette}, Anjali A.~A. and {Moses}, Julianne I.},
        title = "{Carbon-bearing Molecules in a Possible Hycean Atmosphere}",
      journal = {\apjl},
         year = 2023,
        month = oct,
       volume = {956},
       number = {1},
          eid = {L13},
        pages = {L13},
          doi = {10.3847/2041-8213/acf577},
archivePrefix = {arXiv},
       eprint = {2309.05566},
 primaryClass = {astro-ph.EP},
       adsurl = {https://ui.adsabs.harvard.edu/abs/2023ApJ...956L..13M}
}

@ARTICLE{JWST_TOI270,
       author = {{Holmberg}, M{\r{a}}ns and {Madhusudhan}, Nikku},
        title = "{Possible Hycean conditions in the sub-Neptune TOI-270 d}",
      journal = {\aap},
         year = 2024,
        month = mar,
       volume = {683},
          eid = {L2},
        pages = {L2},
          doi = {10.1051/0004-6361/202348238},
archivePrefix = {arXiv},
       eprint = {2403.03244},
 primaryClass = {astro-ph.EP},
       adsurl = {https://ui.adsabs.harvard.edu/abs/2024A&A...683L...2H}
}

@ARTICLE{GIANO_HATP11,
       author = {{Basilicata}, M. and {Giacobbe}, P. and {Bonomo}, A.~S. and {Scandariato}, G. and {Brogi}, M. and {Singh}, V. and {Di Paola}, A. and {Mancini}, L. and {Sozzetti}, A. and {Lanza}, A.~F. and {Cubillos}, P.~E. and {Damasso}, M. and {Desidera}, S. and {Biazzo}, K. and {Bignamini}, A. and {Borsa}, F. and {Cabona}, L. and {Carleo}, I. and {Ghedina}, A. and {Guilluy}, G. and {Maggio}, A. and {Mainella}, G. and {Micela}, G. and {Molinari}, E. and {Molinaro}, M. and {Nardiello}, D. and {Pedani}, M. and {Pino}, L. and {Poretti}, E. and {Southworth}, J. and {Stangret}, M. and {Turrini}, D.},
        title = "{The GAPS Programme at TNG. LV. Multiple molecular species in the atmosphere of HAT-P-11 b and review of the HAT-P-11 planetary system}",
      journal = {\aap},
         year = 2024,
        month = jun,
       volume = {686},
          eid = {A127},
        pages = {A127},
          doi = {10.1051/0004-6361/202347659},
archivePrefix = {arXiv},
       eprint = {2403.01527},
 primaryClass = {astro-ph.EP},
       adsurl = {https://ui.adsabs.harvard.edu/abs/2024A&A...686A.127B}
}

@ARTICLE{CRIRES_WASP107,
       author = {{Boldt-Christmas}, L. and {Rains}, A.~D. and {Piskunov}, N. and {Nortmann}, L. and {Lesjak}, F. and {Cont}, D. and {Kochukhov}, O. and {Hahlin}, A. and {Lavail}, A. and {Marquart}, T. and {Heiter}, U. and {Rengel}, M. and {Shulyak}, D. and {Yan}, F. and {Hatzes}, A. and {Nagel}, E. and {Reiners}, A. and {Seemann}, U.},
        title = "{VLT/CRIRES+ observations of warm Neptune WASP-107 b: Challenges in detecting molecules with ground-based transmission spectroscopy of cooler and cloudy exoplanets}",
      journal = {\aap},
         year = 2026,
        month = jul,
       volume = {711},
          eid = {A254},
        pages = {A254},
          doi = {10.1051/0004-6361/202556677},
archivePrefix = {arXiv},
       eprint = {2508.18964},
 primaryClass = {astro-ph.EP},
       adsurl = {https://ui.adsabs.harvard.edu/abs/2026A&A...711A.254B}
}

@ARTICLE{GJ436_CRIRES,
       author = {{Grasser}, Natalie and {Snellen}, Ignas A.~G. and {Landman}, Rico and {Picos}, Dar{\'\i}o Gonz{\'a}lez and {Gandhi}, Siddharth},
        title = "{Peering above the clouds of the warm Neptune GJ 436 b with CRIRES+}",
      journal = {\aap},
         year = 2024,
        month = aug,
       volume = {688},
          eid = {A191},
        pages = {A191},
          doi = {10.1051/0004-6361/202449932},
archivePrefix = {arXiv},
       eprint = {2407.06707},
 primaryClass = {astro-ph.EP},
       adsurl = {https://ui.adsabs.harvard.edu/abs/2024A&A...688A.191G}
}

@ARTICLE{GJ436_CRIRES_CARMENES,
       author = {{Pel{\'a}ez-Torres}, A. and {S{\'a}nchez-L{\'o}pez}, A. and {Nortmann}, L. and {L{\'o}pez-Puertas}, M. and {Gonz{\'a}lez-{\'A}lvarez}, E. and {Tabernero}, H.~M. and {Jiang}, C. and {Revilla}, D. and {Morello}, G. and {Orell-Miquel}, J. and {Pall{\'e}}, E. and {Amado}, P.~J. and {Caballero}, J.~A. and {Ribas}, I. and {Reiners}, A. and {Quirrenbach}, A. and {Cont}, D. and {Dreizler}, S. and {Fern{\'a}ndez-Mart{\'\i}n}, A. and {Hatzes}, A.~P. and {Henning}, Th. and {Lesjak}, F. and {Montes}, D. and {Schweizer}, A. and {Trifonov}, T. and {Yan}, F.},
        title = "{Tighter constraints on the atmosphere of GJ 436 b from combined high-resolution CARMENES and CRIRES$^{+}$ observations}",
      journal = {\aap},
         year = 2026,
        month = jan,
       volume = {705},
          eid = {A256},
        pages = {A256},
          doi = {10.1051/0004-6361/202557570},
archivePrefix = {arXiv},
       eprint = {2512.04161},
 primaryClass = {astro-ph.EP},
       adsurl = {https://ui.adsabs.harvard.edu/abs/2026A&A...705A.256P}
}

@ARTICLE{GJ436_Keck,
       author = {{Finnerty}, Luke and {Fitzgerald}, Michael P. and {Xuan}, Jerry W. and {Echeverri}, Daniel and {Jovanovic}, Nemanja and {Mawet}, Dimitri and {Blake}, Geoffrey A. and {Baker}, Ashley and {Bartos}, Randall and {Calvin}, Benjamin and {Cetre}, Sylvain and {Delorme}, Jacques-Robert and {Doppmann}, Greg and {Horstman}, Katelyn and {Hsu}, Chih-Chun and {Inglis}, Julie and {Liberman}, Joshua and {L{\'o}pez}, Ronald A. and {Morris}, Evan and {Pezzato-Rovner}, Jacklyn and {Ruffio}, Jean-Baptiste and {Sappey}, Ben and {Schofield}, Tobias and {Skemer}, Andrew and {Wallace}, J. Kent and {Wallack}, Nicole L. and {Wang}, Jason J. and {Wang}, Ji and {Xin}, Yinzi},
        title = "{Possible Stratospheric Emission in the Warm Neptune GJ 436 b from High-resolution Spectroscopy}",
      journal = {\aj},
         year = 2026,
        month = apr,
       volume = {171},
       number = {4},
          eid = {226},
        pages = {226},
          doi = {10.3847/1538-3881/ae47fa},
archivePrefix = {arXiv},
       eprint = {2602.17777},
 primaryClass = {astro-ph.EP},
       adsurl = {https://ui.adsabs.harvard.edu/abs/2026AJ....171..226F}
}

@ARTICLE{GJ1214_CRIRES,
       author = {{Pel{\'a}ez-Torres}, A. and {S{\'a}nchez-L{\'o}pez}, A. and {Jiang}, C. and {Pall{\'e}}, E. and {Orell-Miquel}, J. and {L{\'o}pez-Puertas}, M. and {Parker}, L.~T. and {Diamond-Lowe}, H.},
        title = "{Atmospheric constraints on GJ 1214 b from CRIRES$^{+}$ and prospects for characterisation with ANDES}",
      journal = {\aap},
         year = 2026,
        month = apr,
       volume = {708},
          eid = {A184},
        pages = {A184},
          doi = {10.1051/0004-6361/202558426},
archivePrefix = {arXiv},
       eprint = {2602.14343},
 primaryClass = {astro-ph.EP},
       adsurl = {https://ui.adsabs.harvard.edu/abs/2026A&A...708A.184P}
}

@ARTICLE{GJ1214_CRIRES_Lisa,
       author = {{Nortmann}, L. and {Cont}, D. and {Lesjak}, F. and {Rains}, A.~D. and {Lavail}, A. and {Boldt-Christmas}, L. and {Nagel}, E. and {Reiners}, A. and {Piskunov}, N. and {Yan}, F. and {Hatzes}, A. and {Kochukhov}, O. and {Shulyak}, D. and {Seemann}, U. and {Rengel}, M. and {Hahlin}, A.},
        title = "{Cloudy with a chance of metals: Indications of CO$_{2}$ in the atmosphere of GJ 1214 b from high-resolution K-band spectroscopy}",
      journal = {\aap},
         year = 2026,
        month = may,
       volume = {710},
          eid = {A29},
        pages = {A29},
          doi = {10.1051/0004-6361/202558512},
archivePrefix = {arXiv},
       eprint = {2604.15292},
 primaryClass = {astro-ph.EP},
       adsurl = {https://ui.adsabs.harvard.edu/abs/2026A&A...710A..29N}
}

@ARTICLE{WASP166_ESPRESSO,
       author = {{Lafarga}, M. and {Brogi}, M. and {Gandhi}, S. and {Cegla}, H.~M. and {Seidel}, J.~V. and {Doyle}, L. and {Allart}, R. and {Buchschacher}, N. and {Lendl}, M. and {Lovis}, C. and {Sosnowska}, D.},
        title = "{The hot Neptune WASP-166 b with ESPRESSO - III. A blue-shifted tentative water signal constrains the presence of clouds}",
      journal = {\mnras},
         year = 2023,
        month = may,
       volume = {521},
       number = {1},
        pages = {1233-1252},
          doi = {10.1093/mnras/stad480},
archivePrefix = {arXiv},
       eprint = {2302.04794},
 primaryClass = {astro-ph.EP},
       adsurl = {https://ui.adsabs.harvard.edu/abs/2023MNRAS.521.1233L}
}

@ARTICLE{GJ3470_CARMENES,
       author = {{Dash}, Spandan and {Brogi}, Matteo and {Gandhi}, Siddharth and {Lafarga}, Marina and {Meech}, Annabella and {Bello-Arufe}, Aaron and {Wheatley}, Peter J.},
        title = "{Constraints on atmospheric water abundance and cloud deck pressure in the warm Neptune GJ 3470 b via CARMENES transmission spectroscopy}",
      journal = {\mnras},
         year = 2024,
        month = may,
       volume = {530},
       number = {3},
        pages = {3100-3116},
          doi = {10.1093/mnras/stae997},
archivePrefix = {arXiv},
       eprint = {2404.06648},
 primaryClass = {astro-ph.EP},
       adsurl = {https://ui.adsabs.harvard.edu/abs/2024MNRAS.530.3100D}
}

@ARTICLE{GJ3090_CRIRES,
       author = {{Parker}, Luke T. and {Mendon{\c{c}}a}, Jo{\~a}o M. and {Diamond-Lowe}, Hannah and {Birkby}, Jayne L. and {Meech}, Annabella and {Vaughan}, Sophia R. and {Brogi}, Matteo and {Fisher}, Chloe and {Buchhave}, Lars A. and {Bello-Arufe}, Aaron and {Kreidberg}, Laura and {Dittmann}, Jason},
        title = "{Limits on the atmospheric metallicity and aerosols of the sub-Neptune GJ 3090 b from high-resolution CRIRES+ spectroscopy}",
      journal = {\mnras},
         year = 2025,
        month = apr,
       volume = {538},
       number = {4},
        pages = {3263-3283},
          doi = {10.1093/mnras/staf469},
archivePrefix = {arXiv},
       eprint = {2503.16608},
 primaryClass = {astro-ph.EP},
       adsurl = {https://ui.adsabs.harvard.edu/abs/2025MNRAS.538.3263P}
}

@ARTICLE{LTT_HST_Transit,
       author = {{Edwards}, Billy and {Changeat}, Quentin and {Tsiaras}, Angelos and {Allan}, Andrew and {Behr}, Patrick and {Hagey}, Simone R. and {Himes}, Michael D. and {Ma}, Sushuang and {Stassun}, Keivan G. and {Thomas}, Luis and {Thompson}, Alexandra and {Boley}, Aaron and {Booth}, Luke and {Bouwman}, Jeroen and {France}, Kevin and {Lowson}, Nataliea and {Meech}, Annabella and {Phillips}, Caprice L. and {Vidotto}, Aline A. and {Yip}, Kai Hou and {Bieger}, Michelle and {Gressier}, Am{\'e}lie and {Janin}, Estelle and {Jiang}, Ing-Guey and {Leonardi}, Pietro and {Sarkar}, Subhajit and {Skaf}, Nour and {Taylor}, Jake and {Yang}, Ming and {Ward-Thompson}, Derek},
        title = "{Characterizing a World Within the Hot-Neptune Desert: Transit Observations of LTT 9779 b with the Hubble Space Telescope/WFC3}",
      journal = {\aj},
         year = 2023,
        month = oct,
       volume = {166},
       number = {4},
          eid = {158},
        pages = {158},
          doi = {10.3847/1538-3881/acea77},
archivePrefix = {arXiv},
       eprint = {2306.13645},
 primaryClass = {astro-ph.EP},
       adsurl = {https://ui.adsabs.harvard.edu/abs/2023AJ....166..158E}
}

@ARTICLE{LTT_ESPRESSO_Transit,
       author = {{Ram{\'\i}rez Reyes}, R. and {Jenkins}, J.~S. and {Sedaghati}, E. and {Seidel}, J.~V. and {Pavlenko}, Y. and {Palle}, E. and {L{\'o}pez-Morales}, M. and {Alves}, D. and {Vines}, J. and {Pe{\~n}a R}, P.~A. and {D{\'\i}az}, M.~R. and {Rojo}, P.},
        title = "{A closer look at LTT 9779b:The ESPRESSO endeavour to pierce the atmospheric veil}",
      journal = {\aap},
         year = 2025,
        month = mar,
       volume = {695},
          eid = {A26},
        pages = {A26},
          doi = {10.1051/0004-6361/202451044},
archivePrefix = {arXiv},
       eprint = {2501.17272},
 primaryClass = {astro-ph.EP},
       adsurl = {https://ui.adsabs.harvard.edu/abs/2025A&A...695A..26R}
}

@ARTICLE{LTT_JWST_Transit,
       author = {{Radica}, Michael and {Coulombe}, Louis-Philippe and {Taylor}, Jake and {Albert}, Loic and {Allart}, Romain and {Benneke}, Bj{\"o}rn and {Cowan}, Nicolas B. and {Dang}, Lisa and {Lafreni{\`e}re}, David and {Thorngren}, Daniel and {Artigau}, {\'E}tienne and {Doyon}, Ren{\'e} and {Flagg}, Laura and {Johnstone}, Doug and {Pelletier}, Stefan and {Roy}, Pierre-Alexis},
        title = "{Muted Features in the JWST NIRISS Transmission Spectrum of Hot Neptune LTT 9779b}",
      journal = {\apjl},
         year = 2024,
        month = feb,
       volume = {962},
       number = {1},
          eid = {L20},
        pages = {L20},
          doi = {10.3847/2041-8213/ad20e4},
archivePrefix = {arXiv},
       eprint = {2401.15548},
 primaryClass = {astro-ph.EP},
       adsurl = {https://ui.adsabs.harvard.edu/abs/2024ApJ...962L..20R}
}

@ARTICLE{LTT_Xray,
       author = {{Fern{\'a}ndez Fern{\'a}ndez}, Jorge and {Wheatley}, Peter J. and {King}, George W. and {Jenkins}, James S.},
        title = "{Survival in the Neptune desert: LTT 9779 b kept its atmosphere thanks to an unusually X-ray faint host star}",
      journal = {\mnras},
         year = 2024,
        month = jan,
       volume = {527},
       number = {1},
        pages = {911-918},
          doi = {10.1093/mnras/stad3263},
archivePrefix = {arXiv},
       eprint = {2310.13171},
 primaryClass = {astro-ph.EP},
       adsurl = {https://ui.adsabs.harvard.edu/abs/2024MNRAS.527..911F}
}

@ARTICLE{LTT_CHEOPS,
       author = {{Hoyer}, S. and {Jenkins}, J.~S. and {Parmentier}, V. and {Deleuil}, M. and {Scandariato}, G. and {Wilson}, T.~G. and {D{\'\i}az}, M.~R. and {Crossfield}, I.~J.~M. and {Dragomir}, D. and {Kataria}, T. and {Lendl}, M. and {Ramirez}, R. and {Pe{\~n}a Rojas}, P.~A. and {Vin{\'e}s}, J.~I.},
        title = "{The extremely high albedo of LTT 9779 b revealed by CHEOPS. An ultrahot Neptune with a highly metallic atmosphere}",
      journal = {\aap},
         year = 2023,
        month = jul,
       volume = {675},
          eid = {A81},
        pages = {A81},
          doi = {10.1051/0004-6361/202346117},
       adsurl = {https://ui.adsabs.harvard.edu/abs/2023A&A...675A..81H}
}

@ARTICLE{LTT_JWST_NIRSpec,
       author = {{Ashtari}, Reza and {Collins}, Sean and {Splinter}, Jared and {Stevenson}, Kevin B. and {Parmentier}, Vivien and {Brande}, Jonathan and {Saha}, Suman and {Stamer}, Sarah and {Crossfield}, Ian J.~M. and {Jenkins}, James S. and {Kahle}, K. Angelique and {Lothringer}, Joshua D. and {Mehta}, Nishil and {Cowan}, Nicolas B. and {Dragomir}, Diana and {Kreidberg}, Laura and {L{\'o}pez-Morales}, Mercedes and {Evans-Soma}, Thomas M. and {Esparza-Borges}, Emma and {Weiner Mansfield}, Megan and {Beltz}, Hayley and {Daylan}, Tansu and {Venot}, Olivia and {Pero}, Jason and {Zhang}, Xi},
        title = "{Heat Reveals What Clouds Conceal: Global Carbon and Longitudinally Asymmetric Chemistry on LTT 9779 b}",
      journal = {\aj},
         year = 2026,
        month = apr,
       volume = {171},
       number = {4},
          eid = {215},
        pages = {215},
          doi = {10.3847/1538-3881/ae4494},
archivePrefix = {arXiv},
       eprint = {2510.04863},
 primaryClass = {astro-ph.EP},
       adsurl = {https://ui.adsabs.harvard.edu/abs/2026AJ....171..215A}
}

@ARTICLE{LTT_ESPRESSO_Reflection,
       author = {{Vaughan}, Sophia R. and {Birkby}, Jayne L. and {Batalha}, Natasha E. and {Parker}, Luke T. and {Yu}, Haochuan and {Seidel}, Julia V. and {Radica}, Michael and {Taylor}, Jake and {Kreidberg}, Laura and {Parmentier}, Vivien and {Hoyer}, Sergio and {Jenkins}, James S. and {Meech}, Annabella and {Ram{\'\i}rez Reyes}, Ricardo and {van Sluijs}, Lennart},
        title = "{No TiO detected in the hot-Neptune-desert planet LTT-9779 b in reflected light at high spectral resolution}",
      journal = {\aap},
         year = 2026,
        month = jan,
       volume = {705},
          eid = {A27},
        pages = {A27},
          doi = {10.1051/0004-6361/202557240},
archivePrefix = {arXiv},
       eprint = {2511.08691},
 primaryClass = {astro-ph.EP},
       adsurl = {https://ui.adsabs.harvard.edu/abs/2026A&A...705A..27V}
}

@ARTICLE{LTT_ESPRESSO_Reflection_Detection,
       author = {{Borsa}, F.},
        title = "{High-resolution detection of reflected light from the exo-Neptune LTT-9779 b}",
      journal = {\aap},
         year = 2026,
        month = jul,
       volume = {711},
          eid = {L5},
        pages = {L5},
          doi = {10.1051/0004-6361/202660755},
archivePrefix = {arXiv},
       eprint = {2606.30793},
 primaryClass = {astro-ph.EP},
       adsurl = {https://ui.adsabs.harvard.edu/abs/2026A&A...711L...5B}
}

@ARTICLE{CRIRES_Dorn,
       author = {{Dorn}, R.~J. and {Bristow}, P. and {Smoker}, J.~V. and {Rodler}, F. and {Lavail}, A. and {Accardo}, M. and {van den Ancker}, M. and {Baade}, D. and {Baruffolo}, A. and {Courtney-Barrer}, B. and {Blanco}, L. and {Brucalassi}, A. and {Cumani}, C. and {Follert}, R. and {Haimerl}, A. and {Hatzes}, A. and {Haug}, M. and {Heiter}, U. and {Hinterschuster}, R. and {Hubin}, N. and {Ives}, D.~J. and {Jung}, Y. and {Jones}, M. and {Kaeufl}, H. -U. and {Kirchbauer}, J. -P. and {Klein}, B. and {Kochukhov}, O. and {Korhonen}, H.~H. and {K{\"o}hler}, J. and {Lizon}, J. -L. and {Moins}, C. and {Molina-Conde}, I. and {Marquart}, T. and {Neeser}, M. and {Oliva}, E. and {Pallanca}, L. and {Pasquini}, L. and {Paufique}, J. and {Piskunov}, N. and {Reiners}, A. and {Schneller}, D. and {Schmutzer}, R. and {Seemann}, U. and {Slumstrup}, D. and {Smette}, A. and {Stegmeier}, J. and {Stempels}, E. and {Tordo}, S. and {Valenti}, E. and {Valenzuela}, J.~J. and {Vernet}, J. and {Vinther}, J. and {Wehrhahn}, A.},
        title = "{CRIRES$^{+}$ on sky at the ESO Very Large Telescope. Observing the Universe at infrared wavelengths and high spectral resolution}",
      journal = {\aap},
         year = 2023,
        month = mar,
       volume = {671},
          eid = {A24},
        pages = {A24},
          doi = {10.1051/0004-6361/202245217},
archivePrefix = {arXiv},
       eprint = {2301.08048},
 primaryClass = {astro-ph.IM},
       adsurl = {https://ui.adsabs.harvard.edu/abs/2023A&A...671A..24D}
}

@ARTICLE{CRIRES_Yan,
       author = {{Yan}, F. and {Nortmann}, L. and {Reiners}, A. and {Piskunov}, N. and {Hatzes}, A. and {Seemann}, U. and {Shulyak}, D. and {Lavail}, A. and {Rains}, A.~D. and {Cont}, D. and {Rengel}, M. and {Lesjak}, F. and {Nagel}, E. and {Kochukhov}, O. and {Czesla}, S. and {Boldt-Christmas}, L. and {Heiter}, U. and {Smoker}, J.~V. and {Rodler}, F. and {Bristow}, P. and {Dorn}, R.~J. and {Jung}, Y. and {Marquart}, T. and {Stempels}, E.},
        title = "{CRIRES$^{+}$ detection of CO emissions lines and temperature inversions on the dayside of WASP-18b and WASP-76b}",
      journal = {\aap},
         year = 2023,
        month = apr,
       volume = {672},
          eid = {A107},
        pages = {A107},
          doi = {10.1051/0004-6361/202245371},
archivePrefix = {arXiv},
       eprint = {2302.08736},
 primaryClass = {astro-ph.EP},
       adsurl = {https://ui.adsabs.harvard.edu/abs/2023A&A...672A.107Y}
}

@ARTICLE{molecfit,
       author = {{Smette}, A. and {Sana}, H. and {Noll}, S. and {Horst}, H. and {Kausch}, W. and {Kimeswenger}, S. and {Barden}, M. and {Szyszka}, C. and {Jones}, A.~M. and {Gallenne}, A. and {Vinther}, J. and {Ballester}, P. and {Taylor}, J.},
        title = "{Molecfit: A general tool for telluric absorption correction. I. Method and application to ESO instruments}",
      journal = {\aap},
         year = 2015,
        month = apr,
       volume = {576},
          eid = {A77},
        pages = {A77},
          doi = {10.1051/0004-6361/201423932},
archivePrefix = {arXiv},
       eprint = {1501.07239},
 primaryClass = {astro-ph.IM},
       adsurl = {https://ui.adsabs.harvard.edu/abs/2015A&A...576A..77S}
}

@ARTICLE{sysrem_and_molecfit,
       author = {{Maguire}, Cathal and {Sedaghati}, Elyar and {Gibson}, Neale P. and {Smette}, Alain and {Pino}, Lorenzo},
        title = "{Assessing methods for telluric removal on atmospheric retrievals of high-resolution optical exoplanetary transmission spectra}",
      journal = {\aap},
         year = 2024,
        month = dec,
       volume = {692},
          eid = {A8},
        pages = {A8},
          doi = {10.1051/0004-6361/202451784},
archivePrefix = {arXiv},
       eprint = {2410.19588},
 primaryClass = {astro-ph.EP},
       adsurl = {https://ui.adsabs.harvard.edu/abs/2024A&A...692A...8M}
}

@ARTICLE{KELT9_HARPS,
       author = {{Hoeijmakers}, H. Jens and {Ehrenreich}, David and {Heng}, Kevin and {Kitzmann}, Daniel and {Grimm}, Simon L. and {Allart}, Romain and {Deitrick}, Russell and {Wyttenbach}, Aur{\'e}lien and {Oreshenko}, Maria and {Pino}, Lorenzo and {Rimmer}, Paul B. and {Molinari}, Emilio and {Di Fabrizio}, Luca},
        title = "{Atomic iron and titanium in the atmosphere of the exoplanet KELT-9b}",
      journal = {\nat},
         year = 2018,
        month = aug,
       volume = {560},
       number = {7719},
        pages = {453-455},
          doi = {10.1038/s41586-018-0401-y},
archivePrefix = {arXiv},
       eprint = {1808.05653},
 primaryClass = {astro-ph.EP},
       adsurl = {https://ui.adsabs.harvard.edu/abs/2018Natur.560..453H}
}

@ARTICLE{sysrem,
       author = {{Tamuz}, O. and {Mazeh}, T. and {Zucker}, S.},
        title = "{Correcting systematic effects in a large set of photometric light curves}",
      journal = {\mnras},
         year = 2005,
        month = feb,
       volume = {356},
       number = {4},
        pages = {1466-1470},
          doi = {10.1111/j.1365-2966.2004.08585.x},
archivePrefix = {arXiv},
       eprint = {astro-ph/0502056},
 primaryClass = {astro-ph},
       adsurl = {https://ui.adsabs.harvard.edu/abs/2005MNRAS.356.1466T}
}

@ARTICLE{sysrem2,
       author = {{Snellen}, Ignas A.~G. and {Brandl}, Bernhard R. and {de Kok}, Remco J. and {Brogi}, Matteo and {Birkby}, Jayne and {Schwarz}, Henriette},
        title = "{Fast spin of the young extrasolar planet {\ensuremath{\beta}} Pictoris b}",
      journal = {\nat},
         year = 2014,
        month = may,
       volume = {509},
       number = {7498},
        pages = {63-65},
          doi = {10.1038/nature13253},
       adsurl = {https://ui.adsabs.harvard.edu/abs/2014Natur.509...63S}
}

@ARTICLE{Yan_2020,
       author = {{Yan}, F. and {Pall{\'e}}, E. and {Reiners}, A. and {Molaverdikhani}, K. and {Casasayas-Barris}, N. and {Nortmann}, L. and {Chen}, G. and {Molli{\`e}re}, P. and {Stangret}, M.},
        title = "{A temperature inversion with atomic iron in the ultra-hot dayside atmosphere of WASP-189b}",
      journal = {\aap},
         year = 2020,
        month = aug,
       volume = {640},
          eid = {L5},
        pages = {L5},
          doi = {10.1051/0004-6361/202038294},
archivePrefix = {arXiv},
       eprint = {2007.02716},
 primaryClass = {astro-ph.EP},
       adsurl = {https://ui.adsabs.harvard.edu/abs/2020A&A...640L...5Y}
}

@ARTICLE{Guo,
       author = {{Guo}, B. and {Yan}, F. and {Nortmann}, L. and {Cont}, D. and {Reiners}, A. and {Pall{\'e}}, E. and {Shulyak}, D. and {Molaverdikhani}, K. and {Henning}, Th. and {Chen}, G. and {Stangret}, M. and {Czesla}, S. and {Lesjak}, F. and {L{\'o}pez-Puertas}, M. and {Ribas}, I. and {Quirrenbach}, A. and {Caballero}, J.~A. and {Amado}, P.~J. and {Blazek}, M. and {Montes}, D. and {Morales}, J.~C. and {Nagel}, E. and {Zapatero Osorio}, M.~R.},
        title = "{Detection of Fe and Ti on the dayside of the ultrahot Jupiter MASCARA-1b with CARMENES}",
      journal = {\aap},
         year = 2024,
        month = jul,
       volume = {687},
          eid = {A103},
        pages = {A103},
          doi = {10.1051/0004-6361/202449890},
archivePrefix = {arXiv},
       eprint = {2404.18788},
 primaryClass = {astro-ph.EP},
       adsurl = {https://ui.adsabs.harvard.edu/abs/2024A&A...687A.103G}
}

@ARTICLE{LTT_Discovery,
       author = {{Jenkins}, James S. and {D{\'\i}az}, Mat{\'\i}as R. and {Kurtovic}, Nicol{\'a}s T. and {Espinoza}, N{\'e}stor and {Vines}, Jose I. and {Rojas}, Pablo A. Pe{\~n}a and {Brahm}, Rafael and {Torres}, Pascal and {Cort{\'e}s-Zuleta}, P{\'\i}a and {Soto}, Maritza G. and {Lopez}, Eric D. and {King}, George W. and {Wheatley}, Peter J. and {Winn}, Joshua N. and {Ciardi}, David R. and {Ricker}, George and {Vanderspek}, Roland and {Latham}, David W. and {Seager}, Sara and {Jenkins}, Jon M. and {Beichman}, Charles A. and {Bieryla}, Allyson and {Burke}, Christopher J. and {Christiansen}, Jessie L. and {Henze}, Christopher E. and {Klaus}, Todd C. and {McCauliff}, Sean and {Mori}, Mayuko and {Narita}, Norio and {Nishiumi}, Taku and {Tamura}, Motohide and {de Leon}, Jerome Pitogo and {Quinn}, Samuel N. and {Villase{\~n}or}, Jesus Noel and {Vezie}, Michael and {Lissauer}, Jack J. and {Collins}, Karen A. and {Collins}, Kevin I. and {Isopi}, Giovanni and {Mallia}, Franco and {Ercolino}, Andrea and {Petrovich}, Cristobal and {Jord{\'a}n}, Andr{\'e}s and {Acton}, Jack S. and {Armstrong}, David J. and {Bayliss}, Daniel and {Bouchy}, Fran{\c{c}}ois and {Belardi}, Claudia and {Bryant}, Edward M. and {Burleigh}, Matthew R. and {Cabrera}, Juan and {Casewell}, Sarah L. and {Chaushev}, Alexander and {Cooke}, Benjamin F. and {Eigm{\"u}ller}, Philipp and {Erikson}, Anders and {Foxell}, Emma and {G{\"a}nsicke}, Boris T. and {Gill}, Samuel and {Gillen}, Edward and {G{\"u}nther}, Maximilian N. and {Goad}, Michael R. and {Hooton}, Matthew J. and {Jackman}, James A.~G. and {Louden}, Tom and {McCormac}, James and {Moyano}, Maximiliano and {Nielsen}, Louise D. and {Pollacco}, Don and {Queloz}, Didier and {Rauer}, Heike and {Raynard}, Liam and {Smith}, Alexis M.~S. and {Tilbrook}, Rosanna H. and {Titz-Weider}, Ruth and {Turner}, Oliver and {Udry}, St{\'e}phane and {Walker}, Simon. R. and {Watson}, Christopher A. and {West}, Richard G. and {Palle}, Enric and {Ziegler}, Carl and {Law}, Nicholas and {Mann}, Andrew W.},
        title = "{An ultrahot Neptune in the Neptune desert}",
      journal = {Nature Astronomy},
         year = 2020,
        month = jan,
       volume = {4},
        pages = {1148-1157},
          doi = {10.1038/s41550-020-1142-z},
archivePrefix = {arXiv},
       eprint = {2009.12832},
 primaryClass = {astro-ph.EP},
       adsurl = {https://ui.adsabs.harvard.edu/abs/2020NatAs...4.1148J}
}

@INCOLLECTION{Transit_equation,
       author = {{Winn}, J.~N.},
        title = "{Exoplanet Transits and Occultations}",
    booktitle = {Exoplanets},
         year = 2010,
       editor = {{Seager}, S.},
        pages = {55-77},
        publisher = {University of Arizona Press},
          doi = {10.48550/arXiv.1001.2010},
       adsurl = {https://ui.adsabs.harvard.edu/abs/2010exop.book...55W}
}

@ARTICLE{Gaia_DR2,
       author = {{Gaia Collaboration} and {Brown}, A.~G.~A. and {Vallenari}, A. and {Prusti}, T. and {de Bruijne}, J.~H.~J. and {Babusiaux}, C. and {Bailer-Jones}, C.~A.~L. and {Biermann}, M. and {Evans}, D.~W. and {Eyer}, L. and {Jansen}, F. and {Jordi}, C. and {Klioner}, S.~A. and {Lammers}, U. and {Lindegren}, L. and {Luri}, X. and {Mignard}, F. and {Panem}, C. and {Pourbaix}, D. and {Randich}, S. and {Sartoretti}, P. and {Siddiqui}, H.~I. and {Soubiran}, C. and {van Leeuwen}, F. and {Walton}, N.~A. and {Arenou}, F. and {Bastian}, U. and {Cropper}, M. and {Drimmel}, R. and {Katz}, D. and {Lattanzi}, M.~G. and {Bakker}, J. and {Cacciari}, C. and {Casta{\~n}eda}, J. and {Chaoul}, L. and {Cheek}, N. and {De Angeli}, F. and {Fabricius}, C. and {Guerra}, R. and {Holl}, B. and {Masana}, E. and {Messineo}, R. and {Mowlavi}, N. and {Nienartowicz}, K. and {Panuzzo}, P. and {Portell}, J. and {Riello}, M. and {Seabroke}, G.~M. and {Tanga}, P. and {Th{\'e}venin}, F. and {Gracia-Abril}, G. and {Comoretto}, G. and {Garcia-Reinaldos}, M. and {Teyssier}, D. and {Altmann}, M. and {Andrae}, R. and {Audard}, M. and {Bellas-Velidis}, I. and {Benson}, K. and {Berthier}, J. and {Blomme}, R. and {Burgess}, P. and {Busso}, G. and {Carry}, B. and {Cellino}, A. and {Clementini}, G. and {Clotet}, M. and {Creevey}, O. and {Davidson}, M. and {De Ridder}, J. and {Delchambre}, L. and {Dell'Oro}, A. and {Ducourant}, C. and {Fern{\'a}ndez-Hern{\'a}ndez}, J. and {Fouesneau}, M. and {Fr{\'e}mat}, Y. and {Galluccio}, L. and {Garc{\'\i}a-Torres}, M. and {Gonz{\'a}lez-N{\'u}{\~n}ez}, J. and {Gonz{\'a}lez-Vidal}, J.~J. and {Gosset}, E. and {Guy}, L.~P. and {Halbwachs}, J. -L. and {Hambly}, N.~C. and {Harrison}, D.~L. and {Hern{\'a}ndez}, J. and {Hestroffer}, D. and {Hodgkin}, S.~T. and {Hutton}, A. and {Jasniewicz}, G. and {Jean-Antoine-Piccolo}, A. and {Jordan}, S. and {Korn}, A.~J. and {Krone-Martins}, A. and {Lanzafame}, A.~C. and {Lebzelter}, T. and {L{\"o}ffler}, W. and {Manteiga}, M. and {Marrese}, P.~M. and {Mart{\'\i}n-Fleitas}, J.~M. and {Moitinho}, A. and {Mora}, A. and {Muinonen}, K. and {Osinde}, J. and {Pancino}, E. and {Pauwels}, T. and {Petit}, J. -M. and {Recio-Blanco}, A. and {Richards}, P.~J. and {Rimoldini}, L. and {Robin}, A.~C. and {Sarro}, L.~M. and {Siopis}, C. and {Smith}, M. and {Sozzetti}, A. and {S{\"u}veges}, M. and {Torra}, J. and {van Reeven}, W. and {Abbas}, U. and {Abreu Aramburu}, A. and {Accart}, S. and {Aerts}, C. and {Altavilla}, G. and {{\'A}lvarez}, M.~A. and {Alvarez}, R. and {Alves}, J. and {Anderson}, R.~I. and {Andrei}, A.~H. and {Anglada Varela}, E. and {Antiche}, E. and {Antoja}, T. and {Arcay}, B. and {Astraatmadja}, T.~L. and {Bach}, N. and {Baker}, S.~G. and {Balaguer-N{\'u}{\~n}ez}, L. and {Balm}, P. and {Barache}, C. and {Barata}, C. and {Barbato}, D. and {Barblan}, F. and {Barklem}, P.~S. and {Barrado}, D. and {Barros}, M. and {Barstow}, M.~A. and {Bartholom{\'e} Mu{\~n}oz}, S. and {Bassilana}, J. -L. and {Becciani}, U. and {Bellazzini}, M. and {Berihuete}, A. and {Bertone}, S. and {Bianchi}, L. and {Bienaym{\'e}}, O. and {Blanco-Cuaresma}, S. and {Boch}, T. and {Boeche}, C. and {Bombrun}, A. and {Borrachero}, R. and {Bossini}, D. and {Bouquillon}, S. and {Bourda}, G. and {Bragaglia}, A. and {Bramante}, L. and {Breddels}, M.~A. and {Bressan}, A. and {Brouillet}, N. and {Br{\"u}semeister}, T. and {Brugaletta}, E. and {Bucciarelli}, B. and {Burlacu}, A. and {Busonero}, D. and {Butkevich}, A.~G. and {Buzzi}, R. and {Caffau}, E. and {Cancelliere}, R. and {Cannizzaro}, G. and {Cantat-Gaudin}, T. and {Carballo}, R. and {Carlucci}, T. and {Carrasco}, J.~M. and {Casamiquela}, L. and {Castellani}, M. and {Castro-Ginard}, A. and {Charlot}, P. and {Chemin}, L. and {Chiavassa}, A. and {Cocozza}, G. and {Costigan}, G. and {Cowell}, S. and {Crifo}, F. and {Crosta}, M. and {Crowley}, C. and {Cuypers}, J. and {Dafonte}, C. and {Damerdji}, Y. and {Dapergolas}, A. and {David}, P. and {David}, M. and {de Laverny}, P. and {De Luise}, F.},
        title = "{Gaia Data Release 2. Summary of the contents and survey properties}",
      journal = {\aap},
         year = 2018,
        month = aug,
       volume = {616},
          eid = {A1},
        pages = {A1},
          doi = {10.1051/0004-6361/201833051},
archivePrefix = {arXiv},
       eprint = {1804.09365},
 primaryClass = {astro-ph.GA},
       adsurl = {https://ui.adsabs.harvard.edu/abs/2018A&A...616A...1G}
}

@ARTICLE{EA_Sara,
       author = {{Seager}, Sara and {Deming}, Drake},
        title = "{Exoplanet Atmospheres}",
      journal = {\araa},
         year = 2010,
        month = sep,
       volume = {48},
        pages = {631-672},
          doi = {10.1146/annurev-astro-081309-130837},
archivePrefix = {arXiv},
       eprint = {1005.4037},
 primaryClass = {astro-ph.EP},
       adsurl = {https://ui.adsabs.harvard.edu/abs/2010ARA&A..48..631S}
}

@ARTICLE{WASP178_David,
       author = {{Cont}, D. and {Nortmann}, L. and {Yan}, F. and {Lesjak}, F. and {Czesla}, S. and {Lavail}, A. and {Reiners}, A. and {Piskunov}, N. and {Hatzes}, A. and {Boldt-Christmas}, L. and {Kochukhov}, O. and {Marquart}, T. and {Nagel}, E. and {Rains}, A.~D. and {Rengel}, M. and {Seemann}, U. and {Shulyak}, D.},
        title = "{Exploring the ultra-hot Jupiter WASP-178b. Constraints on atmospheric chemistry and dynamics from a joint retrieval of VLT/CRIRES$^{+}$ and space photometric data}",
      journal = {\aap},
         year = 2024,
        month = aug,
       volume = {688},
          eid = {A206},
        pages = {A206},
          doi = {10.1051/0004-6361/202450064},
archivePrefix = {arXiv},
       eprint = {2406.08166},
 primaryClass = {astro-ph.EP},
       adsurl = {https://ui.adsabs.harvard.edu/abs/2024A&A...688A.206C}
}

@ARTICLE{OH_line,
       author = {{Oliva}, E. and {Origlia}, L. and {Scuderi}, S. and {Benatti}, S. and {Carleo}, I. and {Lapenna}, E. and {Mucciarelli}, A. and {Baffa}, C. and {Biliotti}, V. and {Carbonaro}, L. and {Falcini}, G. and {Giani}, E. and {Iuzzolino}, M. and {Massi}, F. and {Sanna}, N. and {Sozzi}, M. and {Tozzi}, A. and {Ghedina}, A. and {Ghinassi}, F. and {Lodi}, M. and {Harutyunyan}, A. and {Pedani}, M.},
        title = "{Lines and continuum sky emission in the near infrared: observational constraints from deep high spectral resolution spectra with GIANO-TNG}",
      journal = {\aap},
         year = 2015,
        month = sep,
       volume = {581},
          eid = {A47},
        pages = {A47},
          doi = {10.1051/0004-6361/201526291},
archivePrefix = {arXiv},
       eprint = {1506.09004},
 primaryClass = {astro-ph.IM},
       adsurl = {https://ui.adsabs.harvard.edu/abs/2015A&A...581A..47O}
}

@ARTICLE{CCF_2010,
       author = {{Snellen}, Ignas A.~G. and {de Kok}, Remco J. and {de Mooij}, Ernst J.~W. and {Albrecht}, Simon},
        title = "{The orbital motion, absolute mass and high-altitude winds of exoplanet HD209458b}",
      journal = {\nat},
         year = 2010,
        month = jun,
       volume = {465},
       number = {7301},
        pages = {1049-1051},
          doi = {10.1038/nature09111},
archivePrefix = {arXiv},
       eprint = {1006.4364},
 primaryClass = {astro-ph.EP},
       adsurl = {https://ui.adsabs.harvard.edu/abs/2010Natur.465.1049S}
}

@ARTICLE{CCF_2012,
       author = {{Brogi}, Matteo and {Snellen}, Ignas A.~G. and {de Kok}, Remco J. and {Albrecht}, Simon and {Birkby}, Jayne and {de Mooij}, Ernst J.~W.},
        title = "{The signature of orbital motion from the dayside of the planet {\ensuremath{\tau}} Bo{\"o}tis b}",
      journal = {\nat},
         year = 2012,
        month = jun,
       volume = {486},
       number = {7404},
        pages = {502-504},
          doi = {10.1038/nature11161},
archivePrefix = {arXiv},
       eprint = {1206.6109},
 primaryClass = {astro-ph.EP},
       adsurl = {https://ui.adsabs.harvard.edu/abs/2012Natur.486..502B}
}

@ARTICLE{CCF_2013,
       author = {{Birkby}, J.~L. and {de Kok}, R.~J. and {Brogi}, M. and {de Mooij}, E.~J.~W. and {Schwarz}, H. and {Albrecht}, S. and {Snellen}, I.~A.~G.},
        title = "{Detection of water absorption in the day side atmosphere of HD 189733 b  using ground-based high-resolution spectroscopy at 3.2{\ensuremath{\mu}}m.}",
      journal = {\mnras},
         year = 2013,
        month = nov,
       volume = {436},
        pages = {L35-L39},
          doi = {10.1093/mnrasl/slt107},
archivePrefix = {arXiv},
       eprint = {1307.1133},
 primaryClass = {astro-ph.EP},
       adsurl = {https://ui.adsabs.harvard.edu/abs/2013MNRAS.436L..35B}
}

@ARTICLE{CCF_Fe1,
       author = {{Ramkumar}, Swaetha and {Gibson}, Neale P. and {Nugroho}, Stevanus K. and {Maguire}, Cathal and {Fortune}, Mark},
        title = "{High-resolution emission spectroscopy retrievals of MASCARA-1b with CRIRES+: strong detections of CO, H$_{2}$O, and Fe emission lines and a C/O consistent with solar}",
      journal = {\mnras},
         year = 2023,
        month = oct,
       volume = {525},
       number = {2},
        pages = {2985-3005},
          doi = {10.1093/mnras/stad2476},
archivePrefix = {arXiv},
       eprint = {2308.07157},
 primaryClass = {astro-ph.EP},
       adsurl = {https://ui.adsabs.harvard.edu/abs/2023MNRAS.525.2985R}
}

@ARTICLE{W189_Fabio,
       author = {{Lesjak}, F. and {Nortmann}, L. and {Cont}, D. and {Yan}, F. and {Reiners}, A. and {Piskunov}, N. and {Hatzes}, A. and {Boldt-Christmas}, L. and {Czesla}, S. and {Lavail}, A. and {Nagel}, E. and {Rains}, A.~D. and {Rengel}, M. and {Seemann}, U. and {Shulyak}, D.},
        title = "{Retrieving wind properties from the ultra-hot dayside of WASP-189 b with CRIRES$^{+}$}",
      journal = {\aap},
         year = 2025,
        month = jan,
       volume = {693},
          eid = {A72},
        pages = {A72},
          doi = {10.1051/0004-6361/202451391},
archivePrefix = {arXiv},
       eprint = {2411.19662},
 primaryClass = {astro-ph.EP},
       adsurl = {https://ui.adsabs.harvard.edu/abs/2025A&A...693A..72L}
}

@ARTICLE{ltt_jwst_emission,
       author = {{Coulombe}, Louis-Philippe and {Radica}, Michael and {Benneke}, Bj{\"o}rn and {D'Aoust}, {\'E}lyse and {Dang}, Lisa and {Cowan}, Nicolas B. and {Parmentier}, Vivien and {Albert}, Lo{\"\i}c and {Lafreni{\`e}re}, David and {Taylor}, Jake and {Roy}, Pierre-Alexis and {Pelletier}, Stefan and {Allart}, Romain and {Artigau}, {\'E}tienne and {Doyon}, Ren{\'e} and {Jayawardhana}, Ray and {Johnstone}, Doug and {Kaltenegger}, Lisa and {Langeveld}, Adam B. and {MacDonald}, Ryan J. and {Rowe}, Jason F. and {Turner}, Jake D.},
        title = "{Highly reflective white clouds on the western dayside of an exo-Neptune}",
      journal = {Nature Astronomy},
         year = 2025,
        month = apr,
       volume = {9},
        pages = {512-525},
          doi = {10.1038/s41550-025-02488-9},
archivePrefix = {arXiv},
       eprint = {2501.14016},
 primaryClass = {astro-ph.EP},
       adsurl = {https://ui.adsabs.harvard.edu/abs/2025NatAs...9..512C}
}

@ARTICLE{fastchem,
       author = {{Stock}, Joachim W. and {Kitzmann}, Daniel and {Patzer}, A. Beate C. and {Sedlmayr}, Erwin},
        title = "{FastChem: A computer program for efficient complex chemical equilibrium calculations in the neutral/ionized gas phase with applications to stellar and planetary atmospheres}",
      journal = {\mnras},
         year = 2018,
        month = sep,
       volume = {479},
       number = {1},
        pages = {865-874},
          doi = {10.1093/mnras/sty1531},
archivePrefix = {arXiv},
       eprint = {1804.05010},
 primaryClass = {astro-ph.EP},
       adsurl = {https://ui.adsabs.harvard.edu/abs/2018MNRAS.479..865S}
}

@ARTICLE{CO_opacity,
       author = {{Li}, Gang and {Gordon}, Iouli E. and {Rothman}, Laurence S. and {Tan}, Yan and {Hu}, Shui-Ming and {Kassi}, Samir and {Campargue}, Alain and {Medvedev}, Emile S.},
        title = "{Rovibrational Line Lists for Nine Isotopologues of the CO Molecule in the X $^{1}${\ensuremath{\Sigma}}$^{+}$ Ground Electronic State}",
      journal = {\apjs},
         year = 2015,
        month = jan,
       volume = {216},
       number = {1},
          eid = {15},
        pages = {15},
          doi = {10.1088/0067-0049/216/1/15},
       adsurl = {https://ui.adsabs.harvard.edu/abs/2015ApJS..216...15L}
}

@ARTICLE{H2O_opacity,
       author = {{Polyansky}, Oleg L. and {Kyuberis}, Aleksandra A. and {Zobov}, Nikolai F. and {Tennyson}, Jonathan and {Yurchenko}, Sergei N. and {Lodi}, Lorenzo},
        title = "{ExoMol molecular line lists XXX: a complete high-accuracy line list for water}",
      journal = {\mnras},
         year = 2018,
        month = oct,
       volume = {480},
       number = {2},
        pages = {2597-2608},
          doi = {10.1093/mnras/sty1877},
archivePrefix = {arXiv},
       eprint = {1807.04529},
 primaryClass = {astro-ph.EP},
       adsurl = {https://ui.adsabs.harvard.edu/abs/2018MNRAS.480.2597P}
}

@ARTICLE{CO2_opacity,
       author = {{Yurchenko}, S.~N. and {Mellor}, Thomas M. and {Freedman}, Richard S. and {Tennyson}, J.},
        title = "{ExoMol line lists - XXXIX. Ro-vibrational molecular line list for CO$_{2}$}",
      journal = {\mnras},
         year = 2020,
        month = aug,
       volume = {496},
       number = {4},
        pages = {5282-5291},
          doi = {10.1093/mnras/staa1874},
archivePrefix = {arXiv},
       eprint = {2007.02122},
 primaryClass = {astro-ph.EP},
       adsurl = {https://ui.adsabs.harvard.edu/abs/2020MNRAS.496.5282Y}
}

@ARTICLE{OH_opacity,
       author = {{Brooke}, James S.~A. and {Bernath}, Peter F. and {Western}, Colin M. and {Sneden}, Christopher and {Af{\c{s}}ar}, Melike and {Li}, Gang and {Gordon}, Iouli E.},
        title = "{Line strengths of rovibrational and rotational transitions in the X$^{2}$ {\ensuremath{\Pi}} ground state of OH}",
      journal = {\jqsrt},
         year = 2016,
        month = jan,
       volume = {168},
        pages = {142-157},
          doi = {10.1016/j.jqsrt.2015.07.021},
       adsurl = {https://ui.adsabs.harvard.edu/abs/2016JQSRT.168..142B}
}

@ARTICLE{CH4_opacity,
       author = {{Hargreaves}, Robert J. and {Gordon}, Iouli E. and {Rey}, Michael and {Nikitin}, Andrei V. and {Tyuterev}, Vladimir G. and {Kochanov}, Roman V. and {Rothman}, Laurence S.},
        title = "{An Accurate, Extensive, and Practical Line List of Methane for the HITEMP Database}",
      journal = {\apjs},
         year = 2020,
        month = apr,
       volume = {247},
       number = {2},
          eid = {55},
        pages = {55},
          doi = {10.3847/1538-4365/ab7a1a},
archivePrefix = {arXiv},
       eprint = {2001.05037},
 primaryClass = {astro-ph.EP},
       adsurl = {https://ui.adsabs.harvard.edu/abs/2020ApJS..247...55H}
}

@ARTICLE{HCN_opacity,
       author = {{Harris}, G.~J. and {Tennyson}, J. and {Kaminsky}, B.~M. and {Pavlenko}, Ya. V. and {Jones}, H.~R.~A.},
        title = "{Improved HCN/HNC linelist, model atmospheres and synthetic spectra for WZ Cas}",
      journal = {\mnras},
         year = 2006,
        month = mar,
       volume = {367},
       number = {1},
        pages = {400-406},
          doi = {10.1111/j.1365-2966.2005.09960.x},
archivePrefix = {arXiv},
       eprint = {astro-ph/0512363},
 primaryClass = {astro-ph},
       adsurl = {https://ui.adsabs.harvard.edu/abs/2006MNRAS.367..400H}
}
\begin{appendix}
\onecolumn

\section{Additional figures}

\begin{figure}[!ht]
    \centering
    \includegraphics[width=\textwidth]{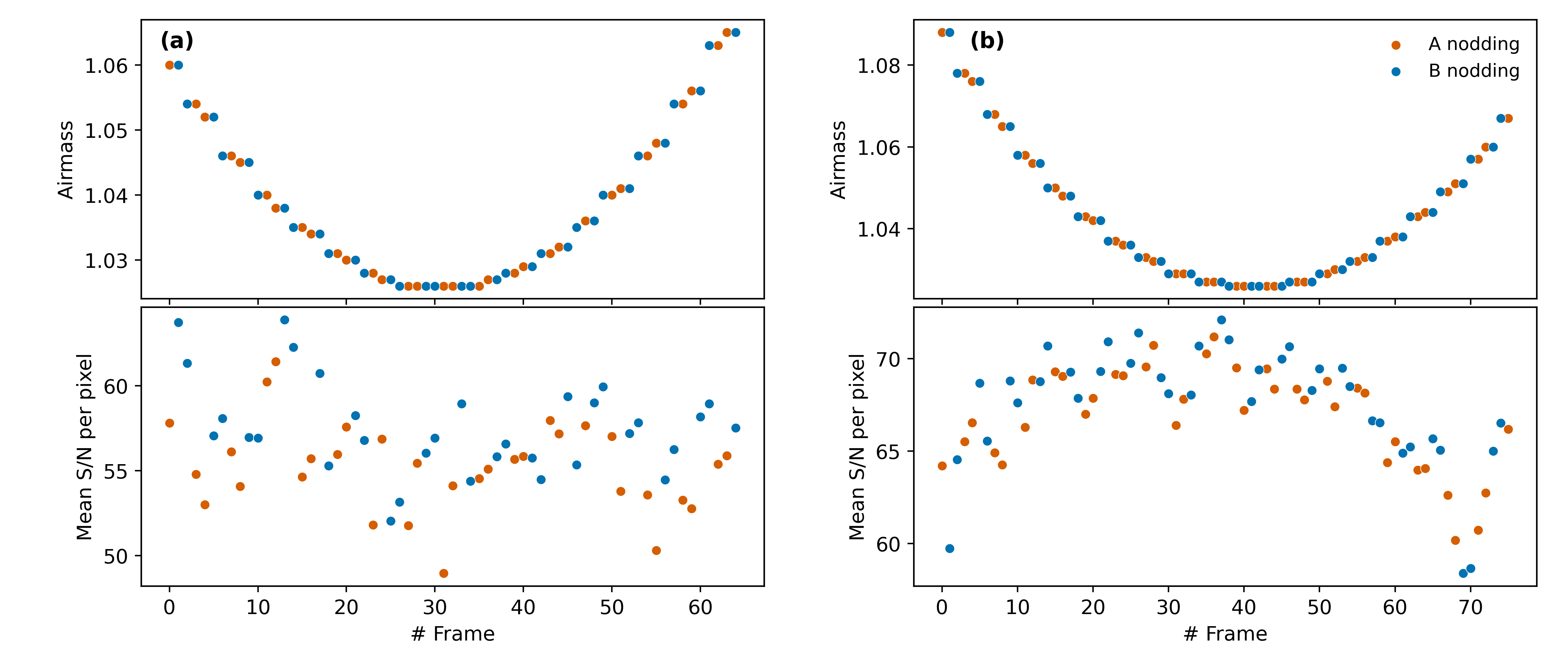}
    \caption{
    Evolution of the airmass and the mean S/N per detector pixel for the transit observation (panel a) and the thermal-emission observation (panel b).
    Upper sub-panels: Evolution of the airmass for each observing sequence. Lower sub-panels: Corresponding mean S/N per detector pixel. Data points corresponding to the A and B nodding positions are shown in orange and blue, respectively.
    }
    \label{fig:airmass_snr}
\end{figure}

\begin{figure}[!ht]
    \centering
    \includegraphics[width=\textwidth]{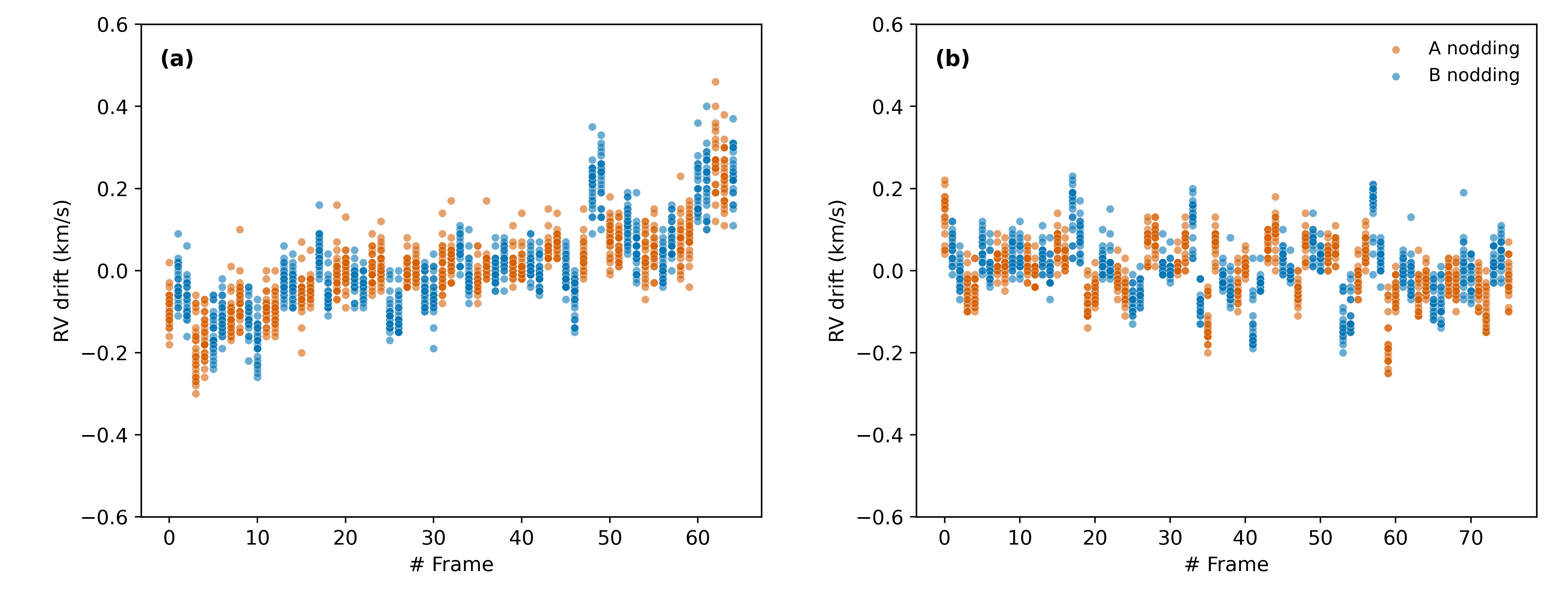}
    \caption{
    RV drift during the acquisition of the spectral time series.
    Each data point corresponds to an individual exposure frame and wavelength segment. Panel (a): Transit observation. Panel (b): Dayside thermal-emission observation. Data points corresponding to the A and B nodding positions are shown in orange and blue, respectively.
    }
    \label{fig:rv_drifts}
\end{figure}

\clearpage
\twocolumn

\section{Mathematical formulation of the \texttt{SYSREM} algorithm}
\label{appendix:sysrem}

The \texttt{SYSREM} algorithm \citep{sysrem} models the dominant systematic trends in the spectral matrix as separable components that vary across wavelength and time. In the classical formulation, the systematic component, $\mathbf{M}_{\mathrm{sys}}$, is represented as the outer product of two vectors:

\begin{equation}
\mathbf{M}_{\mathrm{sys}} = c_i a_j = \mathbf{c}\mathbf{a}^{T},
\end{equation}

where $\mathbf{a} \in \mathbb{R}^{M_{\mathrm{pixel}}}$ describes the wavelength-dependent component across spectral pixels, and $\mathbf{c} \in \mathbb{R}^{N_{\mathrm{frame}}}$ represents its temporal variation across frames.

The optimal coefficients are obtained by minimizing the weighted residual sum of squares:

\begin{equation}
S^2 =
\sum_{i,j}
\frac{
\left(r_{ij} - c_i a_j\right)^2
}{
\sigma_{ij}^2
},
\end{equation}

where $r_{ij}$ denotes the residual spectral matrix and $\sigma_{ij}$ is the uncertainty associated with each data point.

The minimization is performed iteratively using an alternating least-squares procedure. First, fixing $a_j$, the coefficients $c_i$ are updated as

\begin{equation}
c_i =
\frac{
\sum_j
\left(
\frac{r_{ij} a_j}{\sigma_{ij}^2}
\right)
}{
\sum_j
\left(
\frac{a_j^2}{\sigma_{ij}^2}
\right)
}.
\end{equation}

Then, fixing $c_i$, the coefficients $a_j$ are updated as

\begin{equation}
a_j =
\frac{
\sum_i
\left(
\frac{r_{ij} c_i}{\sigma_{ij}^2}
\right)
}{
\sum_i
\left(
\frac{c_i^2}{\sigma_{ij}^2}
\right)
}.
\end{equation}

After convergence, the modeled systematic trend is removed from the residual matrix:

\begin{equation}
r_{ij}
\leftarrow
r_{ij} - c_i a_j.
\end{equation}

This procedure can be repeated iteratively to remove additional systematic components.

\onecolumn
\section{Retrieval}

\begin{table}[!htbp] 
    \large
    \renewcommand\arraystretch{1.35}
    \caption{Parameter values from the retrievals.}
    \label{tab:retrieved_parameter}
    \centering

    \begin{tabular}{l c c c c c}
        \hline\hline
        Parameter & Joint retrieval & JWST-only retrieval & CRIRES$^+$ fixed retrieval & Prior range & Unit\\
        \hline
        $T_\mathrm{1}$ & $2600 \pm 1600$ & $2300 \pm 1400$ & 3114\,(fixed) & 100 to 5000 & K\\
        $T_\mathrm{2}$ & $2300^{+1300}_{-1200}$ & $2100^{+1200}_{-1100}$ & 3151\,(fixed) & 100 to 5000 & K\\
        $T_\mathrm{3}$ & $2100^{+1200}_{-1100}$ & $1880^{+1200}_{-950}$ & 3236\,(fixed) & 100 to 5000 & K\\
        $T_\mathrm{4}$ & $1730^{+1100}_{-870}$ & $1710^{+1000}_{-910}$ & 2524\,(fixed) & 100 to 5000 & K\\
        $T_\mathrm{5}$ & $1510^{+690}_{-770}$ & $1530^{+830}_{-820}$ & 1941\,(fixed) & 100 to 5000 & K\\
        $T_\mathrm{6}$ & $1480^{+510}_{-600}$ & $1440^{+640}_{-650}$ & 1494\,(fixed) & 100 to 5000 & K\\
        $T_\mathrm{7}$ & $1710^{+350}_{-260}$ & $1650^{+330}_{-270}$ & 1868\,(fixed) & 100 to 5000 & K\\
        $T_\mathrm{8}$ & $2480^{+130}_{-150}$ & $2460^{+130}_{-120}$ & 2501\,(fixed) & 100 to 5000 & K\\
        $T_\mathrm{9}$ & $2550^{+750}_{-390}$ & $2510^{+680}_{-380}$ & 2672\,(fixed) & 100 to 5000 & K\\
        $T_\mathrm{10}$ & $2600^{+1400}_{-1300}$ & $2500^{+1200}_{-1300}$ & 3005\,(fixed) & 100 to 5000 & K\\
        $\log \,(\mathrm{CO})$ & $-6.2^{+2.9}_{-2.6}$ & $-9.4^{+4.5}_{-4.4}$ & $-3.59^{+0.72}_{-3.8}$ & -10 to 0 & \\
        $\log \,(\mathrm{H_2O})$ & $-4.92^{+0.57}_{-0.30}$ & $-5.06^{+0.48}_{-0.34}$ & $-4.78^{+0.40}_{-0.42}$ & -10 to 0 & \\
        $\log \,(\mathrm{CO_2})$ & ... & ... & $-7.6^{+1.8}_{-1.7}$ & -10 to 0 & \\
        $\log \,(\mathrm{OH})$ & ... & ... & $-6.6^{+2.5}_{-2.3}$ & -10 to 0 & \\
        $\log \,(\mathrm{CH_4})$ & ... & ... & $-7.6 \pm 1.7$ & -10 to 0 & \\
        $\log \,(\mathrm{HCN})$ & ... & ... & $-7.2^{+1.8}_{-1.9}$ & -10 to 0 & \\
        $\log \,(\mathrm{Fe})$ & ... & ... & $-6.6 \pm 2.3$ & -10 to 0 & \\
        $\log(\mathrm{P_\mathrm{cloud}})$ & $-0.10^{+0.90}_{-0.43}$ & $-0.04^{+1.0}_{-0.38}$ & $0.78^{+0.84}_{-0.89}$ & -8 to 2 & $\log(\mathrm{bar})$\\
        $A_\mathrm{g}$ & $0.37 \pm 0.10$ & $0.39^{+0.10}_{-0.11}$ & ... & 0 to 1 & \\
        $\beta$ & $0.82^{+0.00}_{-0.00}$ & ... & $0.82^{+0.00}_{-0.00}$ & 0 to 10 & \\
        $v_{\rm eq}$ & $4.9^{+4.4}_{-3.0}$ & ... & $2.7^{+2.2}_{-1.8}$ & 0 to 20 & $\mathrm{km\,s^{-1}}$\\
        $\mathrm{\Delta} v$ & $-0.6^{+4.6}_{-6.8}$ & ... & 4\,(fixed) & -20 to 20 & $\mathrm{km\,s^{-1}}$\\
        $K_\mathrm{p}$ & $218.2^{+8.9}_{-11}$ & ... & 213\,(fixed) & 100 to 300 & $\mathrm{km\,s^{-1}}$\\
        \hline
    \end{tabular}
\end{table}
    \clearpage

    \begin{figure*}[htbp]
        \centering
        \includegraphics[width=\textwidth]{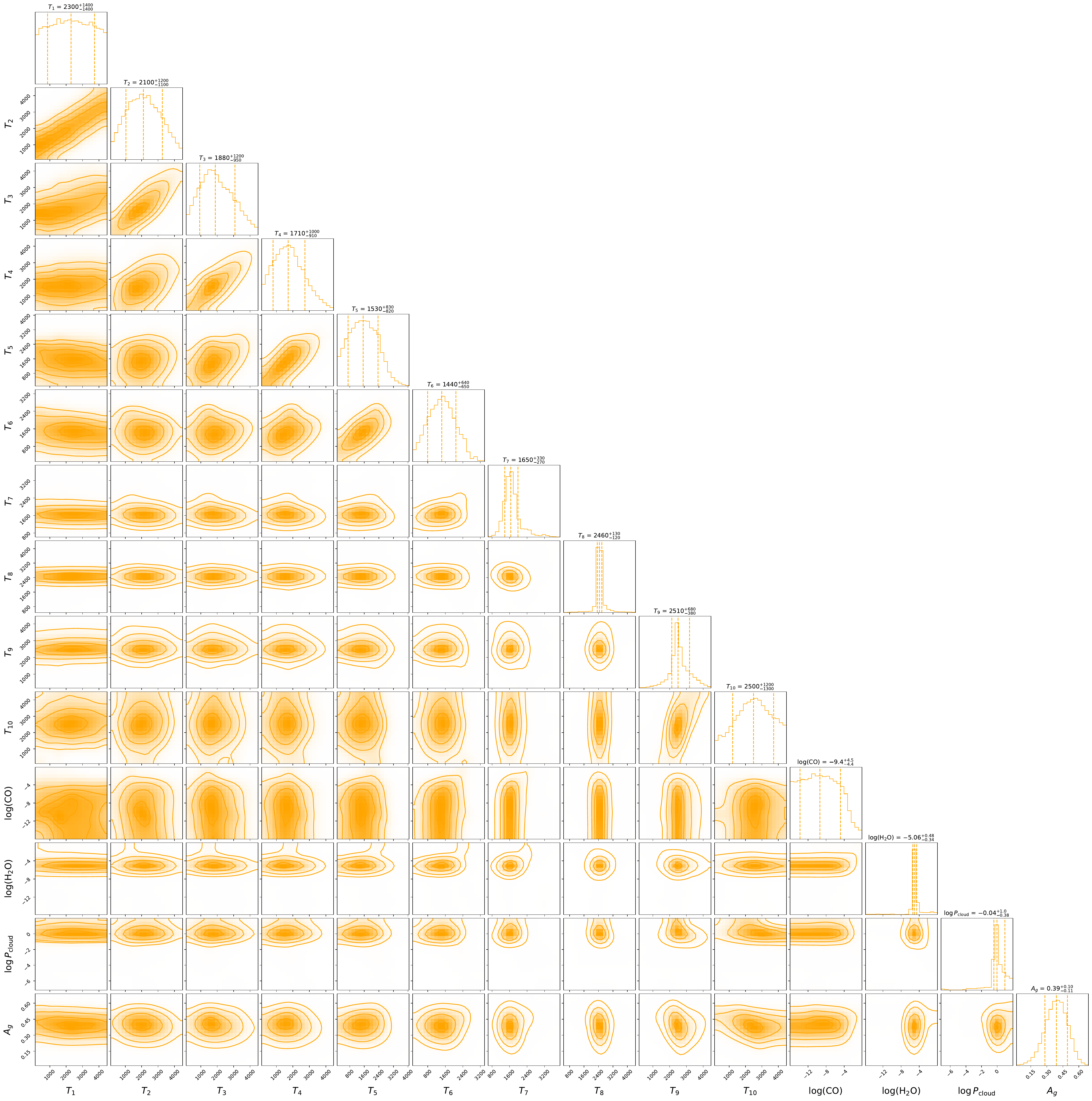}
        \caption{
        Corner plot of the posterior distributions obtained from the retrieval using the JWST NIRISS/SOSS data alone.
        }
        \label{fig:jwst}
    \end{figure*}

    \begin{figure*}[htbp]
        \centering
        \includegraphics[width=\textwidth]{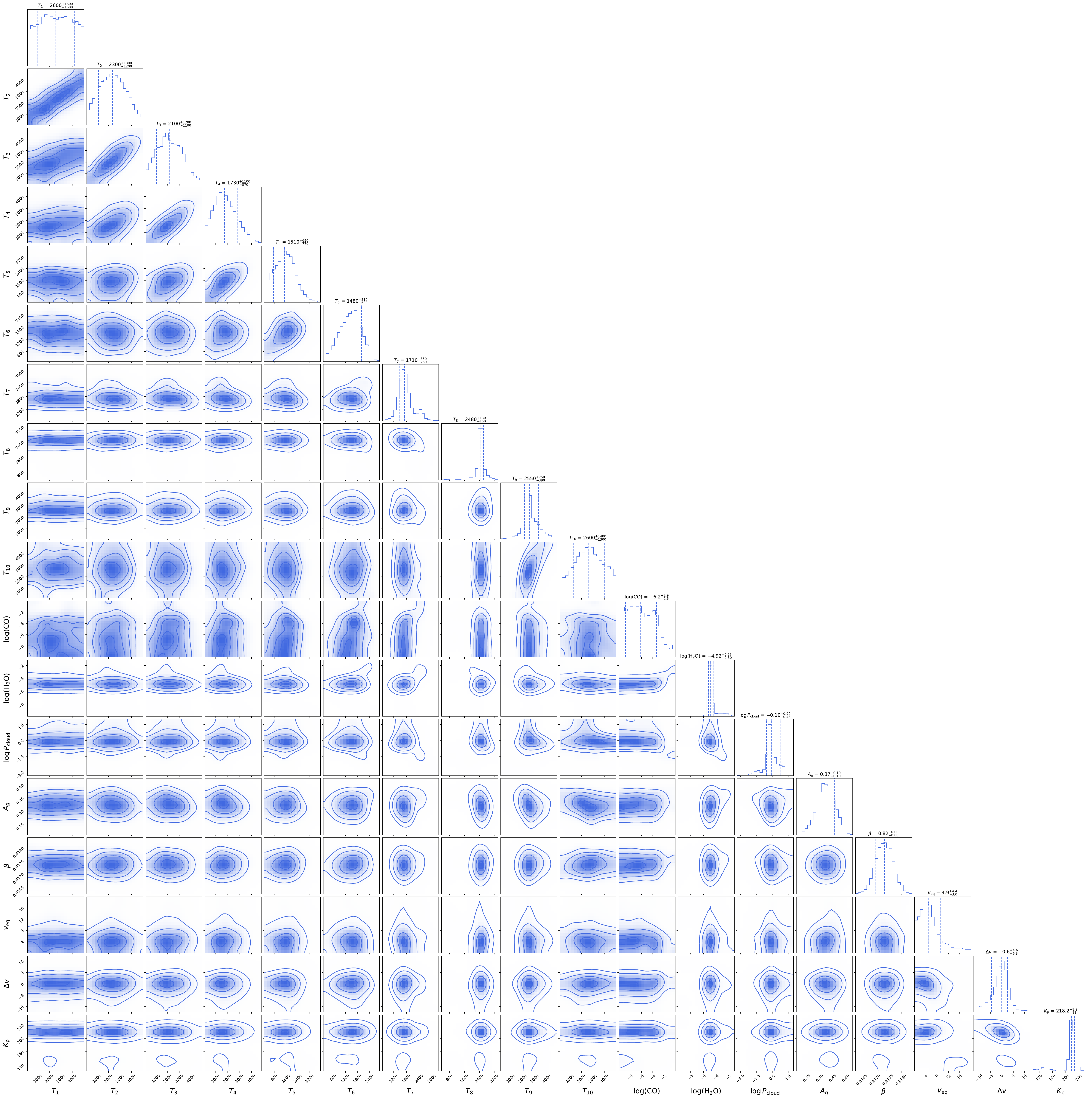}
        \caption{
        Corner plot of the posterior distributions obtained from the joint retrieval.
        }
        \label{fig:joint}
    \end{figure*}
\end{appendix}

\end{document}